\documentclass[a4paper,11pt]{article}
\usepackage{jheppub} % for details on the use of the package, please see the JINST-author-manual
\usepackage{lineno}
\usepackage{tikz}
\usetikzlibrary{decorations.markings, decorations.pathmorphing, arrows, calc, shapes.geometric}
\usepackage{tcolorbox}
\usepackage{cancel}
\usepackage{mathtools}
\usepackage{scalerel}
\usepackage{marginnote}
\newcommand{\negative}{\scalebox{0.75}[1.0]{$-$}}

\newcommand{\avg}[1]{\left\langle#1\right\rangle}
\newcommand{\VolT}{V_{x_{\scaleto{2}{3pt}}y_{\scaleto{2}{3pt}}}}

\newcommand{\Vol}[4]{V_{#1_{\scaleto{#2}{3pt}}#3_{\scaleto{#4}{3pt}}}}
\newcommand{\VolA}[3]{V_{#1 #2_{\scaleto{#3}{3pt}}}}

\definecolor{D1Color}{rgb} {0.960784, 0.756863, 0.752941}
\definecolor{DOverlapColor}{rgb} {0.498,0.498,0.937}
\definecolor{D2Color}{rgb} {0.709,0.98,0.604}
\definecolor{PointColor}{rgb}{0.458824, 0.0784314, 0.0470588}

\def\radius{1pt}

\newcommand{\DiamondOverlapXA}{
\begin{tikzpicture}
[scale=2.0,baseline={([yshift=-.5ex]current bounding box.center)},rounded corners=0.5pt,thick,decoration={
    markings,
    mark=at position 0.5 with {\arrow{>}}}]
    
    \path[clip] (-1,0.9) -- (0.75,0.9) -- (0.75,-0.9) -- (-1,-0.9) -- (-1,0.9);
    
    \def\Diamond{\path (-1/2,0) -- (0,1/2) -- (1/2,0) -- (0,-1/2) --  (-1/2,0)  -- (0,1/2)}
    
   \begin{scope}[scale=1.4]
        \Diamond[fill=D1Color,line width=1.5];
   \end{scope}

    \begin{scope}
        \clip[shift={(0,-1.3)},scale=4] (-1/2,0) -- (0,1/2) -- (1/2,0) -- (0,-1/2) --  (-1/2,0);
        \path[draw,fill=D2Color!70,line width=1.5,shift={(-0.75,0.5)},scale=2] (-1/2,0) -- (0,1/2) -- (1/2,0) -- (0,-1/2) --  (-1/2,0)  -- (0,1/2);
    \end{scope}
    
    \begin{scope}
        \clip[scale=1.4] (-1/2,0) -- (0,1/2) -- (1/2,0) -- (0,-1/2) --  (-1/2,0);
        \path[draw,fill=DOverlapColor,line width=1.5,shift={(-0.75,0.5)},scale=2] (-1/2,0) -- (0,1/2) -- (1/2,0) -- (0,-1/2) --  (-1/2,0)  -- (0,1/2);
    \end{scope}

    \begin{scope}[scale=1.4]
        \Diamond[draw,line width=1.5];
        \path[draw,line width=1.5] (0,1/2) -- (-0.7, 1/2-0.7);
    \end{scope}
    
    \fill[fill=PointColor] (0,1.4*1/2) circle (\radius);
    \fill[fill=PointColor] (0,-1.4*1/2) circle (\radius);
    \fill[fill=PointColor,shift={(-0.75,0.5)}] (0,-1) circle (\radius);

    \node[black] at (0,0.81) {\scriptsize $x$};
    \node[black] at (0,-0.81) {\scriptsize $y_1$};
    \node[black,shift={(-0.75,0.54)}] at (-.35,-0.87) {\scriptsize $y_2$};

\end{tikzpicture}
}

\newcommand{\DiamondOverlapXB}{
\begin{tikzpicture}
[scale=2.0,baseline={([yshift=-.5ex]current bounding box.center)},rounded corners=0.5pt,thick,decoration={
    markings,
    mark=at position 0.5 with {\arrow{>}}}]
    
    \path[draw,fill=D1Color,line width=1.5,scale=1.4] (-1/2,0) -- (0,1/2) -- (1/2,0) -- (0,-1/2) --  (-1/2,0)  -- (0,1/2);
    
    \path[draw,fill=DOverlapColor,line width=1.5,shift={(0,0.32)},scale=0.75] (-1/2,0) -- (0,1/2) -- (1/2,0) -- (0,-1/2) --  (-1/2,0)  -- (0,1/2);
    
    \fill[fill=PointColor] (0,1.4*1/2) circle (\radius);
    \fill[fill=PointColor] (0,-1.4*1/2) circle (\radius);
    \fill[fill=PointColor] (0,-0.05) circle (\radius);

    \node[black] at (0,1.4*0.58-0.01) {\scriptsize $x$};
    \node[black] at (0,-1.4*0.58+0.01) {\scriptsize $y_1$};
    \node[black] at (0.0,-0.15) {\scriptsize $y_2$};

\end{tikzpicture}
}

\newcommand{\DiamondOverlapXC}{
\begin{tikzpicture}
[scale=2.0,baseline={([yshift=-.5ex]current bounding box.center)},rounded corners=0.5pt,thick,decoration={
    markings,
    mark=at position 0.5 with {\arrow{>}}}]
    
    \path[draw,fill=D2Color!70,line width=1.5,scale=1.4] (-1/2,0) -- (0,1/2) -- (1/2,0) -- (0,-1/2) --  (-1/2,0)  -- (0,1/2);
    
    \path[draw,fill=DOverlapColor,line width=1.5,shift={(0,0.32)},scale=0.75] (-1/2,0) -- (0,1/2) -- (1/2,0) -- (0,-1/2) --  (-1/2,0)  -- (0,1/2);
    
    \fill[fill=PointColor] (0,1.4*1/2) circle (\radius);
    \fill[fill=PointColor] (0,-1.4*1/2) circle (\radius);
    \fill[fill=PointColor] (0,-0.05) circle (\radius);

    \node[black] at (0,1.4*0.58-0.01) {\scriptsize $x$};
    \node[black] at (0,-1.4*0.58+0.01) {\scriptsize $y_2$};
    \node[black] at (0.0,-0.15) {\scriptsize $y_1$};

\end{tikzpicture}
}

\newcommand{\DiamondOverlapZeroInsideB}{
\begin{tikzpicture}
[scale=2.0,baseline={([yshift=-.5ex]current bounding box.center)},rounded corners=0.5pt,thick,decoration={
    markings,
    mark=at position 0.5 with {\arrow{>}}}]
    
    \path[draw,fill=D1Color,line width=1.5] (-1/2,0) -- (0,1/2) -- (1/2,0) -- (0,-1/2) --  (-1/2,0)  -- (0,1/2);
    
    \path[draw,fill=D2Color!70,line width=1.5,shift={(0.55,0)}] (-1/2,0) -- (0,1/2) -- (1/2,0) -- (0,-1/2) --  (-1/2,0)  -- (0,1/2);
    
    \begin{scope}
        \clip (-1/2,0) -- (0,1/2) -- (1/2,0) -- (0,-1/2) --  (-1/2,0);
        \path[draw,fill=DOverlapColor,line width=1.5,shift={(0.55,0)}] (-1/2,0) -- (0,1/2) -- (1/2,0) -- (0,-1/2) --  (-1/2,0)  -- (0,1/2);
    \end{scope}
    \begin{scope}
        \clip[shift={(0.55,0)}] (-1/2,0) -- (0,1/2) -- (1/2,0) -- (0,-1/2) --  (-1/2,0);
        \path[draw,line width=1.5] (-1/2,0) -- (0,1/2) -- (1/2,0) -- (0,-1/2) --  (-1/2,0)  -- (0,1/2);
    \end{scope}
    
    \fill[fill=PointColor] (0,1/2) circle (\radius);
    \fill[fill=PointColor] (0,-1/2) circle (\radius);
    \fill[fill=PointColor,shift={(0.55,0)}] (0,1/2) circle (\radius);
    \fill[fill=PointColor,shift={(0.55,0)}] (0,-1/2) circle (\radius);

    \node[black] at (0,0.61) {\scriptsize $x_1$};
    \node[black] at (0,-0.61) {\scriptsize $y_1$};
    \node[black] at (0.55,0.61) {\scriptsize $x_2$};
    \node[black] at (0.55,-0.61) {\scriptsize $y_2$};

\end{tikzpicture}
}

\newcommand{\DiamondOverlapOneInsideD}{
\begin{tikzpicture}
[scale=2.0,baseline={([yshift=-.5ex]current bounding box.center)},rounded corners=0.5pt,thick,decoration={
    markings,
    mark=at position 0.5 with {\arrow{>}}}]
    
    \path[draw,fill=D1Color,line width=1.5,scale=1.4] (-1/2,0) -- (0,1/2) -- (1/2,0) -- (0,-1/2) --  (-1/2,0)  -- (0,1/2);
    
    \path[draw,fill=D2Color!70,line width=1.5,shift={(0.4,0.4)},scale=0.75] (-1/2,0) -- (0,1/2) -- (1/2,0) -- (0,-1/2) --  (-1/2,0)  -- (0,1/2);
    
    \begin{scope}
        \clip[scale=1.4] (-1/2,0) -- (0,1/2) -- (1/2,0) -- (0,-1/2) --  (-1/2,0);
        \path[draw,fill=DOverlapColor,line width=1.5,shift={(0.4,0.4)},scale=0.75] (-1/2,0) -- (0,1/2) -- (1/2,0) -- (0,-1/2) --  (-1/2,0)  -- (0,1/2);
    \end{scope}
    \begin{scope}
        \clip[shift={(0.4,0.4)},scale=0.75] (-1/2,0) -- (0,1/2) -- (1/2,0) -- (0,-1/2) --  (-1/2,0);
        \path[draw,line width=1.5,scale=1.4] (-1/2,0) -- (0,1/2) -- (1/2,0) -- (0,-1/2) --  (-1/2,0)  -- (0,1/2);
    \end{scope}
    
    \fill[fill=PointColor] (0,1.4*1/2) circle (\radius);
    \fill[fill=PointColor] (0,-1.4*1/2) circle (\radius);
    \fill[fill=PointColor,shift={(0.4,0.4)}] (0,0.75*1/2) circle (\radius);
    \fill[fill=PointColor,shift={(0.4,0.4)}] (0,-0.75*1/2) circle (\radius);

    \node[black] at (0,1.4*0.58-0.01) {\scriptsize $x_1$};
    \node[black] at (0,-1.4*0.58+0.01) {\scriptsize $y_1$};
    \node[black,shift={(0.4,0.4)}] at (0.2,0.67) {\scriptsize $x_2$};
    \node[black,shift={(0.4,0.4)}] at (0.2,-0.27) {\scriptsize $y_2$};

\end{tikzpicture}
}

\newcommand{\DiamondOverlapTwoInsideA}{
\begin{tikzpicture}
[scale=2.0,baseline={([yshift=-.5ex]current bounding box.center)},rounded corners=0.5pt,thick,decoration={
    markings,
    mark=at position 0.5 with {\arrow{>}}}]
    
    \path[draw,fill=D1Color,line width=1.5] (-1/2,0) -- (0,1/2) -- (1/2,0) -- (0,-1/2) --  (-1/2,0)  -- (0,1/2);
    
    \path[draw,fill=D2Color!70,line width=1.5,shift={(0,-0.55)}] (-1/2,0) -- (0,1/2) -- (1/2,0) -- (0,-1/2) --  (-1/2,0)  -- (0,1/2);
    
    \begin{scope}
        \clip (-1/2,0) -- (0,1/2) -- (1/2,0) -- (0,-1/2) --  (-1/2,0);
        \path[draw,fill=DOverlapColor,line width=1.5,shift={(0,-0.55)}] (-1/2,0) -- (0,1/2) -- (1/2,0) -- (0,-1/2) --  (-1/2,0)  -- (0,1/2);
    \end{scope}
    \begin{scope}
        \clip[shift={(0, -0.55)}] (-1/2,0) -- (0,1/2) -- (1/2,0) -- (0,-1/2) --  (-1/2,0);
        \path[draw,line width=1.5] (-1/2,0) -- (0,1/2) -- (1/2,0) -- (0,-1/2) --  (-1/2,0)  -- (0,1/2);
    \end{scope}
    
    \fill[fill=PointColor] (0,1/2) circle (\radius);
    \fill[fill=PointColor] (0,-1/2) circle (\radius);
    \fill[fill=PointColor,shift={(0,-0.55)}] (0,1/2) circle (\radius);
    \fill[fill=PointColor,shift={(0,-0.55)}] (0,-1/2) circle (\radius);

    \node[black] at (0,0.61) {\scriptsize $x_1$};
    \node[black] at (0,-0.61) {\scriptsize $y_1$};
    \begin{scope}[shift={(0,-0.55)}]
        \node[black] at (0,0.61) {\scriptsize $x_2$};
        \node[black] at (0,-0.61) {\scriptsize $y_2$};   
    \end{scope}

\end{tikzpicture}
}

\newcommand{\DiamondOverlapTwoInsideC}{
\begin{tikzpicture}
[scale=2.0,baseline={([yshift=-.5ex]current bounding box.center)},rounded corners=0.5pt,thick,decoration={
    markings,
    mark=at position 0.5 with {\arrow{>}}}]
    
    \path[draw,fill=D1Color,line width=1.5,scale=1.4] (-1/2,0) -- (0,1/2) -- (1/2,0) -- (0,-1/2) --  (-1/2,0)  -- (0,1/2);
    
    \path[draw,fill=D2Color!70,line width=1.5,scale=0.75] (-1/2,0) -- (0,1/2) -- (1/2,0) -- (0,-1/2) --  (-1/2,0)  -- (0,1/2);
    
    \begin{scope}
        \clip[scale=1.4] (-1/2,0) -- (0,1/2) -- (1/2,0) -- (0,-1/2) --  (-1/2,0);
        \path[draw,fill=DOverlapColor,line width=1.5,scale=0.75] (-1/2,0) -- (0,1/2) -- (1/2,0) -- (0,-1/2) --  (-1/2,0)  -- (0,1/2);
    \end{scope}
    \begin{scope}
        \clip[scale=0.75] (-1/2,0) -- (0,1/2) -- (1/2,0) -- (0,-1/2) --  (-1/2,0);
        \path[draw,line width=1.5,scale=1.4] (-1/2,0) -- (0,1/2) -- (1/2,0) -- (0,-1/2) --  (-1/2,0)  -- (0,1/2);
    \end{scope}
    
    \fill[fill=PointColor] (0,1.4*1/2) circle (\radius);
    \fill[fill=PointColor] (0,-1.4*1/2) circle (\radius);
    \fill[fill=PointColor] (0,0.75*1/2) circle (\radius);
    \fill[fill=PointColor] (0,-0.75*1/2) circle (\radius);

    \node[black] at (0,1.4*0.58-0.01) {\scriptsize $x_1$};
    \node[black] at (0,-1.4*0.58+0.01) {\scriptsize $y_1$};
    \node[black] at (0.0,0.47) {\scriptsize $x_2$};
    \node[black] at (0.0,-0.47) {\scriptsize $y_2$};

\end{tikzpicture}
}

\newcommand{\DiamondOverlapXYs}{
\begin{tikzpicture}
[scale=0.4,baseline={([yshift=-.5ex]current bounding box.center)},rounded corners=0.5pt,thick,decoration={
    markings,
    mark=at position 0.5 with {\arrow{>}}}]
    
    \path[draw,fill=DOverlapColor,line width=0.4,scale=1.4] (-1/2,0) -- (0,1/2) -- (1/2,0) -- (0,-1/2) --  (-1/2,0)  -- (0,1/2);
    
    \fill[fill=PointColor] (0,1.4*1/2) circle (\radius);
    \fill[fill=PointColor] (0,-1.4*1/2) circle (\radius);
    
\end{tikzpicture}
}

\newcommand{\DiamondOverlapXAs}{
\begin{tikzpicture}
[scale=0.4,baseline={([yshift=-.5ex]current bounding box.center)},rounded corners=0.5pt,thick,decoration={
    markings,
    mark=at position 0.5 with {\arrow{>}}}]
    
    \path[clip] (-1,0.9) -- (0.75,0.9) -- (0.75,-0.9) -- (-1,-0.9) -- (-1,0.9);
    
    \def\Diamond{\path (-1/2,0) -- (0,1/2) -- (1/2,0) -- (0,-1/2) --  (-1/2,0)  -- (0,1/2)}
    
   \begin{scope}[scale=1.4]
        \Diamond[fill=D1Color,line width=0.4];
   \end{scope}

    \begin{scope}
        \clip[shift={(0,-1.3)},scale=4] (-1/2,0) -- (0,1/2) -- (1/2,0) -- (0,-1/2) --  (-1/2,0);
        \path[draw,fill=D2Color!70,line width=0.4,shift={(-0.75,0.5)},scale=2] (-1/2,0) -- (0,1/2) -- (1/2,0) -- (0,-1/2) --  (-1/2,0)  -- (0,1/2);
    \end{scope}
    
    \begin{scope}
        \clip[scale=1.4] (-1/2,0) -- (0,1/2) -- (1/2,0) -- (0,-1/2) --  (-1/2,0);
        \path[draw,fill=DOverlapColor,line width=0.4,shift={(-0.75,0.5)},scale=2] (-1/2,0) -- (0,1/2) -- (1/2,0) -- (0,-1/2) --  (-1/2,0)  -- (0,1/2);
    \end{scope}

    \begin{scope}[scale=1.4]
        \Diamond[draw,line width=0.4];
        \path[draw,line width=0.4] (0,1/2) -- (-0.7, 1/2-0.7);
    \end{scope}
    
    \fill[fill=PointColor] (0,1.4*1/2) circle (\radius);
    \fill[fill=PointColor] (0,-1.4*1/2) circle (\radius);
    \fill[fill=PointColor,shift={(-0.75,0.5)}] (0,-1) circle (\radius);

\end{tikzpicture}
}

\newcommand{\DiamondOverlapXBs}{
\begin{tikzpicture}
[scale=0.4,baseline={([yshift=-.5ex]current bounding box.center)},rounded corners=0.5pt,thin,decoration={
    markings,
    mark=at position 0.5 with {\arrow{>}}}]
    
    \path[draw,fill=D1Color,line width=0.4,scale=1.4] (-1/2,0) -- (0,1/2) -- (1/2,0) -- (0,-1/2) --  (-1/2,0)  -- (0,1/2);
    
    \path[draw,fill=DOverlapColor,line width=0.4,shift={(0,0.32)},scale=0.75] (-1/2,0) -- (0,1/2) -- (1/2,0) -- (0,-1/2) --  (-1/2,0)  -- (0,1/2);
    
    \fill[fill=PointColor] (0,1.4*1/2) circle (\radius);
    \fill[fill=PointColor] (0,-1.4*1/2) circle (\radius);
    \fill[fill=PointColor] (0,-0.05) circle (\radius);

\end{tikzpicture}
}

\newcommand{\DiamondOverlapXCs}{
\begin{tikzpicture}
[scale=0.4,baseline={([yshift=-.5ex]current bounding box.center)},rounded corners=0.5pt,thick,decoration={
    markings,
    mark=at position 0.5 with {\arrow{>}}}]
    
    \path[draw,fill=D2Color!70,line width=0.4,scale=1.4] (-1/2,0) -- (0,1/2) -- (1/2,0) -- (0,-1/2) --  (-1/2,0)  -- (0,1/2);
    
    \path[draw,fill=DOverlapColor,line width=0.4,shift={(0,0.32)},scale=0.75] (-1/2,0) -- (0,1/2) -- (1/2,0) -- (0,-1/2) --  (-1/2,0)  -- (0,1/2);
    
    \fill[fill=PointColor] (0,1.4*1/2) circle (\radius);
    \fill[fill=PointColor] (0,-1.4*1/2) circle (\radius);
    \fill[fill=PointColor] (0,-0.05) circle (\radius);
    
\end{tikzpicture}
}

\newcommand{\DiamondOverlapYAs}{
\begin{tikzpicture}
[scale=0.4,baseline={([yshift=-.5ex]current bounding box.center)},rounded corners=0.5pt,thick,decoration={
    markings,
    mark=at position 0.5 with {\arrow{>}}}]
    
    \path[clip] (-1,0.9) -- (0.75,0.9) -- (0.75,-0.9) -- (-1,-0.9) -- (-1,0.9);
    
    \def\Diamond{\path (-1/2,0) -- (0,1/2) -- (1/2,0) -- (0,-1/2) --  (-1/2,0)  -- (0,1/2)}
    
   \begin{scope}[scale=1.4]
        \Diamond[fill=D1Color,line width=0.4];
   \end{scope}

    \begin{scope}
        \clip[shift={(0,1.3)},scale=4] (-1/2,0) -- (0,1/2) -- (1/2,0) -- (0,-1/2) --  (-1/2,0);
         \path[draw,fill=D2Color!70,line width=0.4,shift={(-0.75,-0.5)},scale=2] (-1/2,0) -- (0,1/2) -- (1/2,0) -- (0,-1/2) --  (-1/2,0)  -- (0,1/2);
    \end{scope}
    
    \begin{scope}
        \clip[scale=1.4] (-1/2,0) -- (0,1/2) -- (1/2,0) -- (0,-1/2) --  (-1/2,0);
        \path[draw,fill=DOverlapColor,line width=0.4,shift={(-0.75,-0.5)},scale=2] (-1/2,0) -- (0,1/2) -- (1/2,0) -- (0,-1/2) --  (-1/2,0)  -- (0,1/2);
    \end{scope}

    \begin{scope}[scale=1.4]
        \Diamond[draw,line width=0.4];
        \path[draw,line width=0.4] (0,-1/2) -- (-0.7, -1/2+0.7);
    \end{scope}
    
    \fill[fill=PointColor] (0,-1.4*1/2) circle (\radius);
    \fill[fill=PointColor] (0,1.4*1/2) circle (\radius);
    \fill[fill=PointColor,shift={(-0.75,-0.5)}] (0,1) circle (\radius);

\end{tikzpicture}
}

\newcommand{\DiamondOverlapYBs}{
\begin{tikzpicture}
[scale=0.4,baseline={([yshift=-.5ex]current bounding box.center)},rounded corners=0.5pt,thick,decoration={
    markings,
    mark=at position 0.5 with {\arrow{>}}}]
    
    \path[draw,fill=D1Color,line width=0.4,scale=1.4] (-1/2,0) -- (0,1/2) -- (1/2,0) -- (0,-1/2) --  (-1/2,0)  -- (0,1/2);
    
    \path[draw,fill=DOverlapColor,line width=0.4,shift={(0,-0.32)},scale=0.75] (-1/2,0) -- (0,1/2) -- (1/2,0) -- (0,-1/2) --  (-1/2,0)  -- (0,1/2);
    
    \fill[fill=PointColor] (0,1.4*1/2) circle (\radius);
    \fill[fill=PointColor] (0,-1.4*1/2) circle (\radius);
    \fill[fill=PointColor] (0,0.05) circle (\radius);
    
\end{tikzpicture}
}

\newcommand{\DiamondOverlapYCs}{
\begin{tikzpicture}
[scale=0.4,baseline={([yshift=-.5ex]current bounding box.center)},rounded corners=0.5pt,thick,decoration={
    markings,
    mark=at position 0.5 with {\arrow{>}}}]
    
    \path[draw,fill=D2Color!70,line width=0.4,scale=1.4] (-1/2,0) -- (0,1/2) -- (1/2,0) -- (0,-1/2) --  (-1/2,0)  -- (0,1/2);
    
    \path[draw,fill=DOverlapColor,line width=0.4,shift={(0,-0.32)},scale=0.75] (-1/2,0) -- (0,1/2) -- (1/2,0) -- (0,-1/2) --  (-1/2,0)  -- (0,1/2);
    
    \fill[fill=PointColor] (0,1.4*1/2) circle (\radius);
    \fill[fill=PointColor] (0,-1.4*1/2) circle (\radius);
    \fill[fill=PointColor] (0,0.05) circle (\radius);    
\end{tikzpicture}
}

\newcommand{\DiamondOverlapZeroInsideAs}{
\begin{tikzpicture}
[scale=0.4,baseline={([yshift=-.5ex]current bounding box.center)},rounded corners=0.5pt,thick,decoration={
    markings,
    mark=at position 0.5 with {\arrow{>}}}]
    
    \path[draw,fill=D1Color,line width=0.4] (-1/2,0) -- (0,1/2) -- (1/2,0) -- (0,-1/2) --  (-1/2,0)  -- (0,1/2);
    
    \path[draw,fill=D2Color!70,line width=0.4,shift={(0.55+0.55,0)}] (-1/2,0) -- (0,1/2) -- (1/2,0) -- (0,-1/2) --  (-1/2,0)  -- (0,1/2);

    \fill[fill=PointColor] (0,1/2) circle (\radius);
    \fill[fill=PointColor] (0,-1/2) circle (\radius);
    \fill[fill=PointColor,shift={(0.55+0.55,0)}] (0,1/2) circle (\radius);
    \fill[fill=PointColor,shift={(0.55+0.55,0)}] (0,-1/2) circle (\radius);
\end{tikzpicture}
}

\newcommand{\DiamondOverlapZeroInsideBs}{
\begin{tikzpicture}
[scale=0.4,baseline={([yshift=-.5ex]current bounding box.center)},rounded corners=0.5pt,thick,decoration={
    markings,
    mark=at position 0.5 with {\arrow{>}}}]
    
    \path[draw,fill=D1Color,line width=0.4] (-1/2,0) -- (0,1/2) -- (1/2,0) -- (0,-1/2) --  (-1/2,0)  -- (0,1/2);
    
    \path[draw,fill=D2Color!70,line width=0.4,shift={(0.55,0)}] (-1/2,0) -- (0,1/2) -- (1/2,0) -- (0,-1/2) --  (-1/2,0)  -- (0,1/2);
    
    \begin{scope}
        \clip (-1/2,0) -- (0,1/2) -- (1/2,0) -- (0,-1/2) --  (-1/2,0);
        \path[draw,fill=DOverlapColor,line width=0.4,shift={(0.55,0)}] (-1/2,0) -- (0,1/2) -- (1/2,0) -- (0,-1/2) --  (-1/2,0)  -- (0,1/2);
    \end{scope}
    \begin{scope}
        \clip[shift={(0.55,0)}] (-1/2,0) -- (0,1/2) -- (1/2,0) -- (0,-1/2) --  (-1/2,0);
        \path[draw,line width=0.4] (-1/2,0) -- (0,1/2) -- (1/2,0) -- (0,-1/2) --  (-1/2,0)  -- (0,1/2);
    \end{scope}
    
    \fill[fill=PointColor] (0,1/2) circle (\radius);
    \fill[fill=PointColor] (0,-1/2) circle (\radius);
    \fill[fill=PointColor,shift={(0.55,0)}] (0,1/2) circle (\radius);
    \fill[fill=PointColor,shift={(0.55,0)}] (0,-1/2) circle (\radius);
\end{tikzpicture}
}

\newcommand{\DiamondOverlapOneInsideAs}{
\begin{tikzpicture}
[scale=0.4,baseline={([yshift=-.5ex]current bounding box.center)},rounded corners=0.5pt,thick,decoration={
    markings,
    mark=at position 0.5 with {\arrow{>}}}]
    
    \path[draw,fill=D2Color!70,line width=0.4,scale=1.4] (-1/2,0) -- (0,1/2) -- (1/2,0) -- (0,-1/2) --  (-1/2,0)  -- (0,1/2);
    
    \path[draw,fill=D1Color,line width=0.4,shift={(0.4,0.4-0.8)},scale=0.75] (-1/2,0) -- (0,1/2) -- (1/2,0) -- (0,-1/2) --  (-1/2,0)  -- (0,1/2);
    
    \begin{scope}
        \clip[scale=1.4] (-1/2,0) -- (0,1/2) -- (1/2,0) -- (0,-1/2) --  (-1/2,0);
        \path[draw,fill=DOverlapColor,line width=0.4,shift={(0.4,0.4-0.8)},scale=0.75] (-1/2,0) -- (0,1/2) -- (1/2,0) -- (0,-1/2) --  (-1/2,0)  -- (0,1/2);
    \end{scope}
    \begin{scope}
        \clip[shift={(0.4,0.4-0.8)},scale=0.8] (-1/2,0) -- (0,1/2) -- (1/2,0) -- (0,-1/2) --  (-1/2,0);
        \path[draw,line width=0.4,scale=1.4] (-1/2,0) -- (0,1/2) -- (1/2,0) -- (0,-1/2) --  (-1/2,0)  -- (0,1/2);
    \end{scope}
    
    \fill[fill=PointColor] (0,1.4*1/2) circle (\radius);
    \fill[fill=PointColor] (0,-1.4*1/2) circle (\radius);
    \fill[fill=PointColor,shift={(0.4,0.4-0.8)}] (0,0.75*1/2) circle (\radius);
    \fill[fill=PointColor,shift={(0.4,0.4-0.8)}] (0,-0.75*1/2) circle (\radius);
\end{tikzpicture}
}

\newcommand{\DiamondOverlapOneInsideBs}{
\begin{tikzpicture}
[scale=0.4,baseline={([yshift=-.5ex]current bounding box.center)},rounded corners=0.5pt,thick,decoration={
    markings,
    mark=at position 0.5 with {\arrow{>}}}]
    
    \path[draw,fill=D2Color!70,line width=0.4,scale=1.4] (-1/2,0) -- (0,1/2) -- (1/2,0) -- (0,-1/2) --  (-1/2,0)  -- (0,1/2);
    
    \path[draw,fill=D1Color,line width=0.4,shift={(0.4,0.4)},scale=0.75] (-1/2,0) -- (0,1/2) -- (1/2,0) -- (0,-1/2) --  (-1/2,0)  -- (0,1/2);
    
    \begin{scope}
        \clip[scale=1.4] (-1/2,0) -- (0,1/2) -- (1/2,0) -- (0,-1/2) --  (-1/2,0);
        \path[draw,fill=DOverlapColor,line width=0.4,shift={(0.4,0.4)},scale=0.75] (-1/2,0) -- (0,1/2) -- (1/2,0) -- (0,-1/2) --  (-1/2,0)  -- (0,1/2);
    \end{scope}
    \begin{scope}
        \clip[shift={(0.4,0.4)},scale=0.75] (-1/2,0) -- (0,1/2) -- (1/2,0) -- (0,-1/2) --  (-1/2,0);
        \path[draw,line width=0.4,scale=1.4] (-1/2,0) -- (0,1/2) -- (1/2,0) -- (0,-1/2) --  (-1/2,0)  -- (0,1/2);
    \end{scope}
    
    \fill[fill=PointColor] (0,1.4*1/2) circle (\radius);
    \fill[fill=PointColor] (0,-1.4*1/2) circle (\radius);
    \fill[fill=PointColor,shift={(0.4,0.4)}] (0,0.75*1/2) circle (\radius);
    \fill[fill=PointColor,shift={(0.4,0.4)}] (0,-0.75*1/2) circle (\radius);
\end{tikzpicture}
}

\newcommand{\DiamondOverlapOneInsideCs}{
\begin{tikzpicture}
[scale=0.4,baseline={([yshift=-.5ex]current bounding box.center)},rounded corners=0.5pt,thick,decoration={
    markings,
    mark=at position 0.5 with {\arrow{>}}}]
    
    \path[draw,fill=D1Color,line width=0.4,scale=1.4] (-1/2,0) -- (0,1/2) -- (1/2,0) -- (0,-1/2) --  (-1/2,0)  -- (0,1/2);
    
    \path[draw,fill=D2Color!70,line width=0.4,shift={(0.4,0.4-0.8)},scale=0.75] (-1/2,0) -- (0,1/2) -- (1/2,0) -- (0,-1/2) --  (-1/2,0)  -- (0,1/2);
    
    \begin{scope}
        \clip[scale=1.4] (-1/2,0) -- (0,1/2) -- (1/2,0) -- (0,-1/2) --  (-1/2,0);
        \path[draw,fill=DOverlapColor,line width=0.4,shift={(0.4,0.4-0.8)},scale=0.75] (-1/2,0) -- (0,1/2) -- (1/2,0) -- (0,-1/2) --  (-1/2,0)  -- (0,1/2);
    \end{scope}
    \begin{scope}
        \clip[shift={(0.4,0.4-0.8)},scale=0.8] (-1/2,0) -- (0,1/2) -- (1/2,0) -- (0,-1/2) --  (-1/2,0);
        \path[draw,line width=0.4,scale=1.4] (-1/2,0) -- (0,1/2) -- (1/2,0) -- (0,-1/2) --  (-1/2,0)  -- (0,1/2);
    \end{scope}
    
    \fill[fill=PointColor] (0,1.4*1/2) circle (\radius);
    \fill[fill=PointColor] (0,-1.4*1/2) circle (\radius);
    \fill[fill=PointColor,shift={(0.4,0.4-0.8)}] (0,0.75*1/2) circle (\radius);
    \fill[fill=PointColor,shift={(0.4,0.4-0.8)}] (0,-0.75*1/2) circle (\radius);
\end{tikzpicture}
}

\newcommand{\DiamondOverlapOneInsideDs}{
\begin{tikzpicture}
[scale=0.4,baseline={([yshift=-.5ex]current bounding box.center)},rounded corners=0.5pt,thick,decoration={
    markings,
    mark=at position 0.5 with {\arrow{>}}}]
    
    \path[draw,fill=D1Color,line width=0.4,scale=1.4] (-1/2,0) -- (0,1/2) -- (1/2,0) -- (0,-1/2) --  (-1/2,0)  -- (0,1/2);
    
    \path[draw,fill=D2Color!70,line width=0.4,shift={(0.4,0.4)},scale=0.75] (-1/2,0) -- (0,1/2) -- (1/2,0) -- (0,-1/2) --  (-1/2,0)  -- (0,1/2);
    
    \begin{scope}
        \clip[scale=1.4] (-1/2,0) -- (0,1/2) -- (1/2,0) -- (0,-1/2) --  (-1/2,0);
        \path[draw,fill=DOverlapColor,line width=0.4,shift={(0.4,0.4)},scale=0.75] (-1/2,0) -- (0,1/2) -- (1/2,0) -- (0,-1/2) --  (-1/2,0)  -- (0,1/2);
    \end{scope}
    \begin{scope}
        \clip[shift={(0.4,0.4)},scale=0.75] (-1/2,0) -- (0,1/2) -- (1/2,0) -- (0,-1/2) --  (-1/2,0);
        \path[draw,line width=0.4,scale=1.4] (-1/2,0) -- (0,1/2) -- (1/2,0) -- (0,-1/2) --  (-1/2,0)  -- (0,1/2);
    \end{scope}
    
    \fill[fill=PointColor] (0,1.4*1/2) circle (\radius);
    \fill[fill=PointColor] (0,-1.4*1/2) circle (\radius);
    \fill[fill=PointColor,shift={(0.4,0.4)}] (0,0.75*1/2) circle (\radius);
    \fill[fill=PointColor,shift={(0.4,0.4)}] (0,-0.75*1/2) circle (\radius);
\end{tikzpicture}
}

\newcommand{\DiamondOverlapTwoInsideAs}{
\begin{tikzpicture}
[scale=0.4,baseline={([yshift=-.5ex]current bounding box.center)},rounded corners=0.5pt,thick,decoration={
    markings,
    mark=at position 0.5 with {\arrow{>}}}]
    
    \path[draw,fill=D1Color,line width=0.4] (-1/2,0) -- (0,1/2) -- (1/2,0) -- (0,-1/2) --  (-1/2,0)  -- (0,1/2);
    
    \path[draw,fill=D2Color!70,line width=0.4,shift={(0,-0.55)}] (-1/2,0) -- (0,1/2) -- (1/2,0) -- (0,-1/2) --  (-1/2,0)  -- (0,1/2);
    
    \begin{scope}
        \clip (-1/2,0) -- (0,1/2) -- (1/2,0) -- (0,-1/2) --  (-1/2,0);
        \path[draw,fill=DOverlapColor,line width=0.4,shift={(0,-0.55)}] (-1/2,0) -- (0,1/2) -- (1/2,0) -- (0,-1/2) --  (-1/2,0)  -- (0,1/2);
    \end{scope}
    \begin{scope}
        \clip[shift={(0, -0.55)}] (-1/2,0) -- (0,1/2) -- (1/2,0) -- (0,-1/2) --  (-1/2,0);
        \path[draw,line width=0.4] (-1/2,0) -- (0,1/2) -- (1/2,0) -- (0,-1/2) --  (-1/2,0)  -- (0,1/2);
    \end{scope}
    
    \fill[fill=PointColor] (0,1/2) circle (\radius);
    \fill[fill=PointColor] (0,-1/2) circle (\radius);
    \fill[fill=PointColor,shift={(0,-0.55)}] (0,1/2) circle (\radius);
    \fill[fill=PointColor,shift={(0,-0.55)}] (0,-1/2) circle (\radius);
\end{tikzpicture}
}

\newcommand{\DiamondOverlapTwoInsideBs}{
\begin{tikzpicture}
[scale=0.4,baseline={([yshift=-.5ex]current bounding box.center)},rounded corners=0.5pt,thick,decoration={
    markings,
    mark=at position 0.5 with {\arrow{>}}}]
    
    \path[draw,fill=D2Color!70,line width=0.4] (-1/2,0) -- (0,1/2) -- (1/2,0) -- (0,-1/2) --  (-1/2,0)  -- (0,1/2);
    
    \path[draw,fill=D1Color,line width=0.4,shift={(0,-0.55)}] (-1/2,0) -- (0,1/2) -- (1/2,0) -- (0,-1/2) --  (-1/2,0)  -- (0,1/2);
    
    \begin{scope}
        \clip (-1/2,0) -- (0,1/2) -- (1/2,0) -- (0,-1/2) --  (-1/2,0);
        \path[draw,fill=DOverlapColor,line width=0.4,shift={(0,-0.55)}] (-1/2,0) -- (0,1/2) -- (1/2,0) -- (0,-1/2) --  (-1/2,0)  -- (0,1/2);
    \end{scope}
    \begin{scope}
        \clip[shift={(0, -0.55)}] (-1/2,0) -- (0,1/2) -- (1/2,0) -- (0,-1/2) --  (-1/2,0);
        \path[draw,line width=0.4] (-1/2,0) -- (0,1/2) -- (1/2,0) -- (0,-1/2) --  (-1/2,0)  -- (0,1/2);
    \end{scope}
    
    \fill[fill=PointColor] (0,1/2) circle (\radius);
    \fill[fill=PointColor] (0,-1/2) circle (\radius);
    \fill[fill=PointColor,shift={(0,-0.55)}] (0,1/2) circle (\radius);
    \fill[fill=PointColor,shift={(0,-0.55)}] (0,-1/2) circle (\radius);
\end{tikzpicture}
}

\newcommand{\DiamondOverlapTwoInsideCs}{
\begin{tikzpicture}
[scale=0.4,baseline={([yshift=-.5ex]current bounding box.center)},rounded corners=0.5pt,thick,decoration={
    markings,
    mark=at position 0.5 with {\arrow{>}}}]
    
    \path[draw,fill=D1Color,line width=0.4,scale=1.4] (-1/2,0) -- (0,1/2) -- (1/2,0) -- (0,-1/2) --  (-1/2,0)  -- (0,1/2);
    
    \path[draw,fill=D2Color!70,line width=0.4,scale=0.75] (-1/2,0) -- (0,1/2) -- (1/2,0) -- (0,-1/2) --  (-1/2,0)  -- (0,1/2);
    
    \begin{scope}
        \clip[scale=1.4] (-1/2,0) -- (0,1/2) -- (1/2,0) -- (0,-1/2) --  (-1/2,0);
        \path[draw,fill=DOverlapColor,line width=0.4,scale=0.75] (-1/2,0) -- (0,1/2) -- (1/2,0) -- (0,-1/2) --  (-1/2,0)  -- (0,1/2);
    \end{scope}
    \begin{scope}
        \clip[scale=0.75] (-1/2,0) -- (0,1/2) -- (1/2,0) -- (0,-1/2) --  (-1/2,0);
        \path[draw,line width=0.4,scale=1.4] (-1/2,0) -- (0,1/2) -- (1/2,0) -- (0,-1/2) --  (-1/2,0)  -- (0,1/2);
    \end{scope}
    
    \fill[fill=PointColor] (0,1.4*1/2) circle (\radius);
    \fill[fill=PointColor] (0,-1.4*1/2) circle (\radius);
    \fill[fill=PointColor] (0,0.75*1/2) circle (\radius);
    \fill[fill=PointColor] (0,-0.75*1/2) circle (\radius);
\end{tikzpicture}
}

\newcommand{\DiamondOverlapTwoInsideDs}{
\begin{tikzpicture}
[scale=0.4,baseline={([yshift=-.5ex]current bounding box.center)},rounded corners=0.5pt,thick,decoration={
    markings,
    mark=at position 0.5 with {\arrow{>}}}]
    
    \path[draw,fill=D2Color!70,line width=0.4,scale=1.4] (-1/2,0) -- (0,1/2) -- (1/2,0) -- (0,-1/2) --  (-1/2,0)  -- (0,1/2);
    
    \path[draw,fill=D1Color,line width=0.4,scale=0.75] (-1/2,0) -- (0,1/2) -- (1/2,0) -- (0,-1/2) --  (-1/2,0)  -- (0,1/2);
    
    \begin{scope}
        \clip[scale=1.4] (-1/2,0) -- (0,1/2) -- (1/2,0) -- (0,-1/2) --  (-1/2,0);
        \path[draw,fill=DOverlapColor,line width=0.4,scale=0.75] (-1/2,0) -- (0,1/2) -- (1/2,0) -- (0,-1/2) --  (-1/2,0)  -- (0,1/2);
    \end{scope}
    \begin{scope}
        \clip[scale=0.75] (-1/2,0) -- (0,1/2) -- (1/2,0) -- (0,-1/2) --  (-1/2,0);
        \path[draw,line width=0.4,scale=1.4] (-1/2,0) -- (0,1/2) -- (1/2,0) -- (0,-1/2) --  (-1/2,0)  -- (0,1/2);
    \end{scope}
    
    \fill[fill=PointColor] (0,1.4*1/2) circle (\radius);
    \fill[fill=PointColor] (0,-1.4*1/2) circle (\radius);
    \fill[fill=PointColor] (0,0.75*1/2) circle (\radius);
    \fill[fill=PointColor] (0,-0.75*1/2) circle (\radius);
\end{tikzpicture}
}

\newcommand{\DiamondOverlapZeroInsideCs}[1]{
\begin{tikzpicture}
[scale=0.4,baseline={([yshift=-.5ex]current bounding box.center)},rounded corners=0.5pt,thick,decoration={
    markings,
    mark=at position 0.5 with {\arrow{>}}}]
    
    \path[draw,fill=D1Color,line width=0.4] (-1/2,0) -- (0,1/2) -- (1/2,0) -- (0,-1/2) --  (-1/2,0)  -- (0,1/2);
    
    \path[draw,fill=D2Color!70,line width=0.4,shift={(0,-1*#1)}] (-1/2,0) -- (0,1/2) -- (1/2,0) -- (0,-1/2) --  (-1/2,0)  -- (0,1/2);

    \fill[fill=PointColor] (0,1/2) circle (\radius);
    \fill[fill=PointColor] (0,-1/2) circle (\radius);
    \fill[fill=PointColor,shift={(0,-1*#1)}] (0,1/2) circle (\radius);
    \fill[fill=PointColor,shift={(0,-1*#1)}] (0,-1/2) circle (\radius);
\end{tikzpicture}
}

\newcommand{\DiamondOverlapZeroInsideC}[1]{
\begin{tikzpicture}
[scale=2.0,baseline={([yshift=-.5ex]current bounding box.center)},rounded corners=0.5pt,thick,decoration={
    markings,
    mark=at position 0.5 with {\arrow{>}}}]
    
    \path[draw,fill=D1Color,line width=1.5] (-1/2,0) -- (0,1/2) -- (1/2,0) -- (0,-1/2) --  (-1/2,0)  -- (0,1/2);
    
    \path[draw,fill=D2Color!70,line width=1.5,shift={(0,-1*#1)}] (-1/2,0) -- (0,1/2) -- (1/2,0) -- (0,-1/2) --  (-1/2,0)  -- (0,1/2);
    
    \fill[fill=PointColor] (0,1/2) circle (\radius);
    \fill[fill=PointColor] (0,-1/2) circle (\radius);
    \fill[fill=PointColor,shift={(0,-1*#1)}] (0,1/2) circle (\radius);
    \fill[fill=PointColor,shift={(0,-1*#1)}] (0,-1/2) circle (\radius);

    \node[black] at (0,0.61*#1) {\scriptsize $y_1$};
    \begin{scope}[shift={(0,-1*#1)}]
        \node[black] at (0.2,0.5*#1) {\scriptsize $y_2$};
        \node[black] at (0,-0.61*#1) {\scriptsize $x$};   
    \end{scope}

\end{tikzpicture}
}

\def\regionOnePath{
    (3.81946, 0.) -- (3.17921, 0.880356) -- (2.89123, 1.84985) -- (2.33689, 2.76657) -- (1.11162, 3.40823) -- (0., 3.41568) -- (-1.08182, 1.99343) -- (-2.34221, 2.58402) -- (-3.04881, 1.72262) -- (-3.27195, 0.476756) -- (-3.76298, 0.) -- (-3.52743, -0.82728) -- (-2.79427, -2.01872) -- (-2.22455, -1.52401) -- (-1.01789, -2.19122) -- (0., -1.41) -- (1.12779, -2.65092) -- (1.88687, -2.3583) -- (2.73412, -0.529107) -- (3.37545, -0.700641) -- cycle
}

\def\regionTwoPath{
   (3.78798, 0.) -- (3.63546, 1.06418) -- (2.96179, 1.86642) -- (2.24901, 3.06697) -- (1.19665, 3.5496) -- (0., 2.81332) -- (-1.11864,  3.12994) -- (-2.33827, 3.17027) -- (-3.2341, 1.64226) -- (-3.68054, 1.04882) -- (-3.85626, 0.) -- (-3.61923, -1.08083) -- (-2.95116, -1.98683) -- (-2.34222, -2.30271) -- (-1.14759, -3.38021) -- (0., -3.53819) -- (1.19231, -2.42098) -- 
(2.32533, -2.46007) -- (3.20926, -2.30163) -- (3.58249, -0.758413) -- cycle
}

\def\regTwoShift{(1.1,0.2)}

\newcommand{\RegionOverlap}{
\begin{tikzpicture}
[scale=2.0,baseline={([yshift=-.5ex]current bounding box.center)},rounded corners=0.5pt,thick,decoration={
    markings,
    mark=at position 0.5 with {\arrow{>}}}]
    
    \path[draw, rounded corners=6pt, fill=D1Color,line width=1.5,scale=0.3]  \regionOnePath;
    
    \path[draw,fill=D2Color!70,rounded corners=5.pt,line width=1.5,shift={\regTwoShift},scale=0.25]  \regionTwoPath;
    
    \begin{scope}
        \clip[scale=0.3, rounded corners=6.pt] \regionOnePath;
        \path[draw,fill=DOverlapColor,line width=1.5, rounded corners=5.pt,shift={\regTwoShift},scale=0.25] \regionTwoPath;
    \end{scope}
    \begin{scope}
        \clip[rounded corners=5.pt,shift={\regTwoShift},scale=0.25] \regionTwoPath;
        \path[draw, rounded corners = 6pt, line width=1.5,scale=0.3] \regionOnePath;
    \end{scope}

    \node[black] at (-0.9, 0.9) {$\mathcal R_1$};
    \node[black] at (2, 0.9) {$\mathcal R_2$};

    \node[black] at (-0.3, 0.15) {$\widetilde{\mathcal R}_a$};
    \node[black] at (0.6, 0.25) {$\widetilde{\mathcal R}_b$};
    \node[black] at (1.4, 0.3) {$\widetilde{\mathcal R}_c$};

\end{tikzpicture}
}

\newcommand{\DiamondOverlapZeroInsideBRegionLabeled}{
\begin{tikzpicture}
[scale=2.5,baseline={([yshift=-.5ex]current bounding box.center)},rounded corners=0.5pt,thick,decoration={
    markings,
    mark=at position 0.5 with {\arrow{>}}}]
    
    \path[draw,fill=D1Color,line width=1.5] (-1/2,0) -- (0,1/2) -- (1/2,0) -- (0,-1/2) --  (-1/2,0)  -- (0,1/2);
    
    \path[draw,fill=D2Color!70,line width=1.5,shift={(0.55,0)}] (-1/2,0) -- (0,1/2) -- (1/2,0) -- (0,-1/2) --  (-1/2,0)  -- (0,1/2);
    
    \begin{scope}
        \clip (-1/2,0) -- (0,1/2) -- (1/2,0) -- (0,-1/2) --  (-1/2,0);
        \path[draw,fill=DOverlapColor,line width=1.5,shift={(0.55,0)}] (-1/2,0) -- (0,1/2) -- (1/2,0) -- (0,-1/2) --  (-1/2,0)  -- (0,1/2);
    \end{scope}
    \begin{scope}
        \clip[shift={(0.55,0)}] (-1/2,0) -- (0,1/2) -- (1/2,0) -- (0,-1/2) --  (-1/2,0);
        \path[draw,line width=1.5] (-1/2,0) -- (0,1/2) -- (1/2,0) -- (0,-1/2) --  (-1/2,0)  -- (0,1/2);
    \end{scope}
    
    \fill[fill=PointColor] (0,1/2) circle (\radius);
    \fill[fill=PointColor] (0,-1/2) circle (\radius);
    \fill[fill=PointColor,shift={(0.55,0)}] (0,1/2) circle (\radius);
    \fill[fill=PointColor,shift={(0.55,0)}] (0,-1/2) circle (\radius);

    \node[black] at (0,0.61) {\scriptsize $x_1$};
    \node[black] at (0,-0.61) {\scriptsize $y_1$};
    \node[black] at (0.55,0.61) {\scriptsize $x_2$};
    \node[black] at (0.55,-0.61) {\scriptsize $y_2$};

    \node[black] at (-0.13, 0) {\scriptsize $\widetilde{\mathcal  R}_a$};
    \node[black] at (0.3, 0) {\scriptsize $\widetilde{\mathcal R}_b$};
    \node[black] at (0.68, 0) {\scriptsize $\widetilde{\mathcal R}_c$};

    \node[black] at (-0.4, 0.34) {\footnotesize $\mathcal R_1$}; 
    \node[black] at (0.9, 0.34) {\footnotesize $\mathcal R_2$};

\end{tikzpicture}
}

\newcommand{\DiamondOverlapOneInsideARegionLabled}{
\begin{tikzpicture}
[scale=2.4,baseline={([yshift=-.5ex]current bounding box.center)},rounded corners=0.5pt,thick,decoration={
    markings,
    mark=at position 0.5 with {\arrow{>}}}]
    
    \path[draw,fill=D2Color!70,line width=1.5,scale=1.4] (-1/2,0) -- (0,1/2) -- (1/2,0) -- (0,-1/2) --  (-1/2,0)  -- (0,1/2);
    
    \path[draw,fill=D1Color,line width=1.5,shift={(0.4,0.4-0.8)},scale=0.75] (-1/2,0) -- (0,1/2) -- (1/2,0) -- (0,-1/2) --  (-1/2,0)  -- (0,1/2);
    
    \begin{scope}
        \clip[scale=1.4] (-1/2,0) -- (0,1/2) -- (1/2,0) -- (0,-1/2) --  (-1/2,0);
        \path[draw,fill=DOverlapColor,line width=1.5,shift={(0.4,0.4-0.8)},scale=0.75] (-1/2,0) -- (0,1/2) -- (1/2,0) -- (0,-1/2) --  (-1/2,0)  -- (0,1/2);
    \end{scope}
    \begin{scope}
        \clip[shift={(0.4,0.4-0.8)},scale=0.8] (-1/2,0) -- (0,1/2) -- (1/2,0) -- (0,-1/2) --  (-1/2,0);
        \path[draw,line width=1.5,scale=1.4] (-1/2,0) -- (0,1/2) -- (1/2,0) -- (0,-1/2) --  (-1/2,0)  -- (0,1/2);
    \end{scope}
    
    \fill[fill=PointColor] (0,1.4*1/2) circle (\radius);
    \fill[fill=PointColor] (0,-1.4*1/2) circle (\radius);
    \fill[fill=PointColor,shift={(0.4,0.4-0.8)}] (0,0.75*1/2) circle (\radius);
    \fill[fill=PointColor,shift={(0.4,0.4-0.8)}] (0,-0.75*1/2) circle (\radius);

    \node[black] at (0,1.4*0.58-0.02) {\scriptsize $x_2$};
    \node[black] at (0,-1.4*0.58+0.01) {\scriptsize $y_2$};
    \node[black,shift={(0.5,0.4-2*0.93)}] at (0.2,0.67) {\scriptsize $x_1$};
    \node[black,shift={(0.5,0.4-2*0.93)}] at (0.2,-0.27) {\scriptsize $y_1$};

    \node[black] at (-0.0, 0) {\scriptsize $\widetilde{\mathcal  R}_c$};
    \node[black] at (0.3, -0.25) {\scriptsize $\widetilde{\mathcal R}_b$};
    \node[black] at (0.5, -0.45) {\scriptsize $\widetilde{\mathcal R}_a$};

    \node[black] at (-0.6, 0.34) {\footnotesize $\mathcal R_2$}; 
    \node[black] at (0.95, -0.4) {\footnotesize $\mathcal R_1$}; 
    
\end{tikzpicture}
}

\newcommand{\DiamondWithzInsideInRegion}{
\begin{tikzpicture}
[scale=2,baseline={([yshift=-.5ex]current bounding box.center)},rounded corners=0.5pt,thick,decoration={
    markings,
    mark=at position 0.5 with {\arrow{>}}}]
    
    \path[draw,fill=D2Color!70,line width=1.5,scale=1.4] (-1/2,0) -- (0,1/2) -- (1/2,0) -- (0,-1/2) --  (-1/2,0)  -- (0,1/2);

    \path[draw,fill=DOverlapColor,rounded corners=0.7pt,line width=1.0,shift={(0.2,0.07)},scale=0.025]  \regionTwoPath;

    \fill[fill=PointColor] (0,1.4*1/2) circle (\radius);
    \fill[fill=PointColor] (0,-1.4*1/2) circle (\radius);
    \fill[fill=PointColor,shift={(0.2,0.4-0.7)}] (0,0.75*1/2) circle (\radius);

    \node[black] at (0,1.4*0.58-0.01) {\scriptsize $x$};
    \node[black] at (0,-1.4*0.58) {\scriptsize $y$};
    \node[black,shift={(-0.2,-0.8)}] at (0.2,0.67) {\footnotesize $\delta R_z$};
 
    \node[black] at (-0.6, 0.34) {\footnotesize $\mathcal R$}; 
\end{tikzpicture}
}

\newcommand{\DiamondWithzInside}[3]{
\begin{tikzpicture}
[scale=2,baseline={([yshift=-.5ex]current bounding box.center)},rounded corners=0.5pt,thick,decoration={
    markings,
    mark=at position 0.5 with {\arrow{>}}}]
    
    \path[draw,fill=D2Color!70,line width=1.5,scale=1.4] (-1/2,0) -- (0,1/2) -- (1/2,0) -- (0,-1/2) --  (-1/2,0)  -- (0,1/2);

    \fill[fill=PointColor] (0,1.4*1/2) circle (\radius);
    \fill[fill=PointColor] (0,-1.4*1/2) circle (\radius);
    \fill[fill=PointColor,shift={(0.2 + #1,0.4-0.7 + #2)}] (0,0.75*1/2) circle (\radius);

    \node[black] at (0,1.4*0.58-0.01) {\footnotesize $x_2$};
    \node[black] at (0,-1.4*0.58) {\footnotesize $y_2$};
    \node[black,shift={(-0.2,-0.8)}] at (0.3 + #1,0.6 + #2) {\footnotesize $x_1$};

    \node[black] at (-0.9,0) {\footnotesize $(#3)$};
\end{tikzpicture}
}

\newcommand{\JJExampleDiagram}{
\begin{tikzpicture}
[scale=2.0,baseline={([yshift=-.5ex]current bounding box.center)},rounded corners=0.5pt,thick,decoration={
    markings,
    mark=at position 0.5 with {\arrow{>}}}]

    \path[draw, line width=1.5,->] (-1.8,0) -- (1.8,0);
    \path[draw, line width=1.5,->] (0,0) -- (0,3);

    \begin{scope}
    \clip (0,0) -- (-2,2) -- (0,4) -- (2,2) -- (0,0);
    \path[draw,fill=D1Color!2,line width=0.0,shift={(-0.5,1.5)}] (-4,-4) -- (0,0) -- (4,-4);
    
    \path[draw,fill=D2Color!2,line width=0.0,shift={(0.5,1.5)}] (-4,-4) -- (0,0) -- (4,-4);

    \path[draw,fill=blue!60,line width=1.5,shift={(-0.0,1.0)}] (-4,-4) -- (0,0) -- (4,-4);

    \path[draw, orange, line width=1.5,shift={(0.5,1.5)}] (-4,-4) -- (0,0);
    \path[draw, orange, line width=1.5,shift={(-0.5,1.5)}] (4,-4) -- (0,0);
    \end{scope}

    \path[draw, line width=1.5] (0,0) -- (-1.5,1.5) -- (0,3) -- (1.5,1.5) -- (0,0);

    \path[draw, red, line width=1.5,->] (0,0) -- (0.8,0.8);
    \path[draw, red, line width=1.5,->] (0,0) -- (-0.8,0.8);
    
    \fill[fill=PointColor] (0,0) circle (\radius); % P
    \fill[fill=PointColor] (0,3) circle (\radius); % Q

    \fill[fill=PointColor] (-0.5,1.5) circle (\radius); % x1
    \fill[fill=PointColor] (0.5,1.5) circle (\radius); % x2

    \node[black] at (0,-0.15) {\scriptsize $p$};
    \node[black] at (0.0,3.15) {\scriptsize $q$};
    \node[black] at (-0.5,1.61) {\scriptsize $x_1$};
    \node[black] at (0.5,1.61) {\scriptsize $x_2$};

    \node[black] at (0.11,1.5) {\scriptsize $t$};
    \path[draw, line width=1.0] (-0.05,1.5) -- (0.05, 1.5); % tick

    \node[black] at (0.2,1.) {\tiny $t-z$};

    \node[red] at (0.75,0.6) {\scriptsize $v$};
    \node[red] at (-0.75,0.6) {\scriptsize $u$};

    \path[draw, dashed, thin] (0.5,1.5) -- (0.5, 0.0); % dash line
    \node[black] at (0.5, -0.1) {\scriptsize $z$};

\end{tikzpicture}
}

\title{Fluctuations and Correlations in Causal Set Theory}
\author[a]{Heidar Moradi,}
\author[b,c]{Yasaman~K.~Yazdi}
\author[d]{and Miguel~Zilh\~{a}o}

\affiliation[a]{Physics and Astronomy, Division of Natural Sciences, University of Kent, Canterbury CT2
7NZ, United Kingdom}
\affiliation[b]{School of Theoretical Physics, Dublin Institute for Advanced Studies, 10 Burlington Road, \newline Dublin 4, Ireland.}
\affiliation[c]{Theoretical  Physics  Group, Blackett  Laboratory,  Imperial  College London, SW7 2AZ, \newline United Kingdom}
\affiliation[d]{Centre for Research and Development in Mathematics and Applications (CIDMA),\\ 
Department of Mathematics, University of Aveiro, 3810-193 Aveiro, Portugal}

\emailAdd{h.moradi@kent.ac.uk, ykyazdi@stp.dias.ie, mzilhao@ua.pt}

\abstract{We study the statistical fluctuations (such as the variance) of causal set quantities, with particular focus on the causal set action. To facilitate calculating such fluctuations, we develop tools to account for correlations between causal intervals with different cardinalities. We present a convenient decomposition of the fluctuations of the causal set action into contributions that depend on different kinds of correlations. This decomposition can be used in causal sets approximated by any spacetime manifold $\mathcal M$. Our work paves the way for investigating a number of interesting discreteness effects, such as certain aspects of the Everpresent $\Lambda$ cosmological model.
}

\makeatletter
\gdef\@fpheader{}
\makeatother

\makeatletter
\newcommand{\boxLabel}[2]{%
   \protected@write \@auxout {}{\string \newlabel {#1}{{#2}{\thepage}{#2}{#1}{}} }%
   \hypertarget{#1}{#2}
}
\makeatother
\begin{document}
\maketitle
\flushbottom

\section{Introduction}
Causal set theory proposes that spacetime is fundamentally discrete and the causal relations among the discrete elements play a prominent role in the physics \cite{PhysRevLett.59.521, Surya:2019ndm}. Fundamental discreteness of spacetime is a promising idea for understanding the finiteness of black hole entropy \cite{Sorkin:2005qx} and resolving the UV divergences of quantum field theories and spacetime singularities. Discreteness is also known to avoid other subtleties such as quantum anomalies~\cite{Nielsen:1983rb, Nielsen:1980rz, Nielsen:1981xu, Nielsen:1981hk, Kravec:2013pua}.
A causal set is made up of elements of spacetime, presumed to be roughly a minimum Planck distance apart, with nothing in between them. This stark difference with continuum spacetime requires us to rethink how familiar smooth quantities can emerge. It also offers new possibilities to probe unknown physics as well as currently unexplained phenomena.

Progress has been made in recognizing and understanding how some continuumlike features can emerge from causal sets at macroscopic scales, i.e., when the number of elements is large. An important result in this direction is that a causal set is well approximated by a continuum spacetime if there is a number-volume correspondence between the causal set and spacetime. This occurs when the number of elements $N$ within an arbitrary spacetime region with spacetime volume $V$ is statistically  proportional to $V$ \cite{Saravani:2014gza}. Such a correspondence is not guaranteed to exist for any causal set, and in fact it does not exist for a large class of causal sets which are said to be \emph{non-manifoldlike} (see e.g.~\cite{Henson:2015fha, Carlip:2024uny}). On the other hand, such a correspondence is guaranteed (with minimal variance~\cite{Saravani:2014gza}) when  the number of causal set elements is randomly distributed according to the Poisson distribution. The proportionality constant between the mean number $\langle N\rangle$ and volume $V$ is the discreteness scale $\rho$, namely $\langle N\rangle= \rho V$,  and the standard deviation $\Delta N=\sqrt{\rho V}$ according to the Poisson distribution. The number-volume correspondence makes it possible to compare discrete quantities with continuum quantities, and the Poisson distribution and the statistics it induces are at the heart of this subject.

As alluded to above, the number-volume correspondence is not exact. There are statistical fluctuations away from the mean. For small deviations from the mean, these fluctuations can be regarded as corrections to continuum geometric quantities and their functions. At the very least, these fluctuations are discrete but still classical corrections. There are also, however, hints as to how such discreteness and fluctuations may relate to and affect quantum phenomena. For example, fundamental discreteness of the kind inherent in causal sets has led to new\footnote{These quantum field theories are defined with respect to spacetime rather than spatial foliations.} and UV regular formulations of quantum field theory~\cite{Johnston:2009fr, X:2023ewv, Albertini:2024srq} and entropy~\cite{Sorkin:2012sn, Yazdi:2022hhv, Homsak:2024tce} on a background causal set. This includes a class of nonlocal d'Alembertians~\cite{Box2d, BDAction}. There is evidence that the means  of these d'Alembertians in causal sets approximated by curved spacetimes agree with the usual continuum d'Alembertian \emph{plus} a term approximated by the Ricci scalar curvature and certain boundary terms. The latter term has been used to define a causal set analogue of the Einstein-Hilbert gravitational action~\cite{BDAction}. 

In a different context and in much earlier work, a cosmological model known as Everpresent $\Lambda$ \cite{originallambda, sorkin1994role, ahmed2004everpresent, Das:2023hbw, Yazdi:2023scl} was introduced whose crucial ingredient is the statistical deviation from the number-volume correspondence. Everpresent $\Lambda$ is a model for a fluctuating cosmological constant which is a direct consequence of the standard deviation of the Poisson distribution and a quantum uncertainty relation between $\Lambda$ and the spacetime volume.  It further takes the quantum uncertainty in $V$ to be the statistical fluctuation in $V$ due to discreteness, thereby implicitly linking the Poisson distribution of the number-volume correspondence to a quantum origin. The model correctly predicts the value of the cosmological constant today and is a candidate solution to current cosmological tensions \cite{Abdalla:2022yfr, Perivolaropoulos:2021jda, Vagnozzi:2023nrq}, making its study especially timely. In this model the value of the cosmological constant fluctuates over cosmic history, but the precise correlation timescale and dynamics governing these fluctuations is not yet known. The dynamical nature of dark energy in models of Everpresent $\Lambda$ make it especially promising as a candidate solution to the Hubble tension \cite{DiValentino:2021izs}. The recent DESI \cite{DESI:2024mwx} observations also favor dynamical dark energy models.

The causal set action mentioned earlier, defined using the nonlocal and discrete d'Alem\-bertian, also has fluctuations about the mean value. A natural and interesting question is: could there be an Everpresent $\Lambda$ term in the fluctuations of the causal set action? In other words, could the fluctuations of the causal set action be the source of Everpresent $\Lambda$? It is also natural to expect there to be a connection between the discrete gravitational action and the cosmological constant, in close analogy with a cosmological constant term often appearing in the continuum gravitational action. While the scope of our work is wider than applications to the causal set action, our focus will be the fluctuations of the action as it paves the way for studying this question. There are several open questions regarding Everpresent $\Lambda$\footnote{One such open question is how to determine the mean about which fluctuations occur. In current models, the mean is assumed to be $0$, but this choice still needs to be motivated. Another open question is how to incorporate spatial inhomogeneities, which are needed in any realistic cosmological model in order to explain structure formation. Current models of Everpresent $\Lambda$ only incorporate temporal inhomogeneities. Yet another important open question, which was already briefly mentioned, is what the dynamics governing the fluctuations of $\Lambda$ are and how the fluctuations at different epochs correlate with one another. At present, slightly modified versions of the Friedmann equation(s) are used, but a better understanding of the correct dynamics to use is needed. This is especially important in order to be able to do a complete cosmological perturbation theory, for example to explain the cosmic microwave background data.} which we would be in a better position to answer if the fluctuations of the action could be used to model it. There are currently two phenomenological models of Everpresent $\Lambda$ (known as Model 1 and Model 2) in the literature. If the causal set action can be used to model Everpresent $\Lambda$, it would provide a third phenomenological model that is independent of Models 1 and 2. Moreover, there are important avenues of investigation for building a path integral dynamics of causal sets. Knowledge of the fluctuations of the action would facilitate a more thorough study of these dynamical models as well as their stability characteristics. When we consider the path sum or partition function $\sum_{\mathcal C} e^{iS(\mathcal C)}$, the action that we use will ultimately not be an averaged one, and hence the higher statistics such as the standard deviation will play a role.

The broad aim of this paper is to formalize the approach to studying and calculating fluctuations within causal set theory. Another major aim of this paper is to provide a comprehensive discussion on the topic of correlations within causal set theory. An example of a correlation would be: the correlation between the probability of the past lightcone of one element containing $N_1$ elements and probability of the past lightcone of another element containing $N_2$ elements, when the two elements have partially overlapping pasts. At a practical level, as we will see below, knowledge of these correlations is needed in order to make the calculation of the fluctuations we are interested in more tractable. More generally, however, these correlations give us a deeper understanding of the causal set itself.  The quantities we work with and whose correlations we discuss depend on causal intervals (of different cardinalities) and such intervals are ubiquitous in causal set theory.

This paper is structured as follows. We begin with a review of causal set theory and Poisson sprinkling in Section \ref{sec:Poisson sprinkling}. In Section \ref{sec:Correlations Between Quantities in Different Regions} we define two useful functions (the cardinality indicator  $\zeta$ and the occupation indicator $\chi$) for accounting for correlations. In the same section we also  discuss the three different types of correlations that result from self and cross correlations between these two functions, i.e., $\zeta-\zeta$, $\chi-\chi$, and $\zeta-\chi$ correlations. Section \ref{sec: causal set action} reviews the definition and properties of the causal set action and its potential connection to Everpresent $\Lambda$. Finally, in Section \ref{sec:Fluctuation of action} we use our formalism to set up, simplify, and calculate the fluctuations of the causal set action. We conclude in Section \ref{sec: Conclusions} with a discussion of the utility of our formalism and future directions of this work. There are also three appendices: Appendix \ref{app:CorrelationFunctions} contains explicit expressions for some of the correlation functions that appear in this paper, and Appendix \ref{app: products of chi} contains further details on products of the $\chi$-function. Finally, Appendix \ref{app:example} contains examples of parametrizations of some of the relevant integrals.

\section{Causal Sets and Poisson Sprinkling} \label{sec:Poisson sprinkling}

For causal sets that are well approximated by continuum spacetimes, the Poisson distribution provides a statistical dictionary between causal set quantities and their counterpart continuum quantities. Below we review the basics of causal set theory and the Poisson distribution and show how to use the Poisson distribution to make statistical statements.

\subsection{Causal Set Theory}

The causal set approach to quantum gravity postulates that spacetime is inherently discrete, with the discrete elements and the causal relations between them serving as fundamental building blocks.
Causal sets obey a set of basic principles which we now outline.
A \emph{causal set} (or \emph{causet}) is a set $\mathcal C$ with a partial order relation $\preceq$ that is
\begin{enumerate}
\item Reflexive: for all $x \in \mathcal C$, $x \preceq x$. \label{axiom1}
\item Antisymmetric: for all $x, y \in \mathcal C$, $x \preceq y$ and $y \preceq x$ implies $x = y$. \label{axiom2}
\item Transitive: for all $x, y, z \in \mathcal C$, $x \preceq y$ and $y \preceq z$ implies $x \preceq z$. \label{axiom3}
\item Locally finite: for all $x, z \in \mathcal C$, the cardinality of the set $\{ y \in \mathcal C \, | \, x \preceq y \preceq z \}$ is finite. \label{axiom4}
\end{enumerate}
We write  $x \prec y$ if $x \preceq y$ and $x \neq y$.
The set $\mathcal C$ corresponds to the collection of spacetime elements, while the order relation $\preceq$ signifies the causal precedence relation between these elements. Conditions 1-3 are satisfied by points in continuum spacetimes as well, while condition 4  entails the discreteness of causal sets. 

Classically, the aim of the causal set approach is that, at large scales, a smooth and continuous manifold will emerge from the underlying discrete causal set, and the Einstein equations would be recovered as an effective description (see e.g.~\cite{Sorkin:2003bx, BDAction, Dowker:2021zel} and references therein).
In general, however, it is not always possible to embed a given causal set in a manifold and it is an open question how \emph{manifoldlike} causal sets would emerge dynamically.\footnote{Some progress in this direction was made in \cite{Loomis:2017jhn, Carlip:2022nsv, Carlip:2023zki}.} Setting aside this question, it is known how to generate (kinematically rather than dynamically) manifoldlike causal sets that are approximated by any given Lorentzian manifold of interest. The recipe for doing so involves the Poisson distribution,  and the process of creating a causal set in this way is referred to as performing a Poisson sprinkling (or sprinkling in short). 

In the sprinkling process, one starts with a known spacetime -- described by a manifold $\mathcal{M}$ with Lorentzian metric $g_{ab}$ -- and according to a \emph{Poisson process} randomly places points in the given spacetime such that the number of elements in any arbitrary region with spacetime volume $V$ follows a Poisson distribution. We elaborate on the properties of the Poisson distribution  in the next subsection. This uniform but random distribution of elements according to the sprinkling is in some sense the most strategic placement of a finite number of elements within a continuum spacetime in order to sample arbitrary volumes well. The sprinkled elements are then endowed with the causal relations according to the causal structure of the continuum spacetime into which they were sprinkled, thus satisfying conditions 1-4 above. 

\subsection{Poisson Sprinkling}

Having established the importance of the Poisson distribution in generating manifoldlike causal sets, let us take a closer look at the definition of this distribution and its consequences for causal set sprinklings produced by it.

As mentioned above, a sprinkling generates a causal set that is well approximated by a continuum spacetime by placing points at random in that spacetime manifold via a Poisson process\footnote{One can think of the Poisson process as  dividing spacetime into small subregions with volume $dV$ and placing at most one point in each  subregion, with probability $\rho dV$. The probability to place a point in each subregion is \emph{independent of other subregions}. Then, the probability of there being $i$ points in a region with volume $V$
is
$P_i(V) = \begin{pmatrix}V/dV \\ i
\end{pmatrix}\,\left(\rho dV\right)^i\,(1-\rho dV)^{V/dV - i}$
which becomes \eqref{poisson_dist} when $dV \rightarrow 0$, according to the Poisson limit theorem \cite{papoulis2002probability, petrov1995limit}.  $\begin{pmatrix}V/dV \\ i
\end{pmatrix}$ is the number of ways to pick $i$ subregions from the total number of subregions which is $V/dV$, $\left(\rho dV\right)^i$ is the probability to have an  element in those $i$ subregions, and the last factor is the probability to not have elements in the remaining $V/dV-i$ subregions. Note that the probability in a submanifold will also converge to the Poisson distribution due to the independence of the point placements in this Poisson process.} such that the probability to find $i$ elements in a region $\mathcal R$ with spacetime volume $V_\mathcal R$ is given by the Poisson distribution \cite{kingman1993poisson}
\begin{equation}
   P_i(V_\mathcal R) = \frac{(\rho V_\mathcal R)^i}{i!}e^{-\rho V_\mathcal R},
\label{poisson_dist}
\end{equation}
where $\rho$ is the average (constant) density of points in the region. We will only work with finite spacetime volumes and regions. This distribution is the defining property of a causal set sprinkling. Much of what we discuss subsequently is either a direct or indirect consequence of \eqref{poisson_dist} being the probability distribution of the number of causal set elements in $V_\mathcal R$. Let us review some of these consequences below.

For any region $\mathcal R\subseteq\mathcal M$, we can define a number operator
\begin{equation}
    \text{Num}_{\mathcal R}(\mathcal C) \equiv \text{Number of elements of $\mathcal C$ in the region $\mathcal R$}.
    \label{numR}
\end{equation}
As a result of and by the definition of the Poisson distribution \eqref{poisson_dist}, the mean and standard deviation of $\text{Num}_{\mathcal R}$ are

\begin{equation}
      \avg{\text{Num}_{\mathcal R}} = \sum_{i=0}^\infty i\,P_i(V_{\mathcal R}) = \rho\, V_{\mathcal R},
      \label{Nmean}
\end{equation}
and 
\begin{equation}
     \Delta \text{Num}_{\mathcal R}
     =\sqrt{\avg{\text{Num}_{\mathcal R}^2}-\avg{\text{Num}_{\mathcal R}}^2} 
     = \sqrt{\sum_{i=0}^\infty i^2\,P_i(V_\mathcal R) - \left(\sum_{i=0}^\infty i\,P_i(V_\mathcal R)\right)^2 }= \sqrt{\rho\, V_{\mathcal R}},
     \label{Nsd}
\end{equation}
 respectively. Hence, the statistics of $\text{Num}_{\mathcal R}$ are straightforward to determine through the Poisson distribution which it follows according to \eqref{poisson_dist}. But what about the statistics of other causal set quantities (in sprinklings)  other than $\text{Num}_{\mathcal R}$? Such quantities will not necessarily be distributed according to the Poisson distribution, but their statistics will nevertheless be a consequence of  $\text{Num}_{\mathcal R}$ following a Poisson distribution. To understand the distributions of more general quantities, it is helpful to reformulate the statistics in terms of an ensemble $\texttt{Sp}[\mathcal M]$ of causal sets produced by repeated Poisson sprinklings (with the same density) of a spacetime manifold $\mathcal M$,
 
 \begin{equation}
    \texttt{Sp}[\mathcal M] = \{\mathcal C_1, \mathcal C_2, \dots\}.
\label{ensemble}
\end{equation}
For simplicity, we assume that the ensemble is very large but finite, such that we capture the properties of the Poisson distribution to a good approximation.\footnote{Note that the ensemble of sprinklings is infinitely large if the Poisson distribution, and some of the properties we will discuss, are to hold exactly.} Given a function $f:\texttt{Sp}[\mathcal M]\longrightarrow\mathbb R$, its average over the ensemble of Poisson sprinklings is\footnote{
For an infinite ensemble, the definition would be
\[
    \avg f \equiv \lim_{n\rightarrow\infty}\frac 1n\sum_{\mathcal C\in\texttt{Sp}^{(n)}[\mathcal M]}f(\mathcal C),
\]
where $\texttt{Sp}^{(n)}[\mathcal M]\equiv\{\mathcal C_1, \cdots\mathcal C_n\}$ is a truncation of the sprinklings to the first $n$ causal sets (according to some ordering).
} 
\begin{equation}
    \label{eq:avg-gen}
    \avg f \equiv \frac 1{|\texttt{Sp}[\mathcal M]|}\sum_{\mathcal C\in\texttt{Sp}[\mathcal M]}f(\mathcal C).
\end{equation}
For example, if $f(\mathcal C)=\text{Num}_{\mathcal R}(\mathcal C)$ in \eqref{eq:avg-gen}, we would recover \eqref{Nmean}. More generally, however, since the probability distributions of generic quantities in sprinklings are not known or do not have closed form expressions, \eqref{eq:avg-gen} describes their statistics.

A useful function which we will use throughout this paper is the Kronecker delta function $\delta_{\text{Num}_{\mathcal R}(\mathcal C),i}$ which evaluates to 1 if there are $i$ elements of the causal set $\mathcal C$ in the spacetime region $\mathcal R$ with spacetime volume $V_\mathcal R$, and evaluates to 0 otherwise. Then the probability of having $i$ elements in the region with volume $V_\mathcal R$ is given by the average of $\delta_{\text{Num}_{\mathcal R},i}$ over the ensemble \eqref{ensemble} 
        \begin{equation}\label{eq:Poisson sprinkling average of delta function}
            \avg{\delta_{\text{Num}_{\mathcal R},i}} = \frac 1{|\texttt{Sp}[\mathcal M]|}\sum_{\mathcal C\in\texttt{Sp}[\mathcal M]}\delta_{\text{Num}_{\mathcal R}(\mathcal C),i} = P_i\left(V_{\mathcal R}\right),
        \end{equation}
i.e.\ the Poisson distribution \eqref{poisson_dist}. Note that from the definitions above, the following operator identities hold
\begin{equation}
    \sum_{i=0}^\infty \,{\delta_{\text{Num}_{\mathcal R}(\mathcal C),i}} = 1,
\end{equation}
 and
\begin{equation}
    \text{Num}_\mathcal R(\mathcal C) = \sum_{i=0}^\infty i\,{\delta_{\text{Num}_{\mathcal R}(\mathcal C),i}}~~.
\end{equation}
The first one implies that $P_i(V_{\mathcal R})$ is normalized, while the second one implies the relation~\eqref{Nmean}.

\section{Correlations Between Quantities in Different Regions} \label{sec:Correlations Between Quantities in Different Regions}
A calculation we will often encounter is to determine the expectation of finding $N_1$ elements in region $\mathcal R_1$ and $N_2$ elements in region $\mathcal R_2$. If $\mathcal R_1$ and $\mathcal R_2$ are overlapping regions, and we will often find that they are, the expectation for there being $N_1$ elements in region $\mathcal R_1$ \emph{depends} on the likelihood of there being $N_2$ elements in region $\mathcal R_2$. As a consequence, to correctly compute the joint expectation, we must take into account the correlation between these two abundances. In the next two subsections we will introduce some notation and machinery that will help us perform computations of such correlation functions. In the third subsection we will apply this machinery to calculate correlations.

\subsection{The $\zeta$-function (Cardinality Indicator)}

In the computation of the fluctuations of the causal set action (see Section \ref{sec:Fluctuation of action}), we will often need an indicator for whether a causal set sprinkling $\mathcal C$ in a particular region $\mathcal R$ has $i$ elements. In this section we will introduce the \textit{Cardinality Indicator} function, $\zeta$, for this purpose. We also explore some of its formal properties, as these will play an important role in our calculations.

The Cardinality Indicator function $\zeta_i$ ($i\geq 0$) for a causal set $\mathcal C$ and region $\mathcal R$ is defined as\footnote{Note that this is merely the delta function we had introduced earlier in \eqref{eq:Poisson sprinkling average of delta function}, but given a new name.}
\begin{equation}
    \zeta^{\mathcal C}_i(\mathcal R) \equiv \delta_{\text{Num}_{\mathcal R}(\mathcal C), i}.
\label{zeta_R}
\end{equation}
It is useful to introduce a special variant when the region is a causal diamond\footnote{Here we define a causal diamond as $I(x,y)=\mathcal J^-(x)\cap \mathcal J^+(y)$, where $\mathcal J^+(x)=\{z|x\prec z\}$ and $\mathcal J^-(x)=\{z|z\prec x\}$. Note that $x\prec y$ also implies $x\neq y$ and therefore $x\not\in \mathcal J^{\pm}(x)$. By abuse of notation, we will often use these regions to either mean submanifolds of $\mathcal M$ or subsets of its sprinkling $\mathcal C$.} $\mathcal R = I(x,y)$

\begin{align}
        \zeta^{\mathcal C}_i(x,y) \equiv 
       \Theta_{x,y}~ \zeta^{\mathcal C}_i(I(x,y)),
       \label{zeta_xy}
\end{align}
where\footnote{Note that in the definition of $\Theta_{x,y}$, $I(x,y)$ is a submanifold of spacetime, i.e.\ not a subset of the causal set.}
\begin{equation}
    \Theta_{x,y} \equiv\begin{cases}
        1 & \text{if }  I(x,y)\not=\emptyset,\\
        0 & \text{otherwise}.
    \end{cases} .
\end{equation}
In other words, \eqref{zeta_xy} tells us if there are $i$ elements in $I(x,y)$ only if the interval $I$ actually exists. Note that in our convention, $x, y\not\in I(x,y)$. If $x$ and $y$ are elements in $\mathcal C$ we can also write 
\begin{gather}\label{eq:definition of zeta funtion}
    \begin{aligned}
        \zeta^{\mathcal C}_i(x,y)
        =\begin{cases}
                            1 & \text{if } y\in \lozenge_{i+1}(x),\\
                            0 & \text{otherwise.}
                    \end{cases},
    \end{aligned}
\end{gather}
where $\lozenge_i(x)$ is the set of elements $y$ that causally precede $x$ ($y\prec x$) and have $i-1$ elements in the causal diamond between $x$ and $y$ (excluding $x$ and $y$ themselves).
%, i.e. $|I(y,x)| = i-1$.
Therefore, $\lozenge_1(x)$ would be a nearest neighbour element. For later convenience we will define $\lozenge_0(x) \equiv \{x\}$, however note that $x\not\in\lozenge_i(x)$ for $i>0$.

With this function we can conveniently define domains such as
\begin{equation}\label{eq:sum diamond to J- using zeta}
    \sum_{y\in\lozenge_{i+1}(x)}f(y) = \sum_{y\in\mathcal J^-(x)}f(y)\,\zeta^{\mathcal C}_i(x, y),
\end{equation}
for $i\geq 0$. This is useful as it is one of the ingredients needed to express sums over causal sets in terms of integrals over manifolds.

\subsubsection{General Properties of $\zeta$}\label{sec: gen properties of zeta}
If a non-empty region $\mathcal R$ is split into $n$ disjoint (possibly empty) subregions\footnote{Any shared boundary between subregions would be considered as part of only one of the subregions.} 
\begin{equation}\label{eq:Decomposition of regions}
    \mathcal R = \bigsqcup_{a=1}^n\mathcal R_a,
\end{equation}
then we have the \textit{Disjoint Decomposition Property}
\begin{equation}\label{zeta_disjoint}
    %\zeta^{\mathcal C}_{i}\left( \mathcal R \right) =
    %
    \zeta^{\mathcal C}_{i}\bigg(\bigsqcup_{a=1}^n\mathcal R_a\bigg) = \sum_{\substack{\alpha_a\geq 0\\\alpha_1 +\dots+ \alpha_n = i}}\prod_{a=1}^n\zeta^{\mathcal C}_{\alpha_a}(\mathcal R_a),
\end{equation}
where the right hand side sums over all possible ways of distributing $i$ elements in the $n$ subregions of $\mathcal R$. Either one of the terms in the sum is $1$ and the rest are $0$, or all are $0$, i.e.\ there is at most one non-zero contribution to the sum \eqref{zeta_disjoint} for a given causal set $\mathcal C$. For example, for two subregions $\mathcal R = \mathcal R_1\sqcup\mathcal R_2$ we have
\begin{gather}
    \begin{aligned}
        \zeta^{\mathcal C}_i(\mathcal R_1\sqcup\mathcal R_2) &= \sum_{\alpha=0}^i \,\zeta^{\mathcal C}_{\alpha}(\mathcal R_1)\zeta^{\mathcal C}_{i-\alpha}(\mathcal R_2), \\
            &= \sum_{\alpha=0}^i \delta_{\text{Num}_{\mathcal R_1}(\mathcal C), \alpha}\,\delta_{\text{Num}_{\mathcal R_2}(\mathcal C), i-\alpha}.
    \end{aligned}
\end{gather}
\\

\noindent Now, we want to address the following problem. Imagine $n$ regions $\mathcal R_1, \dots \mathcal R_n$, which are potentially overlapping. We would like to decompose a product of $\zeta$-functions of the form
\begin{equation}
    \zeta^{\mathcal C}_{i_1}(\mathcal R_1)\cdots \zeta^{\mathcal C}_{i_n}(\mathcal R_n),
\end{equation}
into sums of 
\begin{equation}
    \zeta^{\mathcal C}_{\alpha} \big(\widetilde{\mathcal R}_a\big)
    \zeta^{\mathcal C}_{\beta} \big(\widetilde{\mathcal R}_b\big)
    \zeta^{\mathcal C}_{\gamma} \big(\widetilde{\mathcal R}_c\big)\cdots,
\end{equation}
where the potentially overlapping regions have been decomposed into disjoint regions $\widetilde{\mathcal R}$
\begin{equation}\label{eq: non-overlapping subregions Rtilde}
    \mathcal R_1\cup\cdots\cup \mathcal R_n = \widetilde{\mathcal R}_a \sqcup \widetilde{\mathcal R}_b \sqcup \widetilde{\mathcal R}_c \cdots.
\end{equation}
In order to see how this is done, consider two non-empty regions $\mathcal R_1$ and $\mathcal R_2$ with overlap $\widetilde{\mathcal R}_b = \mathcal R_1\cap\mathcal R_2$. The union can be decomposed into disjoint regions as (see Figure \ref{fig:RegionOverlap})
\begin{equation}\label{eq:R1R1 decomposition}
    \mathcal R_1\cup\mathcal R_2 = \widetilde{\mathcal R}_a\sqcup\widetilde{\mathcal R}_b\sqcup \widetilde{\mathcal R}_c,
\end{equation}
where $\widetilde{\mathcal R}_a = \mathcal R_1\backslash \widetilde{\mathcal R}_b$ and $\widetilde{\mathcal R}_c = \mathcal R_2\backslash \widetilde{\mathcal R}_b$. 
\begin{figure}
    \centering
    \RegionOverlap
    \caption{Two non-empty regions $\mathcal R_1$ and $\mathcal R_2$ with non-empty overlap $\widetilde{\mathcal R}_b = \mathcal R_1\cap\mathcal R_2$. The union $\mathcal R_1\cup\mathcal R_2$ can be decomposed into non-overlapping subregions $\widetilde{\mathcal R}_b$, $\widetilde{\mathcal R}_a = \mathcal R_1\backslash \widetilde{\mathcal R}_b$ and $\widetilde{\mathcal R}_c = \mathcal R_2\backslash \widetilde{\mathcal R}_b$.}
    \label{fig:RegionOverlap}
\end{figure}
Note that some of these regions can be empty\footnote{For example when there is no overlap between regions $\mathcal R_1$ and $\mathcal R_2$.}. %Then we have the identity
%We will also want to compute $\zeta^{\mathcal C}_i(x_1, y_1) \zeta^{\mathcal C}_j(x_2, y_2)$.
The first step is to use the Disjoint Decomposition Property \eqref{zeta_disjoint} for $ \mathcal R_1 = \widetilde{\mathcal R}_a\sqcup\widetilde{\mathcal R}_b $ and $\mathcal R_2 = \widetilde{\mathcal R}_c\sqcup\widetilde{\mathcal R}_b $:
\begin{align}
   \zeta^{\mathcal C}_i(\mathcal R_1)
            &= \sum_{\alpha=0}^i \delta_{\text{Num}_{\widetilde{\mathcal R}_a}(\mathcal C), \alpha}\,\delta_{\text{Num}_{\widetilde{\mathcal R}_b}(\mathcal C), i-\alpha} ,\\
   \zeta^{\mathcal C}_j(\mathcal R_2)
            &= \sum_{\gamma=0}^j \delta_{\text{Num}_{\widetilde{\mathcal R}_c}(\mathcal C), \gamma}\,\delta_{\text{Num}_{\widetilde{\mathcal R}_b}(\mathcal C), j-\gamma}.
\end{align}
Multiplying these two expressions gives us the constraint $i-\alpha = j-\gamma\equiv \beta$, and we can thus write
\begin{gather}\label{eq:Zeta(R1)Zeta(R_2) delta funtion decomposition}
    \begin{aligned}
        \zeta^{\mathcal C}_i(\mathcal R_1) \zeta^{\mathcal C}_j(\mathcal R_2) &=\sum\limits_{\substack{\alpha,\beta,\gamma\geq 0\\ \alpha+\beta=i\\ \beta+\gamma=j }} \delta_{\text{Num}_{\widetilde{\mathcal R}_a}(\mathcal C), \alpha}\,\delta_{\text{Num}_{\widetilde{\mathcal R}_b}(\mathcal C), \beta}\,\delta_{\text{Num}_{\widetilde{\mathcal R}_c}(\mathcal C), \gamma} \\
       & = 
    \sum\limits_{\substack{\alpha,\beta,\gamma\geq 0\\ \alpha+\beta=i\\ \beta+\gamma=j }} \,\zeta^{\mathcal C}_{\alpha}\big(\widetilde{\mathcal R}_a\big)\,\zeta^{\mathcal C}_{\beta}\big(\widetilde{\mathcal R}_b\big)\,\zeta^{\mathcal C}_{\gamma}\big(\widetilde{\mathcal R}_c\big) .
    \end{aligned}
\end{gather}
This is the decomposition we sought. For many overlapping regions, similar decompositions can be derived using the property \eqref{eq:Zeta(R1)Zeta(R_2) delta funtion decomposition}.

\subsection{The $\chi$-function (Occupation Indicator)}
It turns out that besides the $\zeta$-function, we also need the \textit{Occupation Indicator} function, $\chi$, which indicates whether a region $\mathcal R$ contains (\textit{is occupied by}) elements in the sprinkling $\mathcal C$. As we will see below (e.g.~\eqref{eq:sum to integral formula}), this function is useful as it allows us to translate between a continuum manifold and a corresponding sprinkling of it.

 The  $\chi$-function is defined as
\begin{equation}\label{eq:chi_zeta_relation}
    \chi^{\mathcal C}(\mathcal R) \equiv \sum_{i=1}^\infty\zeta^{\mathcal C}_i(\mathcal R) =
                    \begin{cases}
                        1 & \text{if } \text{Num}_{\mathcal R}(\mathcal C)>0,\\
                        0 & \text{otherwise.}
                    \end{cases}
\end{equation}
We are primarily interested in the case where the regions are infinitesimal, $\delta\mathcal R$, and we would like to find an explicit expression for $\chi^{\mathcal C}(\delta \mathcal R)$.

Consider an arbitrary point $x\in\mathcal M$, not necessarily in $\mathcal C$, in the interior of a small region $\Delta \mathcal R_x$ with volume $a^d$. The $\lim_{a\rightarrow 0}\chi^{\mathcal C}(\Delta \mathcal R_x)$ leads to an integration measure $\chi^{\mathcal C}(\delta \mathcal R_x)$ which we can find from
\begin{gather}
    \begin{aligned}
    \chi^{\mathcal C}(\delta \mathcal R_x) \equiv\lim_{a\rightarrow 0}\chi^{\mathcal C}(\Delta \mathcal R_x) &=\lim_{a\rightarrow 0}\sum_{i=1}^{\infty}\frac{\delta_{\text{Num}_{\Delta \mathcal R_x}(\mathcal C),i}}{\Delta V_x}\Delta V_x\\
    &=\lim_{a\rightarrow 0}\frac{\delta_{\text{Num}_{\Delta \mathcal R_x}(\mathcal C),1}}{\Delta V_x}\Delta V_x,
    \end{aligned}
\end{gather}
where in the second line we have used the fact that in the limit $\Delta \mathcal R_x\rightarrow \delta \mathcal R_x$, there can be at most one element in each cell.
The factor $\delta_{\text{Num}_{\Delta \mathcal R_x}(\mathcal C),1}$ is a delta function that checks whether $x$ is in the causal set $\mathcal C$. This can be expressed using Dirac delta functions as
\begin{gather}\label{eq:chi measure written in terms of delta functions}
    \begin{aligned}
    \chi^{\mathcal C}(\delta \mathcal R_x) &= \lim_{a\rightarrow 0}\chi^{\mathcal C}(\Delta \mathcal R_x)
        =\sum_{z\in\mathcal C}\delta^{(d)}(z - x) dV_x.
    \end{aligned}
\end{gather}
By explicit integration, we can readily see that\footnote{Alternatively, this expression can also be derived in a different way. Consider a discretization of a region $\mathcal R\subseteq\mathcal M$, which we will call $\Lambda_a(\mathcal R)$, where $a$ is the lattice spacing. Each cell around the lattice point $x\in \Lambda_a(\mathcal R)$ (Wigner-Seitz cell \cite{WignerSeitz1933}) is denoted  $\Delta \mathcal R_x$, with volume $\Delta V_x=a^d$ in a $d$-dimensional cubic lattice. It is clear that the following relation must hold 
\begin{equation}\label{eq:sum over chi(delta V)}
    \lim_{a\rightarrow 0}\sum_{x\in \Lambda_a(\mathcal R)}\chi^{\mathcal C}(\Delta \mathcal R_x) = \text{Num}_{\mathcal R}(\mathcal C),
\end{equation}
since in the infinitesimal limit $\Delta \mathcal R_x\rightarrow \delta \mathcal R_x$, there can be at most one element in each cell.}
\begin{equation}
    \int_{\mathcal R}\chi^{\mathcal C}(\delta \mathcal R_x) = \text{Num}_\mathcal R(\mathcal C).
    \label{int_chi}
\end{equation}
Essentially, $\chi^{\mathcal C}(\delta \mathcal R_x)$ acts as a microscope that locally detects whether a point $x$ is contained in the sprinkling $\mathcal C$.
This integration measure will be useful for expressing sums of a given Poisson sprinkling of a manifold $\mathcal R\subseteq\mathcal M$ as an integral
\begin{equation}\label{eq:sum to integral formula}
    \int_{\mathcal R}f(x)\chi^{\mathcal C}(\delta \mathcal R_x) = \sum_{z\in\mathcal C}f(z).
\end{equation}
By taking the average of \eqref{int_chi} over many sprinklings and using \eqref{Nmean} we get 
\begin{equation}
   \avg{\int_{\mathcal R}\chi^{\mathcal C}(\delta \mathcal R_x)} = \avg{\text{Num}_\mathcal R(\mathcal C)} = \rho V_{\mathcal R},
\end{equation}
where $V_{\mathcal R}$ is the spacetime volume of $\mathcal R$. In particular, for infinitesimal regions $\mathcal R=\delta\mathcal R_x$ we have 
\begin{equation}\label{mean_chi}
   \avg{\chi^{\mathcal C}(\delta \mathcal R_x)} = \rho\,dV_x,
\end{equation}
where $dV_x$ is the spacetime volume of $\delta\mathcal R_x$.
For simplicity, in the rest of the paper we will often use the notation
\begin{equation}
    \chi^{\mathcal C}(dV_x) \equiv \chi^{\mathcal C}(\delta \mathcal R_x).
\end{equation}

\subsection{Correlation Functions}\label{sec:CorrelationFunctions}
 We are generally interested in computing correlation functions of the form\footnote{From here onwards we omit the superscript $\mathcal{C}$ for notational simplicity.}
\begin{equation}\label{master_correlation}
    \avg{\zeta_{i_1}(x_1,y_1)\cdots \zeta_{i_n}(x_n,y_n)\chi\left(dV_{x_1}\right)\chi\left(dV_{y_1}\right)\cdots \chi\left(dV_{x_n}\right)\chi\left(dV_{y_n}\right)}.
\end{equation}
If all of these functions are uncorrelated, \eqref{master_correlation} will split into
\begin{equation}\label{master_correlation_split}
    \avg{\zeta_{i_1}(x_1,y_1)}\cdots \avg{\zeta_{i_n}(x_n,y_n)}\avg{\chi\left(dV_{x_1}\right)}\avg{\chi\left(dV_{y_1}\right)}\cdots \avg{\chi\left(dV_{x_n}\right)}\avg{\chi\left(dV_{y_n}\right)},
\end{equation}
which can simply be evaluated using
\begin{equation}\label{eq:one point functions of zeta and chi}
    \avg{\zeta_i(x_j,y_j)} = P_i(V_j), \qquad \text{and}\qquad \avg{\chi(dV_x)} = \rho dV_x,
\end{equation}
which were given in \eqref{eq:Poisson sprinkling average of delta function} and \eqref{mean_chi}. Here $V_j$ is the volume of the causal interval $I(x_j,y_j)$ and $P_i(V)$ is \eqref{poisson_dist}.

However, as pointed out in \cite{Dowker:2010pf}, a splitting like \eqref{master_correlation_split} is not always possible as the functions can be correlated. Below we discuss various types of correlations and outline a strategy to compute these correlation functions.

\paragraph{$\zeta$-$\zeta$ correlations.}
Consider the correlation function $\avg{\zeta_{i_1}\cdots\zeta_{i_n}}$, which gives the probability of having $i_1$ elements in $\mathcal R_1$, $i_2$ elements in $\mathcal R_2$ and so on. If we work with non-empty causal diamonds $\mathcal R_1 = I(x_1,y_1), \dots,  \mathcal R_n = I(x_n,y_n)$, such that none of them overlap, i.e.\ pairwise
$\mathcal R_i\cap \mathcal R_j=\emptyset$, then these probabilities are independent.\footnote{This follows from the independence of the Poisson process, and the fact that we are taking the average over a (large enough) ensemble of Poisson sprinklings.} In other words, the $\zeta$ functions are uncorrelated and split in the following manner 
\begin{gather}\label{eq:<zeta...zeta> = <zeta>...<zeta>}
    \begin{aligned}
            \avg{\zeta_{i_1}(x_1,y_1)\cdots\zeta_{i_n}(x_n,y_n)}
        &=\avg{\zeta_{i_1}(x_1,y_1)}\cdots\avg{\zeta_{i_n}(x_n,y_n)}.\\
    \end{aligned}
\end{gather}
However, if the regions overlap then there will be correlations. By using the decomposition properties discussed in Section \ref{sec: gen properties of zeta}, we can decompose any product of $\zeta$-functions into expressions depending only on non-overlapping regions, such as in \eqref{eq:Zeta(R1)Zeta(R_2) delta funtion decomposition}. The correlation function can then be computed by applying \eqref{eq:<zeta...zeta> = <zeta>...<zeta>}, termwise. 

\paragraph{$\chi$-$\chi$ correlations.}
Next consider the correlation function $\avg{\chi(dV_{x_1})\cdots \chi(dV_{x_n})}$. Once again, if all the regions are non-overlapping, then all the $\chi$'s are uncorrelated and we have
\begin{equation}\label{eq:<chi chi> = <chi><chi>}
    \avg{\chi(dV_{x_1})\cdots \chi(dV_{x_n})} = \avg{\chi(dV_{x_1})}\cdots \avg{\chi(dV_{x_n})}.
\end{equation}
This is because sprinklings in neighbourhoods of different points are independent. However, if two points coincide, there are correlations. This can be simply evaluated by noting that $\chi^n = \chi$, since it only takes the values $0$ or $1$. In other words
\begin{equation} \label{chi to the power n}
    \avg{\chi(dV_{x})^n} = \avg{\chi(dV_{x})} \neq \avg{\chi(dV_{x})}^n.
\end{equation}

\paragraph{$\zeta$-$\chi$ correlations.} Finally, let us consider correlations between $\zeta$ and $\chi$. Here we have that
\begin{equation}\label{zeta_chi_correlation}
    \avg{\zeta_i(x,y) \chi(dV_z)} = \begin{cases}
                                    \avg{\zeta_i(x,y)} \avg{\chi(dV_z)} & \text{if } z\not\in I(x,y),\\
                                    \avg{\zeta_{i-1}(x,y)} \avg{\chi(dV_z)} & \text{if } z\in I(x,y).
                                \end{cases}
\end{equation}
This can be seen as follows. The correlation function above computes the probability that $i$ elements are sprinkled in the causal interval $I(x,y)$ and an element at $z$. If $z\not\in I(x,y)$, then the two functions are uncorrelated since the probability of sprinkling in each region is independent. If $z\in I(x,y)$, then the functions are correlated. We can still split the joint probability into a product of probabilities, if we turn the $\zeta$ probability into a conditional probability. For example, in \eqref{zeta_chi_correlation}, in the second line on the right hand side we have expressed this as the probability of there being one element at $z$ times the probability of there being $i$ elements in $I(x,y)$ \emph{given that} there is an element within $I(x,y)$. The latter (to lowest order in $dV_x$) is the probability for there being $i-1$ elements in $I(x,y)$.\footnote{This can also be seen by using \eqref{eq:avg-gen}, with the definitions of $\chi$ and $\zeta$. The goal is to find which fraction of the terms in the sum contribute to the average.} We give a more careful treatment of this case in Section \ref{sec:careful}.

We can also use \eqref{zeta_chi_correlation} to handle more general correlation functions such as 
\begin{equation}
     \avg{\zeta_i(x,y) \chi(dV_z)\cdots} = \avg{\zeta_{i-1}(x,y)\cdots} \avg{\chi(dV_z)},
\end{equation}
where $\cdots$ are any other products of $\zeta$'s and $\chi$'s that are uncorrelated with $\chi(dV_z)$.
\\
\\
\begin{tcolorbox}
    \boxLabel{rulesBox}{}
    To summarize the overall strategy to compute a correlation function \eqref{master_correlation}:
    \begin{enumerate}
        \itemsep0em 
        \item First resolve $\chi-\chi$ correlations using $\chi^n=\chi$, for coinciding points.
        \item Then resolve $\zeta-\chi$ correlations, by pulling out one $\chi$ at a time using \eqref{zeta_chi_correlation}.
        \item Finally, resolve $\zeta-\zeta$ correlations using the decomposition \eqref{eq:Zeta(R1)Zeta(R_2) delta funtion decomposition} and \eqref{eq:<zeta...zeta> = <zeta>...<zeta>}.
    \end{enumerate}
\end{tcolorbox}

\subsubsection{Example Calculations}\label{sec:Correlation function examples}
In this section, we compute correlation functions of the form \eqref{master_correlation} using the strategy outlined above.

Let us consider the $\zeta-\zeta$ correlation for the case illustrated below 
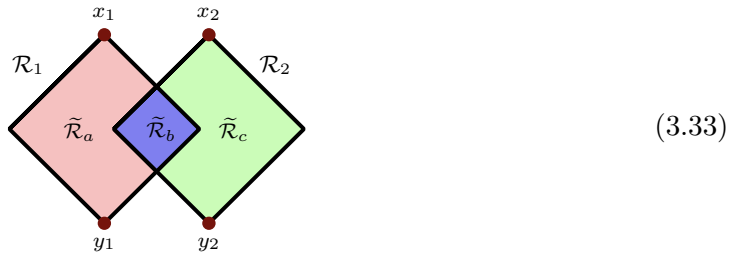
\begin{equation}\label{eq:diamonds overlap with labels}
    \DiamondOverlapZeroInsideBRegionLabeled
\end{equation}
We first use \eqref{eq:Zeta(R1)Zeta(R_2) delta funtion decomposition} to decompose in terms of non-overlapping regions $\widetilde{\mathcal R}_a, \widetilde{\mathcal R}_b$ and $\widetilde{\mathcal R}_c$, and then we use \eqref{eq:<zeta...zeta> = <zeta>...<zeta>} for the non-overlapping regions (step 3 of the \hyperref[rulesBox]{strategy}). We find

\begin{gather}
\label{eq:zetazetacorr}
    \begin{aligned}
        \avg{\zeta_i(x_1,y_1)\zeta_j(x_2,y_2)}
        & =
        \sum\limits_{\substack{\alpha,\beta,\gamma\geq 0\\ \alpha+\beta=i\\ \beta+\gamma=j }} \,
        \avg{
        \zeta_{\alpha}(\widetilde{\mathcal R}_a)\,
        \zeta_{\beta}(\widetilde{\mathcal R}_b)\,\zeta_{\gamma}(\widetilde{\mathcal R}_c)}\\
        &=
        \sum\limits_{\substack{\alpha,\beta,\gamma\geq 0\\ \alpha+\beta=i\\ \beta+\gamma=j }} \,
        \avg{\zeta_{\alpha}(\widetilde{\mathcal R}_a)}
        \avg{\zeta_{\beta}(\widetilde{\mathcal R}_b)}
        \avg{\zeta_{\gamma}(\widetilde{\mathcal R}_c)} \\
        &=\sum\limits_{\substack{\alpha,\beta,\gamma\geq 0\\ \alpha+\beta=i\\ \beta+\gamma=j }} P_\alpha(V_a)P_\beta(V_b)P_\gamma(V_c).
    \end{aligned}
\end{gather}
Here $V_a$, $V_b$ and $V_c$ are the volumes of the regions $\widetilde{\mathcal R}_a$, $\widetilde{\mathcal R}_b $ and $\widetilde{\mathcal R}_c$, respectively.
There can be additional subtleties when both $\chi$ and $\zeta$ functions are present in the correlation functions, due to possible $\zeta-\chi$ correlations. For example, let us consider the $\avg{\zeta_i(x_1,y_1)\zeta_j(x_2,y_2)\chi( dV_{x_1})}$ correlation in the example below

\begin{equation}
    \DiamondOverlapOneInsideARegionLabled
\end{equation}

As a result of $x_1\in \mathcal R_2$, $\chi( dV_{x_1})$ and $\zeta_j(x_2,y_2)$ are correlated. We use step 2 in the \hyperref[rulesBox]{strategy} to resolve the $\chi-\zeta$ correlation, and then step 3 for the remaining $\zeta-\zeta$ correlation.

\begin{gather}
    \begin{aligned}
        \avg{\zeta_i(x_1,y_1)\zeta_j(x_2,y_2)\chi(dV_{x_1})}
        & = \avg{\zeta_i(x_1,y_1)\zeta_{j-1}(x_2,y_2)}\avg{\chi(dV_{x_1})} \\
        &= \sum\limits_{\substack{\alpha,\beta,\gamma\geq 0\\ \alpha+\beta=i\\ \beta+\gamma=j-1 }} \,
        \avg{\zeta_{\alpha}(\widetilde{\mathcal R}_a)}\,
        \avg{\zeta_{\beta}(\widetilde{\mathcal R}_b)}\,
        \avg{\zeta_{\gamma}(\widetilde{\mathcal R}_c)} \,
        \avg{\chi(dV_{x_1})} \\
        &= \rho\sum\limits_{\substack{\alpha,\beta,\gamma\geq 0\\ \alpha+\beta=i\\ \beta+\gamma=j-1 }} P_\alpha(V_a)P_\beta(V_b)P_\gamma(V_c)\,dV_{x_1}.
    \end{aligned}
\end{gather}

Similarly, we can deal with more complicated correlation functions. We can graphically represent the correlation functions in the following way: a causal diamond between $x$ and $y$ represents a $\zeta_i(x,y)$ and each red dot at $z$ represents a $\chi(dV_z)$. For example
\begin{equation}\label{eq:Graphical notation for <zeta zeta ...>}
    \DiamondOverlapXB ~~ \equiv ~~ \avg{\zeta_i(x,y_1)\zeta_j(x,y_2)\chi(dV_x)\chi(dV_{y_1})\chi(dV_{y_2})}.
\end{equation}
The $i$ and $j$ indices are not explicit in this graphical representation. By applying the three steps in our \hyperref[rulesBox]{strategy}, we can compute these correlations. For example, decomposing analogously to the above examples and using \eqref{poisson_dist} we get
\begin{equation}
   \DiamondOverlapXB   ~~ =~~ \frac{(\rho V_a)^{i-j-1}(\rho V_b)^j}{(i-j-1)!\,j!}\, e^{-\rho V_1}\rho^3 dV_{x} dV_{y_1}  dV_{y_2},
\end{equation}
and
\begin{equation}
    \DiamondOverlapTwoInsideA  ~~=~~ \sum\limits_{\substack{\alpha,\beta,\gamma\geq 0\\ \alpha+\beta=i-1\\ \beta+\gamma=j-1 }} \frac{(\rho V_a)^\alpha(\rho V_b)^\beta(\rho V_c)^\gamma}{\alpha!\,\beta!\,\gamma!}\, e^{-\rho |V_1\cup V_2|}\rho^4 dV_{x_1} dV_{y_1} dV_{x_2} dV_{y_2}.
\end{equation}
Note that $V_1$ and $V_2$ are the volumes of the two diamonds $\mathcal R_1$ and $
\mathcal R_2$, as in \eqref{eq:diamonds overlap with labels} (see also Figure \ref{fig:RegionOverlap}).
For convenience, in Appendix \ref{app:CorrelationFunctions} we present explicit expressions for a set of correlation functions that we will need in this work.

\subsubsection{Careful Treatment of $\zeta-\chi$ Correlations}\label{sec:careful}
In order to show the formula \eqref{zeta_chi_correlation}, consider the correlator $\avg{\zeta_i(x,y) \chi(dV_z)}$ where $z\in I(x,y)$
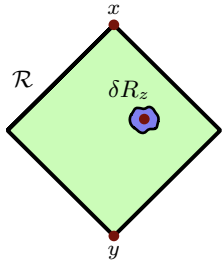
\[  \DiamondWithzInsideInRegion \]
We can evaluate this using the decomposition \eqref{eq:Zeta(R1)Zeta(R_2) delta funtion decomposition} by noting that
\begin{equation}\label{eq:chi is zeta for dV}
    \chi(\delta \mathcal R_z)=\zeta_1(\delta \mathcal R_z)
\end{equation}
for infinitesimal regions, since there can at most be one element there. In other words, only the first term of \eqref{eq:chi_zeta_relation} contributes. In this case, the decomposition into disjoint regions~\eqref{eq:R1R1 decomposition} becomes 
\begin{equation}
    \mathcal R \cup \delta\mathcal R_z = \widetilde{\mathcal R}_a\sqcup\widetilde{\mathcal R}_b\sqcup\widetilde{\mathcal R}_c,
\end{equation}
where $\mathcal R = I(x,y)$ and $\delta\mathcal R_z$ is the infinitesimal region around the point $z$. Here $\widetilde{\mathcal R}_a = R\backslash\delta\mathcal R_z$, $\widetilde{\mathcal R}_b = \delta\mathcal R_z$ and $\widetilde{\mathcal R}_c = \emptyset$.
Using \eqref{eq:chi is zeta for dV} and applying \eqref{eq:Zeta(R1)Zeta(R_2) delta funtion decomposition} for this decomposition we get 
\begin{gather}\label{eq:The derivation of <zeta chi> formula}
    \begin{aligned}
        \avg{\zeta_i(x,y)\chi(\delta \mathcal R_{z})}
         & = \avg{\zeta_i(x,y)\zeta_1(\delta \mathcal R_z)} \\
         & = \sum\limits_{\substack{\alpha,\beta,\gamma\geq 0\\ \alpha+\beta=i\\ \beta+\gamma=1 }} \,
        \avg{\zeta_{\alpha}(\widetilde{\mathcal R}_a)}\,
        \avg{\zeta_{\beta}(\widetilde{\mathcal R}_b)}\,
       \cancelto{1}{ \avg{\zeta_{\gamma}(\widetilde{\mathcal R}_c)}} \\
        &= \avg{\zeta_{i-1}(\mathcal R\backslash\delta \mathcal R_z)} P_1(dV_z),
    \end{aligned}
\end{gather}
where in the second line it was used that since $\widetilde{\mathcal R}_c = \emptyset$, we must have $\gamma=0$. Expanding the Poisson distribution \eqref{poisson_dist} to first order
\begin{equation}
    P_1 (dV) = \rho dV + O(\rho^2 dV^2).
\end{equation}
Similarly $\avg{\zeta_{i-1}(\mathcal R\backslash\delta \mathcal R_z)} = P_{i-1}(V - dV_z) = P_{i-1}(V) + O(\rho dV)$. Therefore, to first order in $\rho dV$, \eqref{eq:The derivation of <zeta chi> formula} becomes \eqref{zeta_chi_correlation}.

\section{Causal Set Action} \label{sec: causal set action}

Thus far we have reviewed the basics of causal set theory and  Poisson sprinklings, and discussed how to compute correlation functions in sprinkled regions. For the remainder of the paper, we will focus on the causal set action and we will apply what we have discussed so far, as well as develop additional tools, specifically for the action. In this section we review the definition and properties of the causal set action.

Finding the correct quantum dynamics for causal sets is an important outstanding question. The sum over histories approach to quantum theory, due to its spacetime nature, is a natural formalism for this. While a fundamentally motivated action for causal sets is not yet at hand, e.g.\ to insert in $\sum_{\mathcal{C}
} e^{iS[\mathcal{C}]}$, there exist a few candidate proposals for actions that possess some of the desired properties. One thing the action and the dynamics that ensue from it must ultimately explain is how manifoldlike causal sets, approximated by solutions to Einstein's equations, emerge macroscopically. This is highly non-trivial given that non-manifoldlike causal sets vastly outnumber manifoldlike ones \cite{Henson:2015fha}. As stated in \cite{BDAction}, the \emph{nonlocality} of causal sets will play a key role in making possible the emergence of causal sets approximated by spacetimes described by General Relativity.

The Benincasa-Dowker-Glaser (BDG) action \cite{BDAction, BDGAction} is one of the most interesting and  useful constructions of an Einstein-Hilbert-like action in terms of quantities intrinsic to causal sets. It was derived somewhat indirectly, through studies of nonlocal analogs of d'Alembertian operators that describe the propagation of scalar fields on causal sets. Studies of these d'Alembertians in causal sets approximated by curved spacetimes, revealed that they (on average) produce a term approximated by the Ricci scalar curvature in addition to a term approximated by the local d'Alembertian. The BDG action \cite{BDAction, BDGAction} in $d$ spacetime dimensions on a causal set $\mathcal C$ is
\begin{equation}
\label{bdgaction}
    S^{(d)}[\mathcal C] = \sum_{x\in\mathcal C}B^{(d)}\phi(x),
\end{equation}
where the expression above needs to be evaluated at the constant\footnote{For notational ease, we keep the $\phi(x)$ unevaluated in some expressions below.} $\phi(x)=-2\ell^2/\alpha_d$,  $\ell$ is the discreteness scale, and $B^{(d)}$ is the d'Alembertian or Box operator \cite{BDAction, BDGAction, Box2d, Glaser:2013xha, Yeats:2024tne} defined as 
\begin{equation}\label{Boxd}
    B^{(d)}\phi(x) = \frac 1{\ell^2}\left(\alpha_d\, \phi(x) + \beta_d \sum_{i=1}^{n_d} c_i \sum_{y\in \lozenge_i(x)}\phi(y)\right),
\end{equation}
 where $c_i$, $\alpha_d$ and $\beta_d$ are constants (fixed to ensure that the continuum limit of the average of $B$ over  sprinklings agrees with $\Box$), and $n_d$ is an integer. We remind the reader that $y\in \lozenge_i(x)$ indicates that $y$ causally precedes $x$ ($y\prec x$) and that there are $i-1$ elements in the causal diamond between $x$ and $y$ (excluding $x$ and $y$ themselves). Therefore, $\lozenge_1(x)$ would be a nearest neighbour. In the continuum limit ($\rho\rightarrow \infty$), the mean of $B$ (in any dimension) over all Poisson sprinklings into a spacetime can be approximated by the usual local d'Alembertian, $\Box$, plus a
term proportional to the Ricci scalar curvature \cite{BDAction, Nasiri:2023iwc}:

\begin{equation}
\label{boxmean}    \displaystyle{\lim_{\rho\rightarrow \infty}}\, \avg{B\, \phi(x)}=\left(\Box-\frac{1}{2}R(x)\right)\, \phi(x).
\end{equation}
 The $1/2$ coefficient of $R$ is universal \cite{Glaser:2013xha} (i.e. independent of spacetime dimension). The right hand side of \eqref{boxmean} is the reason why the BDG action is defined via \eqref{bdgaction}. Recently, a set of higher order curvature invariants were also constructed in a similar manner \cite{deBrito:2023axj}. An explicit example of a nonlocal d'Alembertian operator in $3+1$D is \cite{Dionthesis, Aslanbeigi:2014zva, Belenchia:2015hca}
\begin{equation}
B^{(4)}\phi(x)=\frac{4}{\sqrt{6}\ell^2}\left(-\phi(x)+\left(\sum_{i=1}^3c_i\sum_{y\in \lozenge_i}\right)\phi(y)\right),
\label{box2}
\end{equation}
with coefficients $c_1=1, c_2=-9, c_3=16, c_4=-8$. The values of the coefficients $\{c_i\}$ vary from one definition of nonlocal d'Alembertian to another (see e.g.~\cite{Aslanbeigi:2014zva}). However, a common feature among the different sets of $c_i$'s is that their values alternate in sign for odd and even indices $i$. Thus we can regard the d'Alembertian and action constructed in this way to in some sense obey an inclusion-exclusion principle. It is precisely this inclusion-exclusion principle, via the alternating signs, that creates the necessary cancellations to approximate local quantities using nonlocal inputs. 

As mentioned, the mean of $B$ when averaged over an ensemble of Poisson sprinkled causal sets, is approximated by the usual local d'Alembertian $\Box$ plus a term proportional to $R(x)$. By the same token, the BDG action \eqref{bdgaction}, when averaged over many causal sets, is approximated by the Einstein-Hilbert action. However, there are  fluctuations away from the mean of the action for individual causal set realizations of the same continuum spacetime. These are the fluctuations that we will study below. Partly to tame these fluctuations, a more general one-parameter family of nonlocal BDG actions has also been defined, where a nonlocality scale $\ell_k>\ell$ (recall that $\ell$ is the discreteness scale) damps out the fluctuations below this scale for any single causal set \cite{BDAction, Dionthesis}. This results in an action that is approximated by the Einstein-Hilbert action for any particular sprinkled causal set without the need for averaging. 

The cosmological constant $\Lambda$ and the gravitational action are closely related. The cosmological constant enters the classical gravitational action as
\begin{equation}
S_G=\int\left[\frac{1}{16\pi G}(R-2\Lambda)\right]\sqrt{-g}\,d^dx+\textrm{Boundary terms},
\label{ehs}
\end{equation}
and in the causal set case, the analog of this would be
\begin{equation}
S_\mathcal C=\sum_{x\in \mathcal C}R(x)-2\Lambda V+\textrm{Boundary terms},
\label{css}
\end{equation}
where by $R(x)$ in \eqref{css} we mean the action of $B$ on (an appropriately normalized) constant, and by $V$ we mean $N/\rho$, i.e.\ the expressions in terms of causal set quantities. The cosmological constant term in \eqref{css} is not present in the average of the BDG action over causal sets, but perhaps the fluctuations in $S$ defined on a single causal set could be identified with this term. As mentioned in the introduction, there is a model of a stochastic cosmological constant, known as Everpresent $\Lambda$ \cite{originallambda, ahmed2004everpresent, Das:2023hbw, Das:2023rvg}, that is motivated by principles of  causal set theory. According to the model, the value of the cosmological constant fluctuates over cosmic history, with a standard deviation at each epoch that is of the order of $1/\sqrt{V}$ where $V$ is the spacetime volume at that epoch. The  magnitude of these fluctuations is related to the Poisson distribution that describes the relation between number and volume in manifoldlike causal sets. This Everpresent $\Lambda$ idea did not stem from an action definition. However as we can see from  \eqref{css}, it is very natural for it to be related to  an action. Moreover, since we observe fluctuations in the BDG action \eqref{bdgaction}, and since the cosmological constant in Everpresent $\Lambda$ also fluctuates,  it begs the question of whether there may be  a potential connection between the two fluctuations. Investigating this connection may also  give us deeper insight into the properties of Everpresent $\Lambda$, and furnish some missing pieces in current phenomenological models of it (such as what the value of the mean is about which fluctuations occur). This is one of the main motivations for focusing our investigations on fluctuations of the action, in this paper. The fluctuations of $\Lambda$ in Everpresent $\Lambda$ have a quantum origin but we aim to connect these fluctuations to statistical ones due to the Poisson distribution, in the same spirit as other heuristics on this subject (see e.g. the discussion in Section 3 of \cite{Das:2023hbw}).

We devote the next section to discussing the mean and fluctuations of the BDG family of nonlocal actions, over an ensemble of Poisson sprinkled causal sets.

\section{Fluctuations of the Causal Set Action} \label{sec:Fluctuation of action}

The aim of this section is to compute the fluctuations of the action, given by the variance
\begin{equation}
    (\Delta S)^2 = \langle S^2\rangle - \langle S\rangle^2,
\end{equation}
where the $\langle . \rangle$ are the averages with respect to an ensemble of Poisson sprinklings \eqref{eq:avg-gen} of a spacetime manifold $\mathcal M$.
The expectation value $\avg S$ was computed in \cite{finiterho} and \cite{MachetWang2020}.
We will here briefly review how to compute $\avg S$ using our formalism and then outline how to compute $\avg{S^2}$.

For simplicity of notation we first rewrite the Box operator \eqref{Boxd} as
\begin{equation}
    B^{(d)}\phi(x) = \frac 1{\ell^2}\sum_{i=0}^{n_d} C^{(d)}_i \sum_{y\in \lozenge_i(x)}\phi(y),
\end{equation}
where 
\begin{equation}
    C_i^{(d)} = \begin{cases}
        \alpha_d, & i=0\\
        \beta_d\, c_i^{(d)}, & i\neq 0
    \end{cases}.
\end{equation}

The expectation value of the action \eqref{bdgaction} is then given by
\begin{gather}
    \begin{aligned}
    \avg S &= -\frac 2{\alpha_d}\sum_{i=0}^{n_d} C^{(d)}_i \avg{\sum_{x\in\mathcal C}\sum_{y\in \lozenge_i(x)}}
    = -\frac 2{\alpha_d}\sum_{i=0}^{n_d} C^{(d)}_i\,\mathcal L_{i-1},
    \end{aligned}
\end{gather}
where we have made the substitution $\phi(x)=-2\ell^2/\alpha_d$ above, and where
\begin{equation}
\label{eq:Li_def}
    \mathcal L_i \equiv \avg{\sum_{x\in\mathcal C}\sum_{y\in \lozenge_{i+1}(x)}}.
\end{equation}
Note that the index $i$ in $\mathcal L_i$ spans the range $i=-1, \dots, n_d-1$, as this will turn out to be  more convenient later.
Similarly, for $S^2$ we can express the expectation value as
\begin{gather}
    \begin{aligned}
    \langle S^2\rangle &= \frac 4{\alpha_d^2}\sum_{i, j=0}^{n_d}C^{(d)}_iC^{(d)}_j
    \avg{\sum_{x_1,x_2\in\mathcal C}\sum_{y_1\in \lozenge_i(x_1)}\sum_{y_2\in \lozenge_j(x_2)}},\\
     &= \frac 4{\alpha_d^2}\sum_{i, j=0}^{n_d}C^{(d)}_iC^{(d)}_j
     \, \mathcal K_{i-1\,j-1},
    \end{aligned}
\end{gather}
where we have defined the correlation matrix
\begin{equation}\label{eq:K_ij definition}
    \mathcal K_{ij}\equiv \avg{\sum_{x_1,x_2\in\mathcal C}\sum_{y_1\in \lozenge_{i+1}(x_1)}\sum_{y_2\in \lozenge_{j+1}(x_2)}}.
\end{equation}
This matrix is clearly symmetric
\begin{equation}
    \mathcal K_{ij} = \mathcal K_{ji},
\end{equation}
and we do not need to compute all of its elements. We will from now on only consider the elements $i\leq j$.
Also note that the domains of the sums inside the expectation value are mutually correlated and depend on $\mathcal C$, and therefore cannot be taken out of the expectation value.

In conclusion, the fluctuations are given by
\begin{gather}\label{eq:Fluctuations wrt Mij}
    \left(\Delta S\right)^2 = \frac 4{\alpha_d^2}\sum_{i, j=0}^{n_d}C^{(d)}_{i}C^{(d)}_{j}
     \, \mathcal M_{i-1\,j-1},
     %\bigg(\mathcal K_{ij} - \mathcal L_i\mathcal L_j\bigg)
\end{gather}
where
\begin{equation}
    \mathcal M_{ij} = \mathcal K_{ij} - \mathcal L_i\mathcal L_j.
\end{equation}

\subsection{Integral Formulation}

In order to evaluate the expectation values $
\mathcal L_i$ and $\mathcal K_{ij}$, it is convenient to reformulate them as integrals over various submanifolds of $\mathcal M$. The tools needed to achieve this were developed in Section~\ref{sec:Correlations Between Quantities in Different Regions}.

We start by using~\eqref{eq:sum diamond to J- using zeta} for $i>0$,
\begin{equation}
    \sum_{y\in \lozenge_i(x)} =
    \begin{dcases}
        \text{\hphantom{12}}1 & i=0,\\
        \sum_{y\in \mathcal J^-(x)}\,\zeta^{\mathcal C}_{i-1}(x,y) & i > 0,
    \end{dcases}
\end{equation}
to remove the $i$ (and causal set) dependence in the domain of the sum. Now we can readily use \eqref{eq:sum to integral formula} to express the sums over causal set elements $x$ and $y$ (the left hand side below) as integrals over generic points $x$ and $y$ in the continuum spacetime (the right hand side below)
\begin{equation}\label{eq:x-y sums to integrals}
    \sum_{x\in\mathcal C}\sum_{y\in \lozenge_i(x)} =
    \begin{dcases}
        \int_{\mathcal M}\,\chi^{\mathcal C}(dV_x) & i=0,\\
        \int_{\mathcal M}\int_{\mathcal J^-(x)}\,\zeta^{\mathcal C}_{i-1}(x,y)\, \chi^{\mathcal C}(dV_y) \chi^{\mathcal C}(dV_x) & i > 0.
    \end{dcases}
\end{equation}
Note that the domains of the integrals only depend on the manifold $\mathcal M$, while the dependence on the particular sprinkling $\mathcal C$ is hidden in the measures $\chi^{\mathcal C}(dV_x)$ and $\chi^{\mathcal C}(dV_y)$. This measure can concretely be written as \eqref{eq:chi measure written in terms of delta functions}.

Using \eqref{eq:x-y sums to integrals} in \eqref{eq:Li_def} and \eqref{eq:K_ij definition}, we then find the following integral expressions for the expectations values
\begin{equation}\label{eq: Li vevs}
    \mathcal L_i =
    \begin{dcases}
        \int_{\mathcal M}\,\avg{\chi^{\mathcal C}(dV_x)} & i=-1,\\
        \int_{\mathcal M}\int_{\mathcal J^-(x)}\,\avg{\zeta^{\mathcal C}_i(x,y)\, \chi^{\mathcal C}(dV_y) \chi^{\mathcal C}(dV_x)} & i \geq 0,
    \end{dcases}
\end{equation}
and
\begin{gather}\label{KcorrCases}
    \begin{aligned}
    &\mathcal K_{ij}=\\
    &\begin{dcases}
        \scaleto{\int_{\scaleto{\mathcal M}{4pt}}\int_{\scaleto{\mathcal M}{4pt}}\Big\langle\chi^{\mathcal C}(dV_{x_1})\chi^{\mathcal C}(dV_{x_2})\Big\rangle,}{24pt}
        & \scaleto{i=j=-1\mathstrut}{8pt} \\ 
        % %
        %   \scaleto{\int_{\scaleto{\mathcal M}{4pt}}\int_{\scaleto{\mathcal M}{4pt}}\int_{\scaleto{\mathcal J^-(x_1)}{6pt}}\Big\langle\zeta^{\mathcal C}_{i-1}(x_1, y_1)\chi^{\mathcal C}(dV_{y_1})\chi^{\mathcal C}(dV_{x_1})\chi^{\mathcal C}(dV_{x_2})\Big\rangle,}{24pt}
        % & \scaleto{i>0, j=0\mathstrut}{8pt}\\
        %
          \scaleto{\int_{\scaleto{\mathcal M}{4pt}}\int_{\scaleto{\mathcal M}{4pt}}\int_{\scaleto{\mathcal J^-(x_2)}{6pt}}\Big\langle\zeta^{\mathcal C}_j(x_2, y_2)\chi^{\mathcal C}(dV_{y_2})\chi^{\mathcal C}(dV_{x_1})\chi^{\mathcal C}(dV_{x_2})\Big\rangle,}{24pt}
        & \scaleto{i=-1,\,j\geq 0\mathstrut}{8pt}\\
          \scaleto{\int_{\scaleto{\mathcal M}{4pt}}\int_{\scaleto{\mathcal M}{4pt}}\int_{\scaleto{\mathcal J^-(x_2)}{6pt}}\int_{\scaleto{\mathcal J^-(x_1)}{6pt}}\Big\langle\zeta^{\mathcal C}_i(x_1, y_1)\zeta^{\mathcal C}_j(x_2, y_2)\chi^{\mathcal C}(dV_{y_1})\chi^{\mathcal C}(dV_{y_2})\chi^{\mathcal C}(dV_{x_1})\chi^{\mathcal C}(dV_{x_2})\Big\rangle,}{24pt}
        & \scaleto{i,j\geq 0\mathstrut}{8pt}
    \end{dcases}
    \end{aligned}
\end{gather}
Note that the integration domains are independent of any particular sprinkling and can thus be pulled out of the averaging. The integrands need to be evaluated before the integrals can be performed. This problem can be readily solved using the formalism and techniques developed earlier in this paper, specifically the \hyperref[rulesBox]{strategy} in Section~\ref{sec:CorrelationFunctions}. In particular by using \eqref{zeta_chi_correlation}, \eqref{eq:<chi chi> = <chi><chi>} and \eqref{eq:one point functions of zeta and chi} we find
\begin{equation}\label{eq:L_i integral cases}
    \mathcal L_i =
    \begin{dcases}
        \rho\, V_{\mathcal M} & i=-1,\\
        \frac{\rho^2}{i!}\int_{\mathcal M}\int_{\mathcal J^-(x)}\,\left(\rho V_{xy}\right)^i e^{-\rho V_{xy}}
        %\avg{\zeta^{\mathcal C}_{i-1}(x,y)}
        \, dV_y\,dV_x & i \geq 0,
    \end{dcases}
\end{equation}
where $V_{xy}$ is the spacetime volume of the causal diamond $I(x,y)$. For $i\geq 0$ we actually only need to compute a single integral
\begin{equation}
    \mathcal L \equiv \int_{\mathcal M}\int_{\mathcal J^-(x)} e^{-\rho V_{xy}} dV_{y} dV_{x},
\end{equation}
since
\begin{gather}\label{eq:L_i in terms of L}
    \begin{aligned}
        \mathcal L_i&=\frac{\rho^{i+2}}{i!}\left[-\frac{d}{d\rho}\right]^i \left[\int_{\mathcal M}\int_{\mathcal J^-(x)} e^{-\rho V_{xy}} dV_{y} dV_{x}\right]\\
        &=\frac{\rho^{i+2}}{i!}\left[-\frac{d}{d\rho}\right]^i \mathcal L.
    \end{aligned}
\end{gather}
This observation will turn out to be more general, and dramatically reduce the number of independent integrals needed to be evaluated.

For $\mathcal K_{ij}$ the computation of the integrands is more subtle, as different regions of the integration domain have different kinds of correlations. It is therefore necessary to split the integration domain into submanifolds, based on the type of correlation involved, and compute each separately. The rest of this section, is dedicated to this task.

\subsection{Special Case $\mathcal K_{-1,j}$}
In \eqref{KcorrCases} we split the (symmetric) correlation matrix into three cases. Before dealing with general components of this matrix, it is instructive to separately consider the first two cases, $\mathcal K_{-1,-1}$ and $\mathcal K_{-1,j}$, to understand the nature of these calculations.

\subsubsection{$\mathcal K_{-1, -1}$}
We want to compute
\begin{equation}
    \mathcal K_{-1, -1} = \int_{\mathcal M}\int_{\mathcal M}\Big\langle\chi(dV_{x_1})\chi(dV_{x_2})\Big\rangle.
\end{equation}
As noted in Section \ref{sec:CorrelationFunctions}, there are $\chi-\chi$ correlations only when $x_1 = x_2$. We can thus split the integral into a region where $x_1 = x_2$ and another where $x_1 \neq x_2$, with the integrands reducing to $\avg{\chi(dV_{x})}$ and $\avg{\chi(dV_{x})}\avg{\chi(dV_{x_2})}$, respectively. Using \eqref{eq:one point functions of zeta and chi}, this matrix component thus becomes 
\begin{equation}
    \mathcal K_{-1, -1} = \rho V_{\mathcal M} + \rho^2 V_{\mathcal M}^2.
\end{equation}
The contribution to the fluctuations \eqref{eq:Fluctuations wrt Mij} is then
\begin{equation}
    \mathcal M_{-1, -1} = \rho V_{\mathcal M}.
\end{equation}

\subsubsection{$\mathcal K_{-1, j}$}
We want to compute
\begin{equation}
    \mathcal K_{-1, j} = \int_{\mathcal M}\int_{\mathcal M}\int_{\mathcal J^-(x_2)}\Big\langle\zeta_j(x_2, y_2)\chi(dV_{y_2})\chi(dV_{x_2})\chi(dV_{x_1})\Big\rangle
\end{equation}
for $j\ge 0$.
Now, $\chi-\chi$ correlations will occur when $x_1 = x_2$ or when $x_1 = y_2$. In both of these cases the integral becomes
\begin{equation}
    \int_{\mathcal M}\int_{\mathcal J^-(x)}\avg{\zeta_j(x, y)}\avg{\chi(dV_{y})}\avg{\chi(dV_{x})},
    %=   \frac{\rho^2}{(i-1)!}\int_{\mathcal M}\int_{\mathcal J^-(x)}\,V_{xy}^{i-1}e^{-\rho V_{xy}}\, dV_y\,dV_x.
\end{equation}
which is nothing but $\mathcal L_j$ \eqref{eq:L_i integral cases}.

When $x_1 \neq x_2\neq y_2$, we have two scenarios
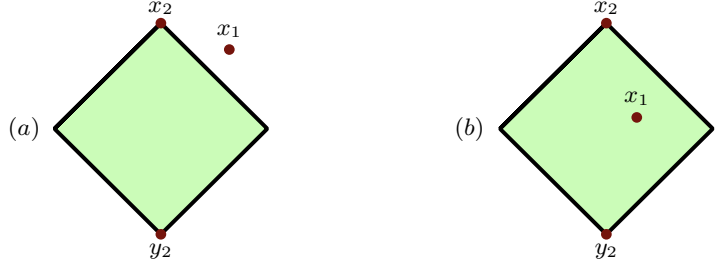
\[  \DiamondWithzInside {0.25}{0.45}a \qquad\qquad\qquad \DiamondWithzInside 00b   \]
(a) when $x_1\not\in I(x_2,y_2)$ and (b) when $x_1\in I(x_2,y_2)$. For (a) we have no $\zeta-\chi$ correlations and the integral becomes 
\begin{equation}
    \int_{\mathcal M}\int_{\mathcal J^-(x_2)}\int_{\mathcal M\backslash I(x_2,y_2)}\avg{\zeta_j(x_2, y_2)}\avg{\chi(dV_{x_1})}\avg{\chi(dV_{y_2})}\avg{\chi(dV_{x_2})},
\end{equation}
while for (b) there are $\zeta-\chi$ correlations and we find 
\begin{equation}\label{eq:x1 in Ix2y2}
    \int_{\mathcal M}\int_{ \mathcal J^-(x_2)}\int_{I(x_2,y_2)}\avg{\zeta_{j-1}(x_2, y_2)}\avg{\chi(dV_{x_1})}\avg{\chi(dV_{y_2})}\avg{\chi(dV_{x_2})},
\end{equation}
using the \hyperref[rulesBox]{rules} in Section \ref{sec:CorrelationFunctions}.
Note that, \eqref{eq:x1 in Ix2y2} only contributes when $j\geq 1$. So for $j=0$ we get
\begin{equation}
    \mathcal K_{-1,0} = 2\mathcal L_j + \rho^3\int_{\mathcal M}\int_{\mathcal J^-(x_2)}\int_{\mathcal M\backslash I(x_2,y_2)}e^{-\rho \VolT}\,dV_{x_1}dV_{y_2}dV_{x_2},
\end{equation}
while for $j>0$ we have
\begin{gather}
    \begin{aligned}
    \mathcal K_{-1, j} = 2\mathcal L_j 
    + \frac{\rho^{3+j}}{j!}\int_{\mathcal M}\int_{\mathcal J^-(x_2)}\int_{\mathcal M\backslash I(x_2,y_2)}\VolT^j e^{-\rho \VolT}\,dV_{x_1}dV_{y_2}dV_{x_2}
    \\
    + \frac{\rho^{3+j-1}}{(j-1)!}\int_{\mathcal M}\int_{\mathcal J^-(x_2)}\int_{I(x_2,y_2)}\VolT^{j-1}e^{-\rho \VolT}\,dV_{x_1}dV_{y_2}dV_{x_2}.
    \end{aligned}
\end{gather}
The contribution of these correlations to fluctuations of the action \eqref{eq:Fluctuations wrt Mij} becomes
\begin{gather}
    \begin{aligned}
    \mathcal M_{-1,j} = \mathcal K_{-1,j} - \rho V_{\mathcal M}\,\mathcal L_j.
    \end{aligned}
\end{gather}

\subsection{General Case $\mathcal K_{ij}$}
Now we will turn to the general case \eqref{KcorrCases} of computing
\begin{equation}\label{eq:kij general}
    \mathcal K_{ij} = \int_{\scaleto{\mathcal M}{4pt}}\int_{\scaleto{\mathcal M}{4pt}}\int_{\scaleto{\mathcal J^-(x_2)}{6pt}}\int_{\scaleto{\mathcal J^-(x_1)}{6pt}}\Big\langle\zeta_i(x_1, y_1)\zeta_j(x_2, y_2)\chi(dV_{y_1})\chi(dV_{y_2})\chi(dV_{x_1})\chi(dV_{x_2})\Big\rangle,
\end{equation}
for $i, j \geq 0$. Just as in the special cases above, we will first deal with the $\chi-\chi$ correlations, which will split the integral \eqref{eq:kij general} into six integrals. In each of these integrals, we will then deal with the $\zeta-\chi$ and $\zeta-\zeta$ correlations using the \hyperref[rulesBox]{strategy} outlined in Section \ref{sec:CorrelationFunctions}. 

\subsubsection{$\chi$-$\chi$ Correlations}
As discussed in Section \ref{sec:CorrelationFunctions}, there will be $\chi-\chi$ correlations in the submanifolds where the $x_i$'s and $y_i$'s coincide. Before accounting for them, it is instructive to revisit why and how these correlations occur.

Let $\mathcal C$ be a causal set that contains the sprinkled element $y_1$. Since $dV_{y_1}$ is a small region surrounding $y_1$, we have by definition \eqref{eq:chi_zeta_relation} that $\chi(dV_{y_1}) = 1$. For any point $y_2\neq y_1$, there is an infinitesimal probability that there is a sprinkled element in an infinitesimal region around $y_2$ in $\mathcal C$. Thus, there is an infinitesimal probability that $\chi(dV_{y_2})=1$.\footnote{In other words, the average of $\chi(dV_{y_2})$ over sprinklings $\mathcal C$ that contain $y_1$, is infinitesimal.} However, if $y_1=y_2$, then the probability for  $\chi(dV_{y_2})=1$ is $1$ and there is therefore a correlation between $\chi(dV_{y_1})$ and $\chi(dV_{y_2})$.

It is thus convenient to decompose the integral into six parts 
\begin{equation}\label{eq:K_ij into J1-J6}
    \mathcal K_{ij} = J^1_{ij}+J^2_{ij}+J^3_{ij}+J^4_{ij}+J^5_{ij}+J^6_{ij},
\end{equation}
depending on which $\chi-\chi$ correlations are present. See Appendix \ref{app: products of chi} for more details on these correlations. Each component in \eqref{eq:K_ij into J1-J6} is integrated over a subregion defined respectively by 
\begin{subequations}\label{eq:J1-J6 definitions}
    \begin{eqnarray}
        J^1&:& x_1=x_2, \indent y_1=y_2, \label{eq:J1 def}\\
        J^2&:& x_1\neq x_2, \indent y_1=y_2, \label{eq:J2 def}\\
        J^3&:& x_1= x_2, \indent y_1\neq y_2, \label{eq:J3 def}\\
        J^4&:& x_1 = y_2, \indent x_2\neq y_1 \label{eq:J4 def}\\
        J^5&:& x_2 = y_1, \indent x_1\neq y_2 \label{eq:J5 def}\\
        % J^r6&:& x_1\neq x_2, \indent y_1\neq y_2,\indent x_1\neq y_2, \indent x_2\neq y_1. \label{eq:J6 def}\\
        J^6&:& x_1\neq x_2\neq y_1\neq y_2.\label{eq:J6 def}
    \end{eqnarray}
\end{subequations}
Using \eqref{chi to the power n}, each component in \eqref{eq:K_ij into J1-J6} is then given by
\begin{align}
    J^{1}_{ij}&=\int_{\mathcal M}\int_{\mathcal J^-(x)}\bigg\langle\zeta_i(x,y)\zeta_j(x,y)\chi(dV_{y})\chi(dV_{x})\bigg\rangle,\label{eq:J1ij}\\
    J^{2}_{ij}&=\int_{\mathcal M}\int_{\mathcal M\backslash\{x_2\}}\int_{\mathcal J^-(x_1)\cap \mathcal J^-(x_2)}\bigg\langle\zeta_i(x_1,y)\zeta_j(x_2,y)\chi(dV_{y})\chi(dV_{x_1})\chi(dV_{x_2})\bigg\rangle,\\
    J^{3}_{ij}&=\int_{\mathcal M}\int_{\mathcal J^-(x)}\int_{\mathcal J^-(x)\backslash\{y_2\}}\bigg\langle\zeta_i(x,y_1)\zeta_j(x,y_2)\chi(dV_{y_1})\chi(dV_{y_2})\chi(dV_{x})\bigg\rangle,\\
    J^{4}_{ij}&=\int_{\mathcal M}\int_{\mathcal J^-(x)}\int_{\mathcal J^-(y_2)}\bigg\langle\zeta_i(y_2,y_1)\zeta_j(x,y_2)\chi(dV_{y_1})\chi(dV_{y_2})\chi(dV_{x})\bigg\rangle,
    \\
    J^{5}_{ij}&=\int_{\mathcal M}\int_{\mathcal J^-(x)}\int_{\mathcal J^-(y_1)}\bigg\langle\zeta_i(x,y_1)\zeta_j(y_1,y_2)\chi(dV_{y_2})\chi(dV_{y_1})\chi(dV_{x})\bigg\rangle,
\end{align}
and
\begin{equation}
    J^{6}_{ij}=\int_{\mathcal M}\int_{\mathcal M\backslash\{x_2\}}\int_{\mathcal J^-(x_2)}\int_{\mathcal J^-(x_1)\backslash\{y_2\}}\bigg\langle\zeta_i(x_1,y_1)\zeta_j(x_2,y_2)\chi(dV_{y_1})\chi( dV_{y_2})\chi(dV_{x_1})\chi(dV_{x_2})\bigg\rangle.
\end{equation}
Note that
\begin{equation}
    J^4_{ij} = J^5_{ji},
\end{equation}
as we can see by relabeling the dummy variables in the integrals. Similarly, there may be other $J^a_{ij}$  components that are equal and therefore the minimal set of independent integrals may be smaller. 

\subsubsection{$\zeta$-$\chi$ Correlations}
Now we will deal with $\zeta-\chi$ correlations, according to the \hyperref[rulesBox]{discussion} in Section \ref{sec:CorrelationFunctions}. This will further split the integration domains \eqref{eq:J1-J6 definitions} into subregions, with different $\zeta-\chi$ correlations and thus different integrands. For this, it is convenient to first define the following notation for the integration domains needed
\begin{subequations}
\label{eq:domains}
\begin{align}
        \mathcal D_1 &= \left\{(x,y)\in\mathcal M\times\mathcal M\;\big|\; y\prec x\right\} \label{eq:IntegrationDomains D1},\\
        \mathcal D_2 &= \left\{(x_1, x_2,y)\in\mathcal M\times\mathcal M\times\mathcal M\;\big|\; (y\prec x_1)\,\wedge\, (y\prec x_2)\,\wedge\, (x_1\neq x_2)\right\}\label{eq:IntegrationDomains D2},\\
        \mathcal
        D_3 &= \left\{(x, y_1,y_2)\in\mathcal M\times\mathcal M\times\mathcal M\;\big|\; (y_1\prec x) \,\wedge\,(y_2\prec x)\,\wedge\, (y_1\neq y_2)\right\}\label{eq:IntegrationDomains D3},\\
        \mathcal
        D_4 &= \left\{(x, y_1,y_2)\in\mathcal M\times\mathcal M\times\mathcal M\;\big|\; (y_2\prec x) \,\wedge\,(y_1\prec y_2)\right\}\label{eq:IntegrationDomains D4},\\
        \mathcal
        D_5 &= \left\{(x, y_1,y_2)\in\mathcal M\times\mathcal M\times\mathcal M\;\big|\; (y_1\prec x) \,\wedge\,(y_2\prec y_1)\right\}\label{eq:IntegrationDomains D5},\\
        \mathcal D_6 &= \big\{(x_1,x_2, y_1,y_2)\in\mathcal M^4\;\big|\; (y_1\prec x_1) \,\wedge\, (y_2\prec x_2) \,\wedge\,(x_1\neq x_2\neq y_1 \neq y_2)\big\} \label{eq:IntegrationDomains D6}
\end{align}
\end{subequations}
In this notation, $\mathcal D_i$ corresponds to the integration domain of the $J^i$ integral. We diagrammatically express the additional splittings of the domains $\mathcal D$ due to the $\zeta-\chi$ correlations as
\begin{equation}\label{eq:D145 diagram}
    \mathcal D_1 = \DiamondOverlapXYs, \qquad \mathcal D_4 = \DiamondOverlapZeroInsideCs {-1}, \qquad \mathcal D_5 = \DiamondOverlapZeroInsideCs 1,
\end{equation}
where the colors are used as in Section \ref{sec:Correlation function examples}, for example see \eqref{eq:diamonds overlap with labels}. 
Similarly we have the decomposition
\begin{equation} \label{eq:D2 diagram}
    \mathcal D_2 = \DiamondOverlapYAs\sqcup \DiamondOverlapYBs \sqcup \DiamondOverlapYCs,
\end{equation}
where each submanifold is
\begin{gather}\label{eq:xxy_regions}
    \begin{aligned}
         \DiamondOverlapYAs &= \left\{(x_1,x_2,y)\in\mathcal D_2\;\big|\; (x_1\not\prec x_2)\;\wedge\; (x_2\not\prec x_1) \right\},\\
        \DiamondOverlapYBs &= \left\{(x_1,x_2,y)\in\mathcal D_2\;\big|\; x_2\prec x_1 \right\},\\
        \DiamondOverlapYCs &= \left\{(x_1,x_2,y)\in\mathcal D_2\;\big|\; x_1\prec x_2\right\},
    \end{aligned}
\end{gather}
and
\begin{equation} \label{eq:D3 diagram}
    \mathcal D_3 = \DiamondOverlapXAs\sqcup\DiamondOverlapXBs\sqcup\DiamondOverlapXCs,
\end{equation}
where
\begin{gather}\label{eq:D3 subdiagrams}
    \begin{aligned}
         \DiamondOverlapXAs &= \left\{(x,y_1,y_2)\in\mathcal D_3\;\big|\; (y_1\not\prec y_2)\;\wedge\; (y_2\not\prec y_1) \right\},\\
        \DiamondOverlapXBs &= \left\{(x,y_1,y_2)\in\mathcal D_3\;\big|\; y_1\prec y_2 \right\},\\
        \DiamondOverlapXCs &= \left\{(x,y_1,y_2)\in\mathcal D_3\;\big|\; y_2\prec y_1 \right\}.
    \end{aligned}
\end{gather}
Finally we have 
\begin{equation}
    \mathcal D_6 = \DiamondOverlapZeroInsideAs \sqcup  \DiamondOverlapZeroInsideBs \sqcup   \DiamondOverlapOneInsideAs \sqcup\DiamondOverlapOneInsideBs \sqcup \DiamondOverlapOneInsideCs \sqcup \DiamondOverlapOneInsideDs \sqcup      \DiamondOverlapTwoInsideAs \sqcup \DiamondOverlapTwoInsideBs \sqcup \DiamondOverlapTwoInsideCs \sqcup \DiamondOverlapTwoInsideDs,
\end{equation}
where the subregions are defined analogously to those shown in \eqref{eq:D145 diagram}-\eqref{eq:D3 subdiagrams}.
For example, there are three cases for $\mathcal D_2$, depending on whether or not one of the $x$'s lies within a diamond. In cases where one of the $x$'s lies within a diamond, a $\zeta-\chi$ correlation is induced. There cannot be any $\zeta-\chi$ correlations in the subdomains $\mathcal D_1$, $\mathcal D_4$, and $\mathcal D_5$, hence these do not have additional splittings, as shown in \eqref{eq:D145 diagram}.

We will next compute the $J^i$ integrals, starting with the relatively simpler cases of $J^1$ and $J^4$, then proceeding to the slightly more involved case of $J^2$, and finally treating $J^6$ which has the most number of splittings of its domain.

\subsubsection{The $J^1$ Integral}
We will begin by computing the first integral $J^1$, \eqref{eq:J1ij}. The $\zeta-\zeta$ correlations, which we have not yet accounted for in this subsection, are easy to account for in this case by using the \hyperref[rulesBox]{rules} of Section \ref{sec:CorrelationFunctions}
\begin{gather}\label{eq:J1ij split vev}
    \begin{aligned}
        J^{1}_{ij} %&=\int_{\mathcal M}\int_{\mathcal J^-(x)}\bigg\langle\zeta_{i-1}(x,y)\zeta_{j-1}(x,y)\chi(dV_{y})\chi(dV_{x})\bigg\rangle,\\
        &=\int_{\mathcal M}\int_{\mathcal J^-(x)}\avg{\zeta_i(x,y)\zeta_j(x,y)}\avg{\chi(dV_{y})}\avg{ \chi(dV_{x})}\\
        &=\delta_{ij}\int_{\mathcal M}\int_{\mathcal J^-(x)}\avg{\zeta_i(x,y)}\avg{\chi(dV_{y})}\avg{ \chi(dV_{x})}.\\
        % &=\rho^2\delta_{ij}\int_{\mathcal M}\int_{\mathcal J^-(x)}\frac{(\rho V(x,y))^{i-1}}{(i-1)!}e^{-\rho V(x,y)} dV_{y} dV_{x}.
    \end{aligned}
\end{gather}
In the second line we used the identity
\begin{equation}
    \zeta_i(x,y)\zeta_j(x,y) = \delta_{ij}\zeta_i(x,y).
\end{equation}
We see that only the diagonal components of $J^1_{ij}$ are non-zero. This is nothing other than $\delta_{ij}$ times \eqref{eq: Li vevs}. Thus
\begin{equation}
    J^1_{ij} = \delta_{ij}\,\mathcal L_i.
\end{equation}

\subsubsection{The $J^4$ Integral}
The next integral we will tackle is $J^4$. Similar to the case of the $J^1$ integral, the absence of $\zeta-\zeta$ correlations in this case reduces the relevant domain $\mathcal D_4$ to a single subregion \eqref{eq:D145 diagram}, simplifying the calculation.

\begin{gather}
    \begin{aligned}
        J^{4}_{ij} & =\int_{\mathcal M}\int_{\mathcal J^-(x)}\int_{\mathcal J^-(y_2)}\avg{\zeta_i(y_2,y_1)}\avg{\zeta_j(x,y_2)}\avg{\chi(dV_{y_1})}\avg{\chi(dV_{y_2})}\avg{\chi(dV_{x})},\\
        & =\rho^3\int_{\mathcal M}\int_{\mathcal J^-(x)}\int_{\mathcal J^-(y_2)}
        \frac{\left(\rho \Vol y2y1\right)^i}{i!}
        \frac{\left(\rho\VolA xy2\right)^j}{j!}\, e^{-\rho \Vol y2y1-\rho \VolA xy2}\,
        dV_{y_1}dV_{y_2}dV_{x}.
    \end{aligned}
\end{gather}
Similar to \eqref{eq:L_i in terms of L}, all of these integrals (i.e. for any indices $i$ and $j$) can be reduced to a single integral
\begin{equation}
    J^4(\rho_1,\rho_2)\equiv\int_{\mathcal M}\int_{\mathcal J^-(x)}\int_{\mathcal J^-(y_2)}
     e^{-\rho_1 \Vol y2y1-\rho_2 \VolA xy2}
        dV_{y_1}dV_{y_2}dV_{x},
\end{equation}
since
\begin{equation}\label{eq:J4ij in terms of J4}
    J^4_{ij} = \frac{\rho^{i+j+3}}{i!j!}\left[-\frac{\partial}{\partial\rho_1}\right]^i\left[-\frac{\partial}{\partial\rho_2}\right]^jJ^4(\rho_1,\rho_2)\bigg|_{\rho_1=\rho_2=\rho}.
\end{equation}
As we will see later, this is a special case of a much more general property that will reduce the number of integrals that need to be computed to a small core set.
For notational convenience, we can adopt a graphical notation for this integral as
\begin{equation}
    J^4_{ij} = \int_{\DiamondOverlapZeroInsideCs {-1}} \equiv \int_{\DiamondOverlapZeroInsideCs {-1}}\avg{\zeta_i(y_2,y_1)\zeta_j(x,y_2)\chi(dV_{y_1})\chi(dV_{y_2})\chi(dV_{x})},
\end{equation}
where the integration domain is defined by the figure and the integrand is defined as a correlation function as discussed in Section \ref{sec:Correlation function examples}, in particular \eqref{eq:J4 correlation function}. In other words, the graphical notation $\DiamondOverlapZeroInsideCs {-1}$ denotes both the domain of integration and the correlation function in the integrand. The $i$ and $j$ indices are not explicit in this notation and must be read from the left hand side, $J^4_{ij}$.

\subsubsection{The $J^2$ Integral}
Let us next consider the $J^2$ integral
\begin{equation}
     J^{2}_{ij}=\int_{\mathcal M}\int_{\mathcal M\backslash\{x_2\}}\int_{\mathcal J^-(x_1)\cap \mathcal J^-(x_2)}\bigg\langle\zeta_i(x_1,y)\zeta_j(x_2,y)\chi(dV_{y})\chi(dV_{x_1})\chi(dV_{x_2})\bigg\rangle.
\end{equation}
In contrast to the previous two cases considered above, $\zeta-\chi$ correlations will occur in this case. We showed in \eqref{eq:D2 diagram} how to split the relevant domain $\mathcal D_2$ into three parts depending on the presence of such correlations.
\begin{equation}
    J^{2}_{ij}=\int_{\DiamondOverlapYBs\bigsqcup \DiamondOverlapYCs\bigsqcup\DiamondOverlapYAs} \bigg\langle\zeta_i(x_1,y)\zeta_j(x_2,y)\chi(dV_{y})\chi(dV_{x_1})\chi(dV_{x_2})\bigg\rangle.\\
\end{equation}
Due to $\zeta-\zeta$ correlations, this can be simplified further. For the diagonal elements $i=j$, the integrand vanishes in the region $\DiamondOverlapYBs\sqcup \DiamondOverlapYCs$ because
\begin{gather}
    \begin{aligned}\label{eq:zeta zeta chi = 0}
        &\avg{\zeta_i(x_1,y)\zeta_i(x_2,y)\chi(dV_{x_1})} = 0, \qquad \text{for}\qquad (x_1,x_2,y)\in\DiamondOverlapYCs, \\
        &\avg{\zeta_i(x_1,y)\zeta_i(x_2,y)\chi(dV_{x_2})} = 0, \qquad \text{for}\qquad (x_1,x_2,y)\in\DiamondOverlapYBs, \\
    \end{aligned}
\end{gather}
as in these cases there will always be at least one more element in one diamond compared to the other. For the off-diagonal elements $i<j$ we have that 
\begin{gather}
    \begin{aligned}
        &\avg{\zeta_i(x_1,y)\zeta_j(x_2,y)\chi(dV_{x_2})} = 0, \qquad \text{for}\qquad (x_1,x_2,y)\in\DiamondOverlapYBs, \\
    \end{aligned}
\end{gather}
since in $\DiamondOverlapYBs$ we have have $x_2 \prec x_1$ and therefore
$i>j$ which contradicts our assumption $i<j$.\footnote{We remind the reader that we are presently only working on the upper triangle of the matrix $J_{ij}$, i.e. where $i < j$.} In summary, the $J^2$ integral reduces to
\begin{equation}
    J^2_{ij} = (1-\delta_{ij})\int_{\DiamondOverlapYCs} + \int_{\DiamondOverlapYAs},
\end{equation}
where again the graphics denote both the integration domain and correlation function in the integrand.

\subsubsection{The $J^6$ Integral}
Finally, we turn to the $J^6$ integral which has the most number of correlations and thus contributions,
\begin{equation}
    J^{6}_{ij}=\int_{\mathcal M}\int_{\mathcal M\backslash\{x_2\}}\int_{\mathcal J^-(x_2)}\int_{\mathcal J^-(x_1)\backslash\{y_2\}}\bigg\langle\zeta_i(x_1,y_1)\zeta_j(x_2,y_2)\chi(dV_{y_1})\chi( dV_{y_2})\chi(dV_{x_1})\chi(dV_{x_2})\bigg\rangle.
\end{equation}
Recall that the relevant domain is
\begin{equation}
    \mathcal D_6 = \DiamondOverlapZeroInsideAs \sqcup  \DiamondOverlapZeroInsideBs \sqcup   \DiamondOverlapOneInsideAs \sqcup\DiamondOverlapOneInsideBs \sqcup \DiamondOverlapOneInsideCs \sqcup \DiamondOverlapOneInsideDs \sqcup      \DiamondOverlapTwoInsideAs \sqcup \DiamondOverlapTwoInsideBs \sqcup \DiamondOverlapTwoInsideCs \sqcup \DiamondOverlapTwoInsideDs,
\end{equation}
due to $\chi-\chi$ correlations. The integral can thus be computed by summing over the integral of each disjoint region, where the integrands are again evaluated as \hyperref[rulesBox]{described} in Section~\ref{sec:CorrelationFunctions} and Appendix~\ref{app:CorrelationFunctions}. Depending on the indices $i$ and $j$ only a subset of the above terms will contribute. For example due to the same logic as in \eqref{eq:zeta zeta chi = 0}, $\DiamondOverlapTwoInsideCs$ never contributes as we are only considering components with  $i\leq j$ whereas $\int_{\scaleto{\DiamondOverlapTwoInsideCs}{12pt}}$ is only non-zero for $i\geq j+2$. Similarly, the region $\DiamondOverlapTwoInsideDs$ will only contribute when $i+2\leq j$. In summary, for the diagonal elements $i\geq0$ we have

\begin{equation}
    J^6_{ii} =
    \begin{dcases}
        \int_{\DiamondOverlapZeroInsideAs} +\int_{\DiamondOverlapZeroInsideBs}, & i = 0,\\
        \int_{\DiamondOverlapZeroInsideAs} +\int_{\DiamondOverlapZeroInsideBs}
        +\int_{\DiamondOverlapOneInsideAs}
        +\int_{\DiamondOverlapOneInsideBs} +\int_{\DiamondOverlapOneInsideCs} +\int_{\DiamondOverlapOneInsideDs} 
        +\int_{\DiamondOverlapTwoInsideAs} +\int_{\DiamondOverlapTwoInsideBs}, & i\geq 1,
    \end{dcases}
\end{equation}
while for the off-diagonal elements ($i,j\geq0$) we have
\begin{equation}\label{eq:J6ij off diagonal}
    J^6_{ij} =
    \begin{dcases}
        \int_{\DiamondOverlapZeroInsideAs} +\int_{\DiamondOverlapZeroInsideBs}
        +\int_{\DiamondOverlapOneInsideAs}
        +\int_{\DiamondOverlapOneInsideBs} , & i=0, j=1,\\
        \int_{\DiamondOverlapZeroInsideAs} +\int_{\DiamondOverlapZeroInsideBs}
        +\int_{\DiamondOverlapOneInsideAs}
        +\int_{\DiamondOverlapOneInsideBs} +\int_{\DiamondOverlapOneInsideCs} +\int_{\DiamondOverlapOneInsideDs} 
        +\int_{\DiamondOverlapTwoInsideAs} +\int_{\DiamondOverlapTwoInsideBs} , & j = i+1, i>0,\\
        \int_{\DiamondOverlapZeroInsideAs} +\int_{\DiamondOverlapZeroInsideBs}
        +\int_{\DiamondOverlapOneInsideAs}
        +\int_{\DiamondOverlapOneInsideBs} +\int_{\DiamondOverlapOneInsideCs} +\int_{\DiamondOverlapOneInsideDs} 
        +\int_{\DiamondOverlapTwoInsideAs} +\int_{\DiamondOverlapTwoInsideBs} +\int_{\DiamondOverlapTwoInsideDs}, & j\geq i+2.
    \end{dcases}
\end{equation}

\subsection{Core Set of Integrals}
In summary, in order to calculate the fluctuations of the action, we see from \eqref{eq:Fluctuations wrt Mij} that we need the matrix $\mathcal M_{ij}$. This was decomposed into $ \mathcal K_{ij} - \mathcal L_i\mathcal L_j$, and $\mathcal K_{ij}$ into $J^1_{ij}, \dots, J^6_{ij}$, and then again these were individually decomposed into integrals over various subregions with associated correlation functions. It thus may appear that many different integrals are needed to compute these fluctuations. However, as we saw in \eqref{eq:L_i in terms of L} and \eqref{eq:J4ij in terms of J4}, computing a single integral is enough to get all the other $i,j$ elements. It turns out that this can be done for any region, and the main reason is due to the following property of the Poisson distribution
\begin{equation}\label{eq:P_i as derivatives of P_0}
    P_i(V) = \frac{\rho^i}{i!} \left[-\frac{d}{d\rho}\right]^iP_0(V)= \frac{\rho^i}{i!} \left[-\frac{d}{d\rho}\right]^ie^{-\rho V}.
\end{equation}
As an example, let us consider the region $\DiamondOverlapZeroInsideBs$. Define the following integral
\begin{gather}\label{eq:int[region] definition}
    \begin{aligned}
    \texttt{int}\left[\DiamondOverlapZeroInsideBs\right] &\equiv \int_{\DiamondOverlapZeroInsideBs}P_0(V_a)_{\rho_a}P_0(V_b)_{\rho_b}P_0(V_c)_{\rho_c}\,dV_{y_1}dV_{y_2}dV_{x_1}dV_{x_2}\\
    &=\int_{\DiamondOverlapZeroInsideBs}e^{-\rho_a V_a-\rho_b V_b-\rho_c V_c}\,dV_{y_1}dV_{y_2}dV_{x_1}dV_{x_2},
    \end{aligned}
\end{gather}
where $P_0(V_a)_{\rho_a} \equiv e^{-\rho_a V_a}$. Here $V_a$, $V_b$ and $V_c$ are the volumes of the three non-overlapping subregions, as used previously (e.g. see Section \ref{sec:Correlation function examples}). By using the property \eqref{eq:P_i as derivatives of P_0} we find
\begin{gather}
    \begin{aligned}
    \texttt{int}_{\alpha\beta\gamma}\left[\DiamondOverlapZeroInsideBs\right]&\equiv 
    \frac{\rho^{\alpha+\beta+\gamma}}{\alpha!\beta!\gamma!}
    \left[-\frac \partial{\partial\rho_a}\right]^\alpha
    \left[-\frac \partial{\partial\rho_b}\right]^\beta
    \left[-\frac \partial{\partial\rho_c}\right]^\gamma
    \bigg |_{\rho_a=\rho_b=\rho_c=\rho}
    \texttt{int}\left[\DiamondOverlapZeroInsideBs\right],\\
    &= \int_{\DiamondOverlapZeroInsideBs}P_\alpha(V_a)P_\beta(V_b)P_\gamma(V_c)\,dV_{y_1}dV_{y_2}dV_{x_1}dV_{x_2}.
    \end{aligned}
\end{gather}
As defined earlier, we can use the notation $\int_{\DiamondOverlapZeroInsideBs}$ to mean an integral over the region $\DiamondOverlapZeroInsideBs$ with the integrand \eqref{eq:DiamondOverlapZeroInsideB correlator}. In other words, we can express it as
\begin{gather}
    \begin{aligned}
        \int_{\DiamondOverlapZeroInsideBs}
        &= \rho^4\sum_{\alpha\beta\gamma} \int_{\DiamondOverlapZeroInsideBs}P_\alpha(V_a)P_\beta(V_b)P_\gamma(V_c)\,dV_{y_1}dV_{y_2}dV_{x_1}dV_{x_2},\\
        &= \rho^4\sum_{\alpha\beta\gamma} \texttt{int}_{\alpha\beta\gamma}\left[\DiamondOverlapZeroInsideBs\right],
    \end{aligned}
\end{gather}
where the domain of the $\alpha\beta\gamma$ sum depends on the specific region; for example $\alpha, \beta,\gamma>0$, $\alpha+\beta = i$ and $\beta+\gamma = j$ for this region. Therefore,  we can compute $\int_{\DiamondOverlapZeroInsideBs}$ for any $i$ and $j$ from a single integral \eqref{eq:int[region] definition}.

As another example, for $\DiamondOverlapYAs$ we only need
\begin{equation}
    \texttt{int}\left[\DiamondOverlapYAs\right] = \int_{\DiamondOverlapYAs}e^{-\rho_a V_a-\rho_b V_b-\rho_c V_c}\,dV_{y_1}dV_{y_2}dV_x,
\end{equation}
since from derivatives of this we can find
\begin{gather}
    \begin{aligned}
        \int_{\DiamondOverlapYAs}
        &= \rho^3\sum_{\alpha\beta\gamma} \int_{\DiamondOverlapYAs}P_\alpha(V_a)P_\beta(V_b)P_\gamma(V_c)\,dV_{y_1}dV_{y_2}dV_x,\\
        &= \rho^3\sum_{\alpha\beta\gamma} \texttt{int}_{\alpha\beta\gamma}\left[\DiamondOverlapYAs\right].
    \end{aligned}
\end{gather}
 The total number of integrals we need to compute for the action fluctuations, is therefore one for each region appearing in our expressions. Everything we need to compute matrices such as $J^6_{ij}$ in \eqref{eq:J6ij off diagonal} for all $i$ and $j$, is a core set of relevant single integrals such as \eqref{eq:int[region] definition}, based on the diagrams appearing in their decompositions. 

In Appendix \ref{app:example} we show explicit examples of parametrizations of some of these integrals.

\section{Conclusions and Outlook}\label{sec: Conclusions}
We have presented a strategy for calculating fluctuations and correlations in causal set theory. These fluctuations are those of causal set quantities with respect to averaging over an ensemble of Poisson sprinklings of a given spacetime manifold. An example of such a fluctuation would be the standard deviation of the number of elements sprinkled into a region of a manifold with spacetime volume $V$; we know the answer in this case is $\Delta N=\sqrt{\rho V}$, but for more general quantities this standard deviation must be computed. The notation and formalism we have laid out in this paper  streamline a wide range of calculations of this kind. 

We paid particular attention to the correlations involved in and the fluctuations of the causal set action. This action, which is a discrete analogue of the Einstein-Hilbert action, depends on causal intervals between causal set elements and on the number of elements within these intervals or diamonds. Therefore the correlations we needed to take into account, for example to compute the standard deviation of the action, included those between pairs of (potentially overlapping) diamonds with different numbers of elements in them. It also turned out to be crucial to carefully treat the expectation of an element lying in an infinitesimal region -- essentially at a point (for example at the endpoints of a causal diamond, in order to form a causal set causal interval). These can induce additional correlations, for example when two diamonds share one or more endpoints.

After careful account of all correlations, we expressed the fluctuations of the action in terms of a core set of integrals that need to be computed. We developed a convenient graphical notation for expressing both the domains of these integrals as well as the correlations they represented. Finally, we showed that for each subregion with a distinct correlation type, there is only one core integral that needs to be computed to obtain the answer for more general index combinations. The fact that we do not need to compute separate integrals for each pair of index values considerably reduces the computational work needed to calculate the fluctuations. 

In future work, we will apply the formalism we have developed in this paper to particular spacetime manifolds, to explicitly evaluate the fluctuations. As a starting application we will consider Minkowski spacetime, where the parametrization of the necessary domains would be relatively simpler than in curved spacetimes. We further aim to apply our formalism in Friedman-Robertson-Walker type cosmological spacetimes to address the main question that motivated this work, which is whether or not the fluctuations of the action can be used to model Everpresent $\Lambda$. Finally, as we have mentioned already, our work opens the door to a large number of similar calculations of fluctuations of generic causal set quantities. Knowledge of these fluctuations will further inform us about the statistical properties of causal sets and can be used in many different contexts such as in questions of dynamics, numerical convergence, and phenomenological model building.

\appendix
\section{Expressions for Correlation Functions}\label{app:CorrelationFunctions}

In Section \ref{sec:Correlation function examples} we developed a simple graphical notation for correlation functions of the type 
\begin{equation}
    \avg{\zeta_i(x_1,y_1)\zeta_j(x_2,y_2)\chi(dV_{x_1})\chi(dV_{y_1})\chi(dV_{x_2})\chi(dV_{y_2})},
\end{equation}
for example see \eqref{eq:Graphical notation for <zeta zeta ...>}. For convenience, in this appendix we will provide the explicit evaluations of some of these correlation functions using the basic steps from Section \eqref{sec:CorrelationFunctions}:
\begin{equation}\label{eq:J4 correlation function}
    \DiamondOverlapZeroInsideC {-1} ~~=~~ \frac{\left(\rho V_1\right)^i\left(\rho V_2\right)^j}{i!\,j!} e^{-\rho |V_1\cup V_2|}\,
       \rho^3  dV_{y_1}dV_{y_2}dV_{x},
\end{equation}

\begin{equation}\label{eq:DiamondOverlapZeroInsideB correlator}
    \DiamondOverlapZeroInsideB  ~~=~~ \sum\limits_{\substack{\alpha,\beta,\gamma\geq 0\\ \alpha+\beta=i\\ \beta+\gamma=j }} \frac{(\rho V_a)^\alpha(\rho V_b)^\beta(\rho V_c)^\gamma}{\alpha!\,\beta!\,\gamma!}\, e^{-\rho |V_1\cup V_2|}\rho^4 dV_{x_1} dV_{y_1} dV_{x_2} dV_{y_2},
\end{equation}

\begin{equation}
    \DiamondOverlapOneInsideD  ~~=~~ \sum\limits_{\substack{\alpha,\beta,\gamma\geq 0\\ \alpha+\beta+1=i\\ \beta+\gamma=j }} \frac{(\rho V_a)^\alpha(\rho V_b)^\beta(\rho V_c)^\gamma}{\alpha!\,\beta!\,\gamma!}\, e^{-\rho |V_1\cup V_2|}
    \,\rho^4 dV_{x_1} dV_{y_1} dV_{x_2} dV_{y_2},
\end{equation}

\begin{equation}
    \DiamondOverlapTwoInsideA  ~~=~~ \sum\limits_{\substack{\alpha,\beta,\gamma\geq 0\\ \alpha+\beta+1=i\\ \beta+\gamma+1=j }} \frac{(\rho V_a)^\alpha(\rho V_b)^\beta(\rho V_c)^\gamma}{\alpha!\,\beta!\,\gamma!}\, e^{-\rho |V_1\cup V_2|}
    \,\rho^4 dV_{x_1} dV_{y_1} dV_{x_2} dV_{y_2},
\end{equation}

\begin{equation} \label{diamond two inside red}
    \DiamondOverlapTwoInsideC   ~~ =~~ \frac{(\rho V_a)^{i-j-2}(\rho V_b)^j}{(i-j-2)!\,j!}\, e^{-\rho V_1}
    \,\rho^4 dV_{x_1} dV_{y_1} dV_{x_2} dV_{y_2}.
\end{equation}
Note that \eqref{diamond two inside red} should be equal to zero when $i<j+2$ as this configuration is not possible in the domain above. We take $\frac 1{n!}$ to mean $\frac 1{\Gamma(n+1)}$, which yields $\Gamma(n+1) = \infty$ when $n$ is a negative integer, making the expression  automatically vanish. Further correlations are given by: 

\begin{equation}
   \DiamondOverlapXA   ~~ =~~ \sum\limits_{\substack{\alpha,\beta,\gamma\geq 0\\ \alpha+\beta=i\\ \beta+\gamma=j }} \frac{(\rho V_a)^\alpha(\rho V_b)^\beta(\rho V_c)^\gamma}{\alpha!\,\beta!\,\gamma!}\, e^{-\rho |V_1\cup V_2|}
   \,\rho^3 dV_{x}  dV_{y_1} dV_{y_2},
\end{equation}

\begin{equation}
   \DiamondOverlapXB   ~~ =~~ \frac{(\rho V_a)^{i-j-1}(\rho V_b)^j}{(i-j-1)!\,j!}\, e^{-\rho V_1}
   \,\rho^3 dV_{x} dV_{y_1}  dV_{y_2},
\end{equation}
and 
\begin{equation}
   \DiamondOverlapXC  ~~ =~~ \frac{(\rho V_b)^i(\rho V_c)^{j-i-1}}{(j-i-1)!\,i!}\, e^{-\rho V_2}
   \,\rho^3 dV_{x} dV_{y_1} dV_{y_2}.
\end{equation}

\section{Products of $\chi$ Functions} \label{app: products of chi}
\noindent The decomposition of \eqref{eq:kij general} into \eqref{eq:K_ij into J1-J6} can be seen more formally in the following way. Using the definition of $\chi$ in \eqref{eq:chi measure written in terms of delta functions}, we can split the product of the measures into a part where $x_1=x_2$ and a part where $x_1\neq x_2$, i.e. 
\begin{gather} \label{eq:chichi measure 1}
    \begin{aligned}
    \chi(dV_{x_1})\chi(dV_{x_2}) &= \sum_{z_1\in\mathcal C}\sum_{z_2\in\mathcal C}\delta^{(d)}(z_1-x_1)\delta^{(d)}(z_2-x_2)dV_{x_1}dV_{x_2}\\
    &=\left[\sum_{z\in\mathcal C}\delta^{(d)}(z-x_1)\delta^{(d)}(z-x_2) + \sum_{z_1\not = z_2\in\mathcal C}\delta^{(d)}(z_1-x_1)\delta^{(d)}(z_2-x_2)\right]dV_{x_1}dV_{x_2},\\
    &= \delta^{(d)}(x_1-x_2)\chi(dV_{x_1})dV_{x_2} + \chi(dV_{x_1},dV_{x_2}),
    \end{aligned}
\end{gather}
and similarly for the $y$ measures
\begin{gather} \label{eq:chichi measure 2}
    \begin{aligned}
    \chi(dV_{y_1})\chi(dV_{y_2}) &= \sum_{z_1\in\mathcal C}\sum_{z_2\in\mathcal C}\delta^{(d)}(z_1-y_1)\delta^{(d)}(z_2-y_2)dV_{y_1}dV_{y_2}\\
    &=\left[\sum_{z\in\mathcal C}\delta^{(d)}(z-y_1)\delta^{(d)}(z-y_2) + \sum_{z_1\not = z_2\in\mathcal C}\delta^{(d)}(z_1-y_1)\delta^{(d)}(z_2-y_2)\right]dV_{y_1}dV_{y_2}\\
    &= \delta^{(d)}(y_1-y_2)\chi(dV_{y_1})dV_{y_2} + \chi(dV_{y_1},dV_{y_2}).
    \end{aligned}
\end{gather}
Here we have defined the measure
\begin{gather}
    \begin{aligned}
        \chi(dV_{x_1},dV_{x_2}) &\equiv \sum_{z_1\not = z_2\in\mathcal C}\delta^{(d)}(z_1-x_1)\delta^{(d)}(z_2-x_2)dV_{x_1}dV_{x_2}\\
            &= \left(1-\delta^{(d)}(x_1-x_2)\right)\chi(dV_{x_1})\chi(dV_{x_2}).
    \end{aligned}\label{pifunc}
\end{gather}
 We can more generally define
\begin{gather}
    \begin{aligned}
        \chi(dV_{x_1},\dots, dV_{x_n}) &\equiv \sum_{z_1\neq \cdots\neq z_n\in\mathcal C}\delta^{(d)}(z_1-x_1)\cdots\delta^{(d)}(z_n,x_n)\,dV_{x_1}\cdots dV_{x_n}.
    \end{aligned}
\end{gather}
Essentially, $  \chi(dV_{x_1},\dots, dV_{x_n})$ represents the product of $\chi$'s while enforcing the points to be distinct. Using these, we can further decompose $\chi(dV_{x_1},dV_{x_2})\chi(dV_{y_1},dV_{y_2})$ into subregions where
$\{x_1=y_2\}$, $\{x_2=y_1\}$, $\{x_1=y_1\}$, $\{x_2=y_2\}$ and $\{x_1\neq x_2\neq y_1\neq y_2\}$. In total $\chi(dV_{x_1})\chi(dV_{x_2})\chi(dV_{y_1})\chi(dV_{y_2})$ splits into 8 terms. By plugging these into \eqref{eq:kij general}, we recover the six terms in $\eqref{eq:K_ij into J1-J6}$ while two of the terms ($\{x_1=y_1\}$ and $\{x_2=y_2\}$) vanish due to the integration domain.\footnote{Note that in these cases, the $\zeta(x,y)$ would also force the integrals to vanish, as the causal diamond between x and y would be the empty set (see \eqref{zeta_xy}).} Finally, we use
\begin{gather}
    \begin{aligned}
        \chi(dV_{x_1},\dots, dV_{x_n})  = \chi(dV_{x_1})\cdots \chi(dV_{x_n}), \qquad \text{when} \quad x_1\neq \cdots\neq x_n,
    \end{aligned}
\end{gather}
to find the integral forms of $J^1, \dots, J^6$. 
\section{Parametrization of Integration Domains}\label{app:example}

In this paper, we reduced the computation of the fluctuations of quantities such as the action to evaluating integrals such as 

\begin{equation}
    \int_{\DiamondOverlapYAs}, \qquad \int_{\DiamondOverlapZeroInsideBs}, \qquad \int_{\DiamondOverlapOneInsideAs}, \qquad \dots
\end{equation}
This notation might seem abstract, hence in this appendix we  illustrate how to evaluate these integrals concretely.

Recall that these diagrams represent all possible submanifolds of $\mathcal M\times\mathcal M\times\mathcal M\times\mathcal M$ where the points $x_1, y_1, x_2, y_2$ are related as in the diagrams. For example the diagram $\DiamondOverlapYAs$ which was defined in \eqref{eq:xxy_regions} can also be written as
\begin{equation}
    \DiamondOverlapYAs = \big\{(x_1,x_2,y)\in\mathcal M^3 \;\big|\; x_1\in \mathcal M, x_2\in \mathcal M\backslash \mathcal J^+(x_1)\cup \mathcal J^-(x_1) \text{ and } y\in \mathcal J^-(x_1)\cap \mathcal J^-(x_2)\big\}.
\end{equation}
In other words, if $x_1$ is chosen to be any point, then $x_2$ can be any point that is causally unrelated to $x_1$ (neither in the past nor future of $x_1$) and $y$ can be any point in the past of both of them. In the above, $\mathcal M\backslash \mathcal J^+(x_1)\cup \mathcal J^-(x_1)$ means the manifold $\mathcal M$ with the submanifold $\mathcal J^+(x_1)\cup \mathcal J^-(x_1)$ carved out.

Performing the integral over $\DiamondOverlapYAs$ requires parametrizing all these submanifolds, for a given $\mathcal M$. However, it turns out that for any diagram one can reduce the needed parametrizations to three types: (1) $\mathcal M$, (2) $\mathcal J^-(x)$ and (3) $\mathcal J^-(x_1)\cap J^-(x_2)$. %Thus no future cones $J^-(x)$ are needed, nor unions or subtractions of manifolds. 
For example
\begin{gather}
    \begin{aligned}
        \int_{\DiamondOverlapYAs} = \int_{\mathcal M}\int_{\mathcal M\backslash \mathcal J^+(x_1)\cup \mathcal J^-(x_1)}
        \int_{\mathcal J^-(x_1)\cap \mathcal J^-(x_2)} dV_{x_1}dV_{x_2}dV_{y},
    \end{aligned}
\end{gather}
can be rewritten as
\begin{gather}
    \begin{aligned}
        \int_{\DiamondOverlapYAs} &= \int_{\mathcal M}\int_{\mathcal M}
   \int_{\mathcal J^-(x_1)\cap \mathcal J^-(x_2)} dV_{x_1}dV_{x_2}dV_{y}\\
    &\qquad-\int_{\mathcal M}\int_{\mathcal J^-(x_1)}
   \int_{\mathcal J^-(x_2)} dV_{x_1}dV_{x_2}dV_{y}
    -\int_{\mathcal M}\int_{\mathcal J^-(x_2)}
   \int_{\mathcal J^-(x_1)} dV_{x_2}dV_{x_1}dV_{y}.
     \end{aligned}
\end{gather}
Here we first take $x_1$ and $x_2$ to be unconstrained, then subtract all the contributions from when they are causally related. Note that the last two integrals are equal if the integrand is symmetric in $x_1$ and $x_2$. Other diagrams can also be written in this form.

For concreteness, below we will consider examples of explicit parametrizations of $\mathcal J^-(x)$ and $\mathcal J^-(x_1)\cap \mathcal J^-(x_2)$. In general $\mathcal J^-(x_1)\cap \mathcal J^-(x_2)$ is harder to parametrize than $\mathcal J^-(x)$. For  simplicity, we will consider a $1+1$ dimensional example for $\mathcal J^-(x_1)\cap \mathcal J^-(x_2)$, and a $3+1$ dimensional example for $\mathcal J^-(x)$.

\subsection{Example: Parametrization of $\mathcal J^-(x_1)\cap \mathcal J^-(x_2)$ in a $1+1$D Minkowski Diamond}

Consider the case of a causal diamond $\mathcal M = I(q,p)$ in $1+1$ dimensional Minkowski spacetime.
\begin{figure}[htb!]
    \centering
    \JJExampleDiagram
    \caption{Setup of $\mathcal J^-(x_1)\cap \mathcal J^-(x_2)$ in a $1+1$D Minkowski Diamond.}
    \label{fig:J(x1) intersect with J(x2)}
\end{figure}
For simplicity, we restrict to pairs of points that are at the same Cartesian time and equal spatial distance (but with opposite sign) from the origin, namely $x_1=(t,z)$ and $x_2=(t,-z)$.
Let us also place $p$ at the origin and $q$ at height $h$, $q=(h,0)$. See Figure \ref{fig:J(x1) intersect with J(x2)} for the setup.  It is convenient to work in lightcone coordinates $u$ and $v$ where
\begin{equation}
    u=\frac{1}{\sqrt2} (t-z),~~~~~~     v=\frac{1}{\sqrt2} (t+z).
\end{equation}
In this coordinate system our integral becomes
\begin{equation}
    \int_{\mathcal J ^-(x_1)\cap\mathcal J^-(x_2)} f(y) d^2y=\int_0^{\frac{t-z}{\sqrt2}}du\int_0^{\frac{t-z}{\sqrt2}}dv\, f(u,v)
\end{equation}

\subsection{Example: Parametrization of $\mathcal J^-(x)$ in a $3+1$D Minkowski Diamond}

Consider a causal diamond $\mathcal M = I(q,p)$ with local Cartesian coordinates (chart) $Y = (Y^0, \mathbf Y)$ and metric
\begin{equation}
    ds^2 = \eta_{\mu\nu}dY^\mu dY^{\nu},
\end{equation}
where $\eta = \text{diag}(1,\negative 1,\negative 1, \negative 1)$.
In this coordinate system, $h$ is the height between the top and bottom corners, $q$ and $p$. Their coordinates are  $P=(0, \mathbf 0)_Y$ and $Q=(h, \mathbf0)_Y$.

Let $x$ be a point in $\mathcal M$ with coordinates $X = (X^0, \mathbf X)_Y$. For any $x\in\mathcal M$, we are interested in the submanifold $\mathcal J^-(x)=I(x,p)\subset\mathcal M$. In this subsection we will parametrize this submanifold as a function of the points $p, q$ and $x$.\footnote{The treatment in this subsection is similar to the one in \cite{finiterho}.}
\\

\noindent
The proper time $\tau_{px}$ between the points $p$ and $x$, is given by
\begin{equation}
    \tau^2_{px} = (X^0)^2 - \left\|\mathbf X\right\|^2.
\end{equation}
We want to map the point $y\in I(x,p)$ from $Y$ coordinates into $\tilde Y$ coordinates such that the interval $I(x,p)$  in this coordinate system takes the form of a standard upright diamond i.e. $\tilde X = (\tilde X^0, \mathbf 0)_{\tilde Y}$, $\tilde P = (\tilde P^0, \mathbf 0)_{\tilde Y}$ and
\begin{equation}
    \tau^2_{px} = \left( \tilde X^0 -\tilde P^0 \right)^2.
\end{equation}
The mapping from $Y$ to $\tilde Y$ will only use transformations in the Poincar\'e group $\mathbb P = \mathbb R^4\rtimes SO(3,1)$ and the metric will therefore remain of the Minkowski form
\begin{equation}
    ds^2 = \eta_{\mu\nu}d\tilde Y^\mu d\tilde Y^{\nu}.
\end{equation}
In these coordinates, null geodesics will remain straight lines at 45 degrees.
\\

\noindent First we bring the origin to $x$, using a translation
\begin{equation}
    Y^\mu \longrightarrow Y^\mu-X^\mu,
\end{equation}
mapping the points $p$ and $x$ to
\begin{equation}
    P\longrightarrow P-X= (-X^0, -\mathbf X)\qquad \text{and}\qquad X\longrightarrow X-X=(0, \mathbf 0).
\end{equation}
We will now use $SO(3,1)$ Lorentz-group transformations to bring $p$ to the form $\tilde P = (\tilde P^0, \mathbf 0)_{\tilde Y}$. Such transformations will keep $\tilde X = (0, \mathbf 0)_{\tilde Y}$ and thus give us the coordinate system we are seeking.

\noindent First we need a rotation $R\in SO(3)\subset SO(3,1)$ such that $p$ has only one non-zero spatial component (say, the $x$-direction)
\begin{equation}
P \longrightarrow R(P-X) \overset != (-X^0, \|\mathbf X\|,0,0).
\end{equation}
This rotation matrix is given by
\begin{equation}
    R = \exp(\theta\, \mathbf n\cdot\mathbf L),
\end{equation}
where
\begin{equation}
    \mathbf n = -\frac{\mathbf X}{\|\mathbf X\|}\times (1, 0, 0), \qquad \cos(\theta) = -\frac{X^1}{\|\mathbf X\|},
\end{equation}
and $\mathbf L = (L_x, L_y, L_z)$ are the infinitesimal rotations along the three axes (as $4\times 4$ anti-symmetric matrices).
% Explicitly, this rotation is given by
% \begin{equation}
%     R = 
%     \begin{pmatrix}
%         1 & 0 & 0 & 0\\
%         0 & -\frac{X^1}{\|\mathbf X\|}   &   -\frac{X^2}{\|\mathbf X\|}    &    -\frac{X^3}{\|\mathbf X\|}\\
%        0 & \frac{X^2}{\|\mathbf X\|} & \frac{-X^1(X^2)^2 + (X^3)^2\|\mathbf X\|}{\left[(X^2)^2 + (X^3)^2\right]\|\mathbf X\|} & -\frac{X^2 X^3\left(X^1 + \|\mathbf X\|\right)}{\left[(X^2)^2 + (X^3)^2\right]\|\mathbf X\|}   \\
%        0 & \frac{X^3}{\|\mathbf X\|}  &  -\frac{X^2 X^3\left(X^1 +\|\mathbf X\|\right)}{\left[(X^2)^2 + (X^3)^2\right]\|\mathbf X\|} & \frac{-X^1(X^3)^2 + (X^2)^2\|\mathbf X\|}{\left[(X^2)^2 + (X^3)^2\right]\|\mathbf X\|}
%     \end{pmatrix}.
% \end{equation}
Next we need a boost $B\in SO(3,1)\subset\mathbb P$, in the $x$-direction and parametrized by $w$
\begin{equation}
    B\cdot
        \begin{pmatrix}
            t \\ 
            x \\
            y \\z
        \end{pmatrix}
        =
        \begin{pmatrix}
            \gamma (t-xw) \\
            \gamma (x-tw) \\ 
            y \\
            z
        \end{pmatrix}
\end{equation}
in order to eliminate the remaining non-zero spatial component. The transformation leads to
\begin{equation}
    BR(P - X) =  \gamma\begin{pmatrix}
                 -X^0 -\|\mathbf X\|w\\
                 \|\mathbf X\|+X^0 w \\ 
                0 \\
                0
            \end{pmatrix}.
\end{equation}
We choose the velocity $w$ such that the spatial component vanishes
\begin{equation}
    w  = -\frac{\|\mathbf X\|}{X^0}.
\end{equation}
Thus finally the coordinate $P$ is mapped to
\begin{equation}
    \tilde P = BR(P-X) = \left(\tilde P^0,\mathbf 0\right)_{\tilde Y},
\end{equation}
where
\begin{equation}
    \tilde P^0 = \gamma\left(-X^0+\frac{\|\mathbf X\|^2}{X^0}\right).
\end{equation}
The coordinate transformation is thus
\begin{equation}
    \tilde Y=BR(Y-X).
\end{equation}
In this coordinate system the region $\mathcal J^-(x) = I(x,p)$ will look like a standard upright diamond between the points $\tilde P = (\tilde P^0, \mathbf 0)_{\tilde Y}$ and $\tilde X = (0, \mathbf 0)_{\tilde Y}$. We will now parametrize $\mathcal J^-(x)$ using radial lightcone coordinates. First we use spherical coordinates $(\tilde Y^0, \tilde Y^1, \tilde Y^2, \tilde Y^3)\longrightarrow (\tilde Y^0, r, \theta, \phi)$ where
\begin{gather}
    \begin{aligned}
          \tilde Y^1 &= r\cos\theta,\\
          \tilde Y^2 &= r\cos\phi\sin\theta,\\
          \tilde Y^3 &= r\sin\phi\sin\theta.
    \end{aligned}
\end{gather}
In particular, $r = \|\tilde{\mathbf Y}\|$. Then we use the radial lightcone coordinates $(\tilde Y^0, r, \theta, \phi)\longrightarrow (u, v, \theta, \phi)$
\begin{gather}
    \begin{aligned}
          u &= \frac 1{\sqrt 2}\left(\tilde Y^0 - \|\tilde{\mathbf Y}\|\right),\\
          v &= \frac 1{\sqrt{2}}\left(\tilde Y^0 + \|\tilde{\mathbf Y}\|\right).
    \end{aligned}
\end{gather}
The combined coordinate transformation is thus
\begin{gather}\label{ytildeuv}
    \begin{aligned}
          \tilde Y^0 &= \frac{(v+u)}{\sqrt 2},\\
          \tilde Y^1 &= \frac{(v-u)}{\sqrt 2}\cos\theta,\\
    \end{aligned}
    \qquad
    \begin{aligned}
          \tilde Y^2 &= \frac{(v-u)}{\sqrt 2}\cos\phi\sin\theta,\\
          \tilde Y^3 &= \frac{(v-u)}{\sqrt 2}\sin\phi\sin\theta.
    \end{aligned}
\end{gather}
The volume form is given by
\begin{gather}
    \begin{aligned}
        d^4y &= dY^0dY^1dY^2dY^3\\
             &=d\tilde Y^0d\tilde Y^1d\tilde Y^2d\tilde Y^3,\\
             &=r^2 d\tilde Y^0 dr\,d^2\Omega,\\
             &=\frac 12\left(v-u\right)^2 du\,dv\,d^2\Omega.
    \end{aligned}
\end{gather}

\begin{equation}
    \mathcal J^-(x) = \left\{(u,v,\theta, \phi)\,\, \Big|\,\, u\in\left[0,-\frac 1{\sqrt 2} \tau_{px}\right]\wedge
            v\in\left[0,u\right]\wedge
            \theta\in\left[0,\pi\right]\wedge
            \phi\in\left[0,2\pi\right]\right\},
\end{equation}
where $\tau_{px} = \tilde P^0$. The mapping $Y = R^{-1}B^{-1}\tilde Y + X$ together with equations \eqref{ytildeuv}, allows us to express $Y = Y(u,v, \theta, \phi)$. In particular, we can compute the integrals over $\mathcal J^-(x)$ as
\begin{equation}
    \int_{\mathcal J^-(x)} f(Y)d^4Y = \int_{S^2}d^2\Omega \int_0^{-\frac{\tau_{px}}{\sqrt 2}}du\int_0^udv\;\frac 12(v-u)^2\: f(Y(u,v, \theta, \phi))
\end{equation}

\acknowledgments

We acknowledge the support of the European Consortium for Astroparticle Theory (EuCAPT) in the form of an Exchange Travel Grant held at CERN. HM is supported by the Leverhulme Trust Early Career Fellowship. YY acknowledges financial support from Research Ireland under Grant number 22/PATH-S/10704, as well as support from a Leverhulme Trust Research Project Grant.  
MZ acknowledges financial support by the Center for Research and Development in Mathematics and Applications (CIDMA) through the Portuguese Foundation for Science and Technology (FCT -- Fundaç\~ao para a Ci\^encia e a Tecnologia) through projects: UIDB/04106/2020 (with DOI identifier \url{https://doi.org/10.54499/UIDB/04106/2020}); UIDP/04106/2020 (DOI identifier \url{https://doi.org/10.54499/UIDP/04106/2020});  PTDC/FIS-AST/3041/2020 (DOI identifier \url{http://doi.org/10.54499/PTDC/FIS-AST/3041/2020}); CERN/FIS-PAR/0024/2021 (DOI identifier \url{http://doi.org/10.54499/CERN/FIS-PAR/0024/2021}); 2022.04560.PTDC (DOI identifier \url{https://doi.org/10.54499/2022.04560.PTDC}); and 2022.00721.CEECIND (DOI identifier \url{https://doi.org/10.54499/2022.00721.CEECIND/CP1720/CT0001}).

\bibliography{references}

\providecommand{\href}[2]{#2}\begingroup\raggedright\begin{thebibliography}{10}

\bibitem{PhysRevLett.59.521}
L.~Bombelli, J.~Lee, D.~Meyer and R.D.~Sorkin, \emph{Space-time as a causal
  set}, \href{https://doi.org/10.1103/PhysRevLett.59.521}{\emph{Phys. Rev.
  Lett.} {\bfseries 59} (1987) 521}.

\bibitem{Surya:2019ndm}
S.~Surya, \emph{{The causal set approach to quantum gravity}},
  \href{https://doi.org/10.1007/s41114-019-0023-1}{\emph{Living Rev. Rel.}
  {\bfseries 22} (2019) 5} [\href{https://arxiv.org/abs/1903.11544}{{\ttfamily
  1903.11544}}].

\bibitem{Sorkin:2005qx}
R.D.~Sorkin, \emph{{Ten theses on black hole entropy}},
  \href{https://doi.org/10.1016/j.shpsb.2005.02.002}{\emph{Stud. Hist. Phil.
  Sci. B} {\bfseries 36} (2005) 291}
  [\href{https://arxiv.org/abs/hep-th/0504037}{{\ttfamily hep-th/0504037}}].

\bibitem{Nielsen:1983rb}
H.B.~Nielsen and M.~Ninomiya, \emph{{The Adler-Bell-Jackiw anomaly and Weyl
  fermions in a crystal}},
  \href{https://doi.org/10.1016/0370-2693(83)91529-0}{\emph{Phys. Lett. B}
  {\bfseries 130} (1983) 389}.

\bibitem{Nielsen:1980rz}
H.B.~Nielsen and M.~Ninomiya, \emph{{Absence of Neutrinos on a Lattice. 1.
  Proof by Homotopy Theory}},
  \href{https://doi.org/10.1016/0550-3213(82)90011-6}{\emph{Nucl. Phys. B}
  {\bfseries 185} (1981) 20}.

\bibitem{Nielsen:1981xu}
H.B.~Nielsen and M.~Ninomiya, \emph{{Absence of Neutrinos on a Lattice. 2.
  Intuitive Topological Proof}},
  \href{https://doi.org/10.1016/0550-3213(81)90524-1}{\emph{Nucl. Phys. B}
  {\bfseries 193} (1981) 173}.

\bibitem{Nielsen:1981hk}
H.B.~Nielsen and M.~Ninomiya, \emph{{No Go Theorem for Regularizing Chiral
  Fermions}}, \href{https://doi.org/10.1016/0370-2693(81)91026-1}{\emph{Phys.
  Lett. B} {\bfseries 105} (1981) 219}.

\bibitem{Kravec:2013pua}
S.M.~Kravec and J.~McGreevy, \emph{{A gauge theory generalization of the
  fermion-doubling theorem}},
  \href{https://doi.org/10.1103/PhysRevLett.111.161603}{\emph{Phys. Rev. Lett.}
  {\bfseries 111} (2013) 161603}
  [\href{https://arxiv.org/abs/1306.3992}{{\ttfamily 1306.3992}}].

\bibitem{Saravani:2014gza}
M.~Saravani and S.~Aslanbeigi, \emph{{On the Causal Set-Continuum
  Correspondence}},
  \href{https://doi.org/10.1088/0264-9381/31/20/205013}{\emph{Class. Quant.
  Grav.} {\bfseries 31} (2014) 205013}
  [\href{https://arxiv.org/abs/1403.6429}{{\ttfamily 1403.6429}}].

\bibitem{Henson:2015fha}
J.~Henson, D.P.~Rideout, R.D.~Sorkin and S.~Surya, \emph{{Onset of the
  Asymptotic Regime for Finite Orders}},
  \href{https://arxiv.org/abs/1504.05902}{{\ttfamily 1504.05902}}.

\bibitem{Carlip:2024uny}
S.~Carlip, \emph{{Causal sets and an emerging continuum}},
  \href{https://doi.org/10.1007/s10714-024-03281-1}{\emph{Gen. Rel. Grav.}
  {\bfseries 56} (2024) 95} [\href{https://arxiv.org/abs/2405.14059}{{\ttfamily
  2405.14059}}].

\bibitem{Johnston:2009fr}
S.~Johnston, \emph{{Feynman Propagator for a Free Scalar Field on a Causal
  Set}}, \href{https://doi.org/10.1103/PhysRevLett.103.180401}{\emph{Phys. Rev.
  Lett.} {\bfseries 103} (2009) 180401}
  [\href{https://arxiv.org/abs/0909.0944}{{\ttfamily 0909.0944}}].

\bibitem{X:2023ewv}
N.~X., \emph{{Quantum Field Theory on Causal Sets}},  (2024),
  \href{https://doi.org/10.1007/978-981-19-3079-9_80-1}{DOI}
  [\href{https://arxiv.org/abs/2306.04800}{{\ttfamily 2306.04800}}].

\bibitem{Albertini:2024srq}
E.~Albertini, F.~Dowker, A.~Nasiri and S.~Zalel, \emph{{In-in correlators and
  scattering amplitudes on a causal set}},
  \href{https://doi.org/10.1103/PhysRevD.109.106014}{\emph{Phys. Rev. D}
  {\bfseries 109} (2024) 106014}
  [\href{https://arxiv.org/abs/2402.08555}{{\ttfamily 2402.08555}}].

\bibitem{Sorkin:2012sn}
R.D.~Sorkin, \emph{{Expressing entropy globally in terms of (4D)
  field-correlations}},
  \href{https://doi.org/10.1088/1742-6596/484/1/012004}{\emph{J. Phys. Conf.
  Ser.} {\bfseries 484} (2014) 012004}
  [\href{https://arxiv.org/abs/1205.2953}{{\ttfamily 1205.2953}}].

\bibitem{Yazdi:2022hhv}
Y.K.~Yazdi, \emph{{Entanglement Entropy and Causal Set Theory}},
  \href{https://arxiv.org/abs/2212.13586}{{\ttfamily 2212.13586}}.

\bibitem{Homsak:2024tce}
V.~Hom\v{s}ak and S.~Veroni, \emph{{Boltzmannian state counting for black hole
  entropy in causal set theory}},
  \href{https://doi.org/10.1103/PhysRevD.110.026015}{\emph{Phys. Rev. D}
  {\bfseries 110} (2024) 026015}
  [\href{https://arxiv.org/abs/2404.11670}{{\ttfamily 2404.11670}}].

\bibitem{Box2d}
R.D.~Sorkin, \emph{Does locality fail at intermediate length-scales},
  \href{https://arxiv.org/abs/gr-qc/0703099}{{\ttfamily gr-qc/0703099}}.

\bibitem{BDAction}
D.M.T.~Benincasa and F.~Dowker, \emph{Scalar curvature of a causal set},
  \href{https://doi.org/10.1103/physrevlett.104.181301}{\emph{Physical Review
  Letters} {\bfseries 104} (2010) }
  [\href{https://arxiv.org/abs/1001.2725}{{\ttfamily 1001.2725}}].

\bibitem{originallambda}
R.D.~Sorkin, \emph{{A Modified Sum-Over-Histories for Gravity reported in the
  article by D. Brill and L. Smolin: “Workshop on quantum gravity and new
  directions”}},  in \emph{Highlights in gravitation and cosmology:
  Proceedings of the International Conference on Gravitation and Cosmology,
  Goa, India, 14–19 December 1987}, B.R.~Iyer, A.~Kembhavi, J.V.~Narlikar and
  C.V.~Vishveshwara, eds., pp.~184--186, 1988.

\bibitem{sorkin1994role}
R.D.~Sorkin, \emph{Role of time in the sum-over-histories framework for
  gravity}, {\emph{International journal of theoretical physics} {\bfseries 33}
  (1994) 523}.

\bibitem{ahmed2004everpresent}
M.~Ahmed, S.~Dodelson, P.B.~Greene and R.~Sorkin, \emph{{Everpresent
  $\Lambda$}}, {\emph{Physical Review D} {\bfseries 69} (2004) 103523}
  [\href{https://arxiv.org/abs/astro-ph/0209274}{{\ttfamily
  astro-ph/0209274}}].

\bibitem{Das:2023hbw}
S.~Das, A.~Nasiri and Y.K.~Yazdi, \emph{{Aspects of Everpresent
  \ensuremath{\Lambda}. Part I. A~fluctuating cosmological constant from
  spacetime discreteness}},
  \href{https://doi.org/10.1088/1475-7516/2023/10/047}{\emph{JCAP} {\bfseries
  10} (2023) 047} [\href{https://arxiv.org/abs/2304.03819}{{\ttfamily
  2304.03819}}].

\bibitem{Yazdi:2023scl}
Y.K.~Yazdi, \emph{{Everything you always wanted to know about how causal set
  theory can help with open questions in cosmology, but were afraid to ask}},
  \href{https://doi.org/10.1142/S0217732323300033}{\emph{Mod. Phys. Lett. A}
  {\bfseries 39} (2024) 2330003}
  [\href{https://arxiv.org/abs/2311.14881}{{\ttfamily 2311.14881}}].

\bibitem{Abdalla:2022yfr}
E.~Abdalla et~al., \emph{{Cosmology intertwined: A review of the particle
  physics, astrophysics, and cosmology associated with the cosmological
  tensions and anomalies}},
  \href{https://doi.org/10.1016/j.jheap.2022.04.002}{\emph{JHEAp} {\bfseries
  34} (2022) 49} [\href{https://arxiv.org/abs/2203.06142}{{\ttfamily
  2203.06142}}].

\bibitem{Perivolaropoulos:2021jda}
L.~Perivolaropoulos and F.~Skara, \emph{{Challenges for
  \ensuremath{\Lambda}CDM: An update}},
  \href{https://doi.org/10.1016/j.newar.2022.101659}{\emph{New Astron. Rev.}
  {\bfseries 95} (2022) 101659}
  [\href{https://arxiv.org/abs/2105.05208}{{\ttfamily 2105.05208}}].

\bibitem{Vagnozzi:2023nrq}
S.~Vagnozzi, \emph{{Seven Hints That Early-Time New Physics Alone Is Not
  Sufficient to Solve the Hubble Tension}},
  \href{https://doi.org/10.3390/universe9090393}{\emph{Universe} {\bfseries 9}
  (2023) 393} [\href{https://arxiv.org/abs/2308.16628}{{\ttfamily
  2308.16628}}].

\bibitem{DiValentino:2021izs}
E.~Di~Valentino, O.~Mena, S.~Pan, L.~Visinelli, W.~Yang, A.~Melchiorri et~al.,
  \emph{{In the realm of the Hubble tension\textemdash{}a review of
  solutions}}, \href{https://doi.org/10.1088/1361-6382/ac086d}{\emph{Class.
  Quant. Grav.} {\bfseries 38} (2021) 153001}
  [\href{https://arxiv.org/abs/2103.01183}{{\ttfamily 2103.01183}}].

\bibitem{DESI:2024mwx}
{\scshape DESI} collaboration, \emph{{DESI 2024 VI: Cosmological Constraints
  from the Measurements of Baryon Acoustic Oscillations}},
  \href{https://arxiv.org/abs/2404.03002}{{\ttfamily 2404.03002}}.

\bibitem{Sorkin:2003bx}
R.D.~Sorkin, \emph{{Causal sets: Discrete gravity}},  in \emph{{School on
  Quantum Gravity}}, pp.~305--327, 9, 2003,
  \href{https://doi.org/10.1007/0-387-24992-3_7}{DOI}
  [\href{https://arxiv.org/abs/gr-qc/0309009}{{\ttfamily gr-qc/0309009}}].

\bibitem{Dowker:2021zel}
F.~Dowker and J.~Butterfield, \emph{{Recovering General Relativity from a
  Planck scale discrete theory of quantum gravity}},
  \href{https://arxiv.org/abs/2106.01297}{{\ttfamily 2106.01297}}.

\bibitem{Loomis:2017jhn}
S.P.~Loomis and S.~Carlip, \emph{{Suppression of non-manifold-like sets in the
  causal set path integral}},
  \href{https://doi.org/10.1088/1361-6382/aa980b}{\emph{Class. Quant. Grav.}
  {\bfseries 35} (2018) 024002}
  [\href{https://arxiv.org/abs/1709.00064}{{\ttfamily 1709.00064}}].

\bibitem{Carlip:2022nsv}
P.~Carlip, S.~Carlip and S.~Surya, \emph{{Path integral suppression of badly
  behaved causal sets}},
  \href{https://doi.org/10.1088/1361-6382/acc50c}{\emph{Class. Quant. Grav.}
  {\bfseries 40} (2023) 095004}
  [\href{https://arxiv.org/abs/2209.00327}{{\ttfamily 2209.00327}}].

\bibitem{Carlip:2023zki}
P.~Carlip, S.~Carlip and S.~Surya, \emph{{The Einstein-Hilbert Action for
  Entropically Dominant Causal Sets}},
  \href{https://arxiv.org/abs/2311.18238}{{\ttfamily 2311.18238}}.

\bibitem{papoulis2002probability}
A.~Papoulis and S.~Pillai, \emph{Probability, Random Variables, and Stochastic
  Processes}, McGraw-Hill series in electrical engineering: Communications and
  signal processing, Tata McGraw-Hill (2002).

\bibitem{petrov1995limit}
V.~Petrov, \emph{Limit Theorems of Probability Theory: Sequences of Independent
  Random Variables}, Oxford science publications, Clarendon Press (1995).

\bibitem{kingman1993poisson}
J.~Kingman, \emph{Poisson Processes}, Oxford science publications, Clarendon
  Press (1993).

\bibitem{WignerSeitz1933}
E.~Wigner and F.~Seitz, \emph{On the constitution of metallic sodium},
  \href{https://doi.org/10.1103/PhysRev.43.804}{\emph{Phys. Rev.} {\bfseries
  43} (1933) 804}.

\bibitem{Dowker:2010pf}
F.~Dowker, J.~Henson and R.~Sorkin, \emph{{Discreteness and the transmission of
  light from distant sources}},
  \href{https://doi.org/10.1103/PhysRevD.82.104048}{\emph{Phys. Rev. D}
  {\bfseries 82} (2010) 104048}
  [\href{https://arxiv.org/abs/1009.3058}{{\ttfamily 1009.3058}}].

\bibitem{BDGAction}
F.~Dowker and L.~Glaser, \emph{Causal set d’alembertians for various
  dimensions},
  \href{https://doi.org/10.1088/0264-9381/30/19/195016}{\emph{Classical and
  Quantum Gravity} {\bfseries 30} (2013) 195016}
  [\href{https://arxiv.org/abs/1305.2588}{{\ttfamily 1305.2588}}].

\bibitem{Glaser:2013xha}
L.~Glaser, \emph{{A closed form expression for the causal set
  d\textquoteright{}Alembertian}},
  \href{https://doi.org/10.1088/0264-9381/31/9/095007}{\emph{Class. Quant.
  Grav.} {\bfseries 31} (2014) 095007}
  [\href{https://arxiv.org/abs/1311.1701}{{\ttfamily 1311.1701}}].

\bibitem{Yeats:2024tne}
K.~Yeats, \emph{{Combinatorial interpretation of the coefficients of the causal
  set d'Alembertian}},  \href{https://arxiv.org/abs/2412.14036}{{\ttfamily
  2412.14036}}.

\bibitem{Nasiri:2023iwc}
A.~Nasiri, \emph{{Synge's world function applied to causal diamonds and causal
  sets}},  \href{https://arxiv.org/abs/2304.00088}{{\ttfamily 2304.00088}}.

\bibitem{deBrito:2023axj}
G.P.~de~Brito, A.~Eichhorn and C.~Pfeiffer, \emph{{Higher-order curvature
  operators in causal set quantum gravity}},
  \href{https://doi.org/10.1140/epjp/s13360-023-04202-y}{\emph{Eur. Phys. J.
  Plus} {\bfseries 138} (2023) 592}
  [\href{https://arxiv.org/abs/2301.13525}{{\ttfamily 2301.13525}}].

\bibitem{Dionthesis}
D.M.T.~Benincasa, \emph{The Action of a Causal Set}, Ph.D. thesis, 2013.

\bibitem{Aslanbeigi:2014zva}
S.~Aslanbeigi, M.~Saravani and R.D.~Sorkin, \emph{{Generalized causal set
  d`Alembertians}}, \href{https://doi.org/10.1007/JHEP06(2014)024}{\emph{JHEP}
  {\bfseries 06} (2014) 024} [\href{https://arxiv.org/abs/1403.1622}{{\ttfamily
  1403.1622}}].

\bibitem{Belenchia:2015hca}
A.~Belenchia, D.M.T.~Benincasa and F.~Dowker, \emph{{The continuum limit of a
  4-dimensional causal set scalar d\textquoteright{}Alembertian}},
  \href{https://doi.org/10.1088/0264-9381/33/24/245018}{\emph{Class. Quant.
  Grav.} {\bfseries 33} (2016) 245018}
  [\href{https://arxiv.org/abs/1510.04656}{{\ttfamily 1510.04656}}].

\bibitem{Das:2023rvg}
S.~Das, A.~Nasiri and Y.K.~Yazdi, \emph{{Aspects of Everpresent $\Lambda$ (II):
  Cosmological Tests of Current Models}},
  \href{https://arxiv.org/abs/2307.13743}{{\ttfamily 2307.13743}}.

\bibitem{finiterho}
F.~Dowker, \emph{{Boundary contributions in the causal set action}},
  \href{https://doi.org/10.1088/1361-6382/abc2fd}{\emph{Gen. Rel. Grav.}
  {\bfseries 38} (2021) 075018}
  [\href{https://arxiv.org/abs/2007.13206}{{\ttfamily 2007.13206}}].

\bibitem{MachetWang2020}
L.~Machet and J.~Wang, \emph{On the continuum limit of
  benincasa–dowker–glaser causal set action},
  \href{https://doi.org/10.1088/1361-6382/abc274}{\emph{Classical and Quantum
  Gravity} {\bfseries 38} (2020) 015010}
  [\href{https://arxiv.org/abs/2007.13192}{{\ttfamily 2007.13192}}].

\end{thebibliography}\endgroup



\providecommand{\href}[2]{#2}\begingroup\raggedright\endgroup
\bibliographystyle{JHEP}

\end{document}